\documentclass[12pt, draftcls]{IEEEtran}
\onecolumn
\pdfoutput=1
\ifCLASSINFOpdf
\else
\fi
\usepackage{setspace}
\usepackage{amssymb}
\usepackage[usenames]{color}
\usepackage[cmex10]{amsmath}
\usepackage[pdftex]{graphicx}
\usepackage{cite}

\begin{document}
%
\title{Optimal Sensor Placement for Intruder Detection}
%
%
%

\author{Waseem A. Malik, Nuno C. Martins, and Ananthram Swami}

%
%


\maketitle

%
\IEEEpeerreviewmaketitle

\section{Introduction}
%
%
%
%
\IEEEPARstart{T}he fields of detection, resource allocation, and security have seen tremendous research activity in the past few decades. Different frameworks have been designed to tackle various problems using a host of techniques from statistics, engineering, and economics. The fields of centralized and decentralized detection, see \cite{Tsitsiklis} for a review, have been a major research focus in the communications, controls, and statistics communities and are now considered mature fields of research. Our main focus in this paper is to develop and pursue research ideas in the intersection of these exciting fields. In this paper, we operate within the centralized detection framework under which complete observations are available to the decision makers. This preference for a relatively simple and more developed framework enables us to fully extract the benefits offered by the diversity of these research ideas.
\par
We consider the centralized detection of an intruder, whose location is modeled as uniform across a specified set of points. A team of sensors is tasked to make observations which are completely reported to a centralized decision making authority. The challenge is to optimally place the sensors, before measurements are made, and to compute an optimal decision rule for detecting the location of the intruder. Simplifying assumptions like a uniform prior distribution and conditional independence of observations are made. It should be noted that sensors observations are noisy as a result of their limited accuracy. The limited accuracy of the sensors makes the conditional independence assumption practically reasonable \cite{Francolin}.         
\par
We exploit majorization theory, see \cite{Arnold} and \cite{Olkin} for a review, within the framework of this paper to establish general laws governing the jointly optimal detection and placement policies. Majorization is an important mathematical technique for partial ordering of vectors of real numbers. It is widely used in mathematical statistics and has recently been applied to solve challenging problems in controls \cite{Lipsa} and communications \cite{Hajek}. 
\par
\textbf{The following notation is adopted:}
\begin{itemize}
\item Vectors are represented as $\bar{x} = (x_{_{1}},...,x_{_{k}})$, where the vector length would be obvious from the context. 
\item The observation vector received by the fusion center is denoted by $\displaystyle \bar{y} = (y_{_{1}},...,y_{_{M}})$, the position vector of the sensors is denoted by $\displaystyle \bar{u} = (u_{_{1}},...,u_{_{M}})$, and the placement vector for the sensors is denoted by $\displaystyle \bar{v} = (v_{_{1}},...,v_{_{N}})$. 
\item Random variables are represented by bold face capital letters, such as \textbf{X} used to represent the position of the intruder. Realizations are represented using small letters.
\item The probability of error is denoted by $\displaystyle P_{e}$, the prior distribution by $\pi$, the probability density of a random variable $\textbf{A}$ if it exists by $p(a)$, the joint probability density of \textbf{A} and \textbf{B} by $p(a,b)$, and the conditional probability density of \textbf{A} given \textbf{B} by $p_{_{b}}(a)$. 
\item The observation space is denoted by $\Omega$.
\item Capital letters $M$ and $N$ are used to denote the number of sensors and the number of placement points respectively.
\end{itemize}
\par
The paper is organized into seven main sections. In Section II we provide the technical preliminaries and in Section III we provide the problem formulation. The main results of this paper are presented in Section IV and Section V. In Section IV, we use mathematical induction to establish general principles regarding the strict optimality of the uniform placement. In Section V, majorization theory is exploited to formalize and conjecture important properties of the optimal solution. Simulations and conclusions are presented in Section VI and Section VII.  
\section{Technical Preliminaries}
In this section, we present definitions of the main concepts to be utilized in the sequel. 
\par
\textit{Definition 2.1:} \textbf{(Partition Function)} The partition function $f(N)$ of a positive integer $N$ is defined as the number of ways in which $N$ can be written as a sum of positive integers. For example $4$ can be written as $ 4,3+1,2+2,2+1+1,1+1+1+1$ and $f(4)=5$. 
\par   
\textit{Definition 2.2:} \textbf{(Majorization}\cite{Olkin}\textbf{)} For $\bar{x}, \bar{y} \in \Re^{N}$, $\bar{x}$ is said to be majorized by $\bar{y}$, $\bar{y} \succ \bar{x}$, provided that: ~
$\displaystyle \sum_{j=1}^{n}x_{_{[j]}} =\sum_{j=1}^{n}y_{_{[j]}}$,~$\displaystyle \sum_{j=1}^{k} x_{_{[j]}} \leq \sum_{j=1}^{k}y_{_{[j]}}, k=1,...,n-1$
where $\displaystyle x_{_{[j]}},j=1,...,n$ represents the elements of $\bar{x}$ in nonincreasing order, $x_{_{[1]}} \geq,...,\geq x_{_{[n]}}$.
\\
\par
\textit{Definition 2.3:} \textbf{(Majorization Based Placement Scale)} Sensor placements $\bar{\alpha},\bar{\beta},$ and $\bar{\gamma}$ can be placed on a majorization based scale $(\bar{\alpha},\bar{\beta},\bar{\gamma})$ if the following majorization ordering exists: $\bar{\alpha} \succ \bar{\beta} \succ \bar{\gamma}$. The placement $\bar{\alpha}$ is said to be at the highest level on this scale followed by $\bar{\beta}$ and $\bar{\gamma}$. 

\section{Problem Formulation}
\par
In this section a rigorous problem formulation along with a precise problem statement is presented.
\subsection{Sensor Placement, Data Collection, and Performance Criterion}
\par
Consider a team of $M$ identical sensors tasked with detecting the location of an intruder $\textbf{X}$ which can occur on a set of points $\displaystyle \{1,..,N\}$, with a uniform distribution. It is assumed that $M \leq N$. The sensors are placed using a specific sensor position vector (or placement vector), sensors make observations, and report them directly to a fusion center. The observations made by the sensors are assumed to be conditionally independent given the true position of the intruder. The fusion center collects these observations and computes an estimate of the position of the intruder. It should be noted that no quantization of information occurs at the sensors and observations are completely reported to the fusion center. The performance criterion minimized at the fusion center is the probability of error in detecting the actual location of the intruder. The main problem is to design jointly optimal location detection and sensor placement policies.
\subsection{Observation Model}
\par
Let \textbf{Y}$_{k}$ be the measurement obtained by sensor $k$, $k = 1,...,M$. The random variable \textbf{Y}$_{k}$ either takes the value $1$ (Intruder present) or the value $0$ (Intruder not present). The observation model is presented below:
\begin{equation}
Pr(\textbf{Y}_{k}=1|{\textbf{X}=u_{k}}) = P_{_{D}} \nonumber
\end{equation}
\begin{equation}
Pr(\textbf{Y}_{k}=1|{\textbf{X}\neq u_{k}}) = P_{_{F}} \nonumber
\end{equation}
\par
where $P_{_{D}}$ and $P_{_{F}}$ are the detection and false alarm probabilities of the sensors. We assume that these values have been provided by the manufacturer of these sensors and are well known. It should be noted that the sensors can only make measurements at the point on which they are placed and cannot infer any information about the presence of the intruder at ther points.
\subsection{M-ary Hypothesis Testing}
The problem is modeled as an M-ary hypothesis testing problem \cite{Trees} under the Bayesian formulation. Under a uniform prior distribution, $\pi_{i} = \frac{1}{N}$, the probability of error at the fusion center for a specific sensor position vector $(u_{_{1}},...,u_{_{M}})$ is given by \cite{Poor}:
\begin{equation}
\displaystyle P_{e}(u_{_{1}},...,u_{_{M}}) =\min_{\delta}\frac{1}{N} \sum_{i=1}^{N} \sum_{\bar{y} \in \Gamma_{i}} \sum_{j=1,j \neq i}^{N} p_{j}(\bar{y}) \nonumber
\end{equation}
where $\Gamma_{i}, i=1,...,N$ defines a partition of the observation space $\Omega$ such that $\displaystyle \Omega =  \bigcup_{i=1}^{N}\Gamma_{i}$ where $\Gamma_{i} = \{ \bar{y} \in \Omega | \delta(y) = H_{i}\}$. Here $p_{j}(\bar{y})$ is the joint probability mass function (pmf) for the random variables (\textbf{Y}$_{1}$,...,\textbf{Y}$_{M}$) given that \textbf{X}$=j$. It should be noted that the joint pmf depends on the specific sensor position vector (or placement vector). The sensor position vector, of length $M$, indicates the points at which each sensor is placed whereas the sensor placement vector, of length $N$, indicates how many sensors are placed at each point. It should be noted that:
\begin{equation}
\displaystyle 1\leq u_{_{j}} \leq N, ~j=1,...,M \hspace{2cm}0\leq v_{_{k}} \leq M, ~k=1,...,N \nonumber
\end{equation}
Since we consider identical sensors using the same observation model it is sufficient to consider the number of sensors which are placed at each point. The placement vector for the sensors is more convenient to use than the sensor position vector. In the sequel we will utilize the sensor placement vector $\bar{v} = (v_{_{1}},...,v_{_{N}})$ instead of the sensor position vector $\bar{u} = (u_{_{1}},...,u_{_{M}})$ for the calculation of the probabilities of error.
\subsection{Calculation of Conditional Probabilities}
Since the observations are conditionally independent:
\begin{equation}
\displaystyle p_{j}(\bar{y}) = \prod_{i=1}^{M}p_{j}(y_{i}), ~~ j = 1,...,N \nonumber
\end{equation}
\par
We will only consider placements of the form $v_{_{1}} \geq v_{_{2}}\geq ....\geq v_{_{N}}$. For the case M=N=5, consider the two sensor placements $(v_{_{1}},...,v_{_{5}}) = (2,1,1,1,0)$ and $(v_{_{1}},...,v_{_{5}})=(1,2,1,1,0)$. Since the intruder occurs at any of the five points with a uniform distribution so clearly $P_{e}$(2,1,1,1,0) = $P_{e}$(1,2,1,1,0). Therefore, we will only consider (2,1,1,1,0) and not other similar permutations of the placement vector in which one point has two sensors, three points have one sensor each, and one point has no sensor placed on it. For notational convenience we assume that during deployment the sensors are placed in an ascending order, i-e first $v_{_{1}}$ sensors are deployed on point 1 followed by $v_{_{2}}$ sensors on point 2 and, so on. We hasten to add that this deployment policy is really not restrictive; it merely amounts to a relabeling of the observation areas or locations: location 1 is the one with the most number of sensors, and so on. Thus location $j$ need not be `adjacent' to location $j-1$ or $j+1$. This deployment policy ensures that observations $(y_{_{1}},...,y_{v_{_{1}}})$ are made by sensors operating at point 1 and similarly we can interpret the observations $(y_{_{v_{_{1}}+1}},...,y_{_{M}})$. With $M \leq N$ and the aforementioned assumptions in place, it is clear that $(v_{_{M+1}},...,v_{_{N}}) = (0,...,0)$. Note that depending on the values of the $v_{i}$'s it is possible that $v_{_{k}}=0$, for $k = k_{0},...,M$, as well. The conditional probability is calculated as follows:
\begin{equation}
\displaystyle p_{j}(\bar{y}) = \big\{(P_{_{F}})^{^{(y_{_{1}}+...+y_{_{M}})}}\times(1-P_{_{F}})^{^{M - (y_{_{1}}+...+y_{_{M}})}}\chi_{(v_{_{j}} = 0)}\big\} ~+~ \big\{(P_{_{D}})^{^{a_{_{j}}}}(1-P_{_{D}})^{^{v_{_{j}}-a_{_{j}}}}(P_{_{F}})^{^{(y_{_{1}}+...+y_{_{M}}) - a_{_{j}}}}\times
\nonumber 
\end{equation}
\begin{equation}
\label{pj} 
\displaystyle (1-P_{_{F}})^{^{M -(y_{_{1}}+...+y_{_{M}})-(v_{_{j}}-a_{_{j}})}}\chi_{(v_{_{j}} \neq 0)}\big\} \hspace{3cm}
\end{equation}
where $a_{_{j}}$ is the number of accurate `alarmed' sensors at location $j$ and is given by: 
\begin{equation}
\displaystyle a_{_{j}}= y_{_{1}} + ... + y_{_{v_{{j}}}}, j = 1 \nonumber
\end{equation}
\begin{equation}
\displaystyle a_{_{j}} = y_{_{(v_{_{1}}+...+v_{_{j-1}}) + 1}}+...+~y_{_{(v_{_{1}}+...+v_{_{j-1}}) + v_{_{j}}}}, j=2,...,N \nonumber
\end{equation}
\textit{E. Problem Statement}
\par
Given $M$ sensors and $N$ placement points we want to solve the following optimization problem:
\begin{equation}
P_{e}^{*} =\min_{\delta}\min_{v_{_{1}},...,v_{_{M}}}\frac{1}{N} \sum_{i=1}^{N} \sum_{\bar{y} \in \Gamma_{i}} \sum_{j=1,j \neq i}^{N} p_{j}(\bar{y}) \nonumber
\end{equation}
subject to the constraint:
\begin{equation}
(v_{_{1}},...,v_{_{M}}) \in \Lambda^{M} \nonumber
\end{equation}
where the set $\Lambda^{M}$ is given by an integer partition of $M$. Utilizing the notion of the partition function defined in Section II, the size of the constraint set $\Lambda^{M}$ is given by:
\begin{equation}
\displaystyle |\Lambda^{M}| = f(M) \nonumber
\end{equation}
The partition function $f(M)$ of an integer $M$ can be expressed asymptotically as \cite{Ramanujan}:
\begin{equation}
f(M)\approx\frac{1}{4M\sqrt{3}}\times e^{{\pi\sqrt{\frac{2M}{3}}}}~ as ~M \rightarrow \infty \nonumber
\end{equation}
This expression was first proved by S. Ramanujan and G. Hardy in 1918. The form of the cost function and the exponential complexity of the constraint set make this problem difficult to solve for a general (M,N). We prove important solution properties for this problem in Section IV and Section V. Such properties are the basis of any suboptimal scheme designed to solve this problem. For any placement $(v_{_{1}},...,v_{_{M}})$ the probability of error is given by:
\begin{equation}
\displaystyle P_{e}(v_{_{1}},...,v_{_{M}}) =\min_{\delta}\frac{1}{N} \sum_{i=1}^{N} \sum_{\bar{y} \in \Gamma_{i}} \sum_{j=1,j \neq i}^{N} p_{j}(\bar{y}) \nonumber
\end{equation}
Minimization is carried out by optimally selecting $\Gamma_{i}, i=1,...,N$:
\begin{equation}
\Gamma_{i}^{*} = \bigg\{ \bar{y} \in \Omega | \sum_{j=1,j\neq i}^{N} p_{j}(\bar{y})\leq\sum_{j=1,j\neq k}^{N} p_{j}(\bar{y}), i\neq k \bigg\} \nonumber
\end{equation}
which can be simplified to give:
\begin{equation}
\label{Gi*}
\Gamma_{i}^{*} = \bigg\{ \bar{y} \in \Omega ~|~ p_{i}(\bar{y}) = \max_{k=1,...,N} p_{k}(\bar{y})\bigg\}
\end{equation}  
Using (\ref{Gi*}) and the uniform distribution of the priors the optimal decision rule is given by:
\begin{equation}
\label{delta*}
\delta^{*}(\bar{y}) = \left\{ H_{i} ~|~ \pi_{\bar{y}}({H_{i}}) = \max_{k=1,..,N} \pi_{\bar{y}}({H_{k}})\right\}, \bar{y} \in \Omega
\end{equation}
where $\pi_{\bar{y}}({H})$ is the posterior distribution. Note that the optimal decision rule in (\ref{delta*}) is the maximum a posteriori probability (MAP) rule. Since $\delta^{*}$ is optimal for any arbitrary placement so the optimization problem can be rewritten as follows:
\begin{equation}
\label{Pe*}
P_{e}^{*} =\min_{v_{_{1}},...,v_{_{M}}}\frac{1}{N} \sum_{i=1}^{N} \sum_{\bar{y} \in \Gamma_{i}^{*}} \sum_{j=1,j \neq i}^{N} p_{j}(\bar{y})
\end{equation}
subject to the constraint:
\begin{equation}
(v_{_{1}},...,v_{_{M}}) \in \Lambda^{M} \nonumber
\end{equation}
\section{Characterization of the Optimal Solution}
We first provide a definition of the strict optimality of sensor placement and then present the results of this section.
\par
\textit{Definition 4.1:} \textbf{(Strict Optimality of Placement)} We call a placement $(v_{_{1}}^{'},...,v_{_{M}}^{'})$ strictly optimal on the $(P_{_{F}},P_{_{D}})$ plane, if there exist $P_{_{D}}^{'},P_{_{F}}^{'}$ such that $\displaystyle P_{e}(v_{_{1}},...,v_{_{M}})\big|_{P_{_{D}}=P_{_{D}}^{'}, P_{_{F}}=P_{_{F}}^{'}} > P_{e}(v_{_{1}}^{'},...,v_{_{M}}^{'})\big|_{P_{_{D}}=P_{_{D}}^{'}, P_{_{F}}=P_{_{F}}^{'}}$ for any $(v_{_{1}},...,v_{_{M}}) \in \Lambda^{M}$.  
\\
\par
It should be noted that an optimal placement for (\ref{Pe*}) is an optimal solution to the aforementioned problem. Using  Definition 4.1, the main result of this section is described as follows:
\par
\textbf{\textit{Theorem 4.1:}} For the case $(M=N)$, the uniform sensor placement $(v_{_{1}},...,v_{_{M}}) = (1,...,1)$ is not strictly optimal for any value of $P_{_{D}},P_{_{F}}$ on the $(P_{_{F}},P_{_{D}})$ plane.
\par
\begin{proof}
We will prove this result by using induction. For notational convenience we will use the following form of $P_{e}(v_{_{1}},...,v_{_{M}})$ in the proof:
\begin{equation}
\label{Pewf}
\displaystyle P_{e}(v_{_{1}},...,v_{_{M}})= \frac{1}{N} \sum_{\bar{y}\in \{0,1\}^{M}}\min_{i=1,..,N} \{\sum_{j=1,j\neq i}^{N} p_{j}(\bar{y})\}
\end{equation}
For $M$ sensors we have $2^{M}$ observations so $\{0,1\}^{M}$ is used to denote the observation space $\Omega$. 
\par
First consider the case M=N=2. In this case (\ref{Pewf}) can be simplified to give:
\begin{multline}
\label{Pev1v2}
\displaystyle P_{e}(v_{_{1}}, v_{_{2}})=\frac{1}{2}[\min\{p_{_{1}}(0,0),p_{_{2}}(0,0)\} + \min\{p_{_{1}}(0,1),\\
p_{_{2}}(0,1)\}+\min\{p_{_{1}}(1,0),p_{_{2}}(1,0)\} + \min\{p_{_{1}}(1,1),p_{_{2}}(1,1)\}]
\end{multline}
Using equation (\ref{pj}) the conditional probabilities for the $(v_{_{1}},v_{_{2}})=(1,1)$ and $(v_{_{1}},v_{_{2}})=(2,0)$ placements are given as follows: 
\begin{align}
\label{p12v1v2}
p_{_{1}}(y_{_{1}},y_{_{2}})\big|_{(1,1)}&=(P_{_{D}})^{y_{_{1}}}(1-P_{_{D}})^{1-y_{_{1}}}(P_{_{F}})^{y_{_{2}}}(1-P_{_{F}})^{1-y_{_{2}}}\nonumber\\
p_{_{2}}(y_{_{1}},y_{_{2}})\big|_{(1,1)}&=(P_{_{D}})^{y_{_{2}}}(1-P_{_{D}})^{1-y_{_{2}}}(P_{_{F}})^{y_{_{1}}}(1-P_{_{F}})^{1-y_{_{1}}}\nonumber\\
p_{_{1}}(y_{_{1}},y_{_{2}})\big|_{(2,0)}&=(P_{_{D}})^{y_{_{1}}+y_{_{2}}}(1-P_{_{D}})^{2-y_{_{1}}-y_{_{2}}}\nonumber\\
p_{_{2}}(y_{_{1}},y_{_{2}})\big|_{(2,0)}&=(P_{_{F}})^{y_{_{1}}+y_{_{2}}}(1-P_{_{F}})^{2-y_{_{1}}-y_{_{2}}}
\end{align}
Plugging (\ref{p12v1v2}) into (\ref{Pev1v2}) we get:
\begin{align}
\label{Pe11Pe20}
\displaystyle P_{e}(1,1)& = \frac{1}{2}\bigg[(1-P_{_{D}})(1-P_{_{F}}) + P_{_{D}}P_{_{F}}    +2\min\{P_{_{F}}-P_{_{D}}P_{_{F}},P_{_{D}}-P_{_{D}}P_{_{F}}\}\bigg]\nonumber\\
\displaystyle P_{e}(2,0)& = \frac{1}{2} \bigg[\min\{(1-P_{_{D}})^{2},(1-P_{_{F}})^{2}\} +\min\{P_{_{D}}^{2},P_{_{F}}^{2}\} +2\min\{P_{_{D}}(1-P_{_{D}}),P_{_{F}}(1-P_{_{F}})\}\bigg]
\end{align}
\\
First consider the case $\bf P_{_{D}}>P_{_{F}}$:
\par
For $P_{_{D}} > P_{_{F}}, P_{_{D}}(1-P_{_{D}}) > P_{_{F}}(1-P_{_{F}})$:
\begin{align}
\displaystyle P_{e}(1,1)& = \frac{1}{2}\big[1 + P_{_{F}}-P_{_{D}}\big]\nonumber \\
\displaystyle P_{e}(2,0)& = \big[\frac{1}{2} + (P_{_{F}}-P_{_{D}}) + \frac{(P_{_{D}}^{2}-P_{_{F}}^{2})}{2}\big] \nonumber
\end{align}
where $(P_{_{D}}^{2}-P_{_{F}}^{2})-(P_{_{D}}-P_{_{F}})<0 \Rightarrow P_{e}(2,0) < P_{e}(1,1)$
\par
For $P_{_{D}} > P_{_{F}}, P_{_{D}}(1-P_{_{D}}) = P_{_{F}}(1-P_{_{F}})$:
\begin{equation}
P_{e}(1,1) = P_{e}(2,0) = \frac{1}{2}\big[1 + P_{_{F}}-P_{_{D}}\big] \nonumber
\end{equation}
\par
For $P_{_{D}} > P_{_{F}}, P_{_{D}}(1-P_{_{D}}) < P_{_{F}}(1-P_{_{F}})$:
\begin{equation}
\displaystyle P_{e}(1,1) = \frac{1}{2}\big[1 + (P_{_{F}}-P_{_{D}})\big],P_{e}(2,0) = \frac{1}{2}\big[1 + (P_{_{F}}^{2}-P_{_{D}}^{2})\big] \nonumber
\end{equation}
where $(P_{_{F}}^{2} - P_{_{D}}^{2}) < (P_{_{F}} - P_{_{D}}) \Rightarrow P_{e}(2,0) < P_{e}(1,1)$ 
\\
\newline
For the case $\bf P_{_{D}}=P_{_{F}}$ we have:
\begin{equation}
\displaystyle P_{e}(2,0) = P_{e}(1,1) = \frac{1}{2}  \nonumber
\end{equation}
For the purposes of this proof it is sufficient to only consider the case $P_{_{D}} \geq P_{_{F}}$. Further details regarding this sufficiency are provided at the end of the proof. This proves that uniform placement is not strictly optimal for $M$=$N$=$2$. 
\par
Assume that the result holds for $M$=$N$=$k$: 
\begin{equation}
\label{Istep}
\displaystyle P_{e}\big(\underbrace{1,...,1}_{k}\big) \geq P_{e}\big(\underbrace{2,1,...,1,0}_{k}\big)
\end{equation}
\par
\textbf{Calculation of} $\bf \displaystyle P_{e}\big(\underbrace{1,...,1}_{k}\big)$:
\par
Using equations (\ref{pj}) and (\ref{Pewf}) we get the following expression:
\begin{align}
\label{Pek}
\displaystyle P_{e}\big(\underbrace{1,...,1}_{k}\big) =\frac{1}{k}\bigg[\sum_{\bar{y}\in\{0,1\}^{k}} \min_{i=1,..,k} \big\{ A_{i}(\bar{y}) \big\}\bigg]\hspace{3cm}\nonumber \\
\displaystyle  A_{i}(\bar{y}) = \sum_{j=1,j\neq i}^{k}\big\{ (P_{_{D}})^{y_{_{j}}}(1-P_{_{D}})^{1-y_{_{j}}}(P_{_{F}})^{(y_{_{1}}+...+y_{_{k}})-y_{_{j}}}(1-P_{_{F}})^{k-1-(y_{_{1}}+...+y_{_{k}}) +y_{_{j}}}\big\}
\end{align}
\par
\textbf{Calculation of} $\bf \displaystyle P_{e}\big(\underbrace{2,1,...,1,0}_{k}\big)$:
\par
Using equations (\ref{pj}) and (\ref{Pewf}) we get the following expression:
\begin{equation}
\label{Pe2k}
\displaystyle P_{e}\big(\underbrace{2,1,...,1,0}_{k}\big) =\frac{1}{k}\bigg[\sum_{\bar{y}\in\{0,1\}^{k}} \min_{i=1,..,k} \big\{ B_{i}(\bar{y}) \big\}\bigg]
\end{equation}
where $B_{i}(\bar{y}), i=1,..,k$ is given by:
\begin{align}
\label{Bis}
\displaystyle & B_{1}(\bar{y}) =\bigg[\sum_{j=2}^{k-1} (P_{_{D}})^{y_{_{j+1}}}(1-P_{_{D}})^{1-y_{_{j+1}}}(P_{_{F}})^{(y_{_{1}}+...+y_{_{k}}) - y_{_{j+1}}}(1-P_{_{F}})^{k-1-(y_{_{1}}+...+y_{_{k}}) + y_{_{j+1}}}\bigg] \nonumber \\  
\displaystyle &+(P_{_{F}})^{(y_{_{1}}+..+y_{_{k}})}(1-P_{_{F}})^{k-(y_{_{1}}+..+y_{_{k}})} \nonumber\\
\displaystyle & B_{i}(\bar{y}) = \bigg[\sum_{j=2,j\neq i}^{k-1} (P_{_{D}})^{y_{_{j+1}}}(1-P_{_{D}})^{1-y_{_{j+1}}}(P_{_{F}})^{(y_{_{1}}+...+y_{_{k}}) - y_{_{j+1}}}(1-P_{_{F}})^{k-1-(y_{_{1}}+...+y_{_{k}}) + y_{_{j+1}}}\bigg] \nonumber \\
\displaystyle &+ (P_{_{F}})^{(y_{_{1}}+..+y_{_{k}})}(1-P_{_{F}})^{k-(y_{_{1}}+..+y_{_{k}})}+(P_{_{D}})^{y_{_{1}}+y_{_{2}}}(1-P_{_{D}})^{2-y_{_{1}}-y_{_{2}}}(P_{_{F}})^{(y_{_{3}}+...+y_{_{k}})}(1-P_{_{F}})^{k-2-(y_{_{3}}+...+y_{_{k}})}\nonumber \\
\displaystyle &,i=2,...,k-1 \nonumber\\
\displaystyle & B_{k}(\bar{y}) = \bigg[\sum_{j=2}^{k-1} (P_{_{D}})^{y_{_{j+1}}}(1-P_{_{D}})^{1-y_{_{j+1}}}(P_{_{F}})^{(y_{_{1}}+...+y_{_{k}}) - y_{_{j+1}}}(1-P_{_{F}})^{k-1-(y_{_{1}}+...+y_{_{k}}) + y_{_{j+1}}}\bigg] + \nonumber \\
\displaystyle &  +(P_{_{D}})^{y_{_{1}}+y_{_{2}}}(1-P_{_{D}})^{2-y_{_{1}}-y_{_{2}}}(P_{_{F}})^{(y_{_{3}}+...+y_{_{k}})}(1-P_{_{F}})^{k-2-(y_{_{3}}+...+y_{_{k}})}
\end{align}
Using equation (\ref{Istep}) we can conclude that:
\begin{equation}
\displaystyle \bigg[\sum_{\bar{y}\in\{0,1\}^{k}} \min_{i=1,..,k} \big\{ A_{i}(\bar{y}) \big\} - \sum_{\bar{y}\in\{0,1\}^{k}} \min_{i=1,..,k} \big\{ B_{i}(\bar{y}) \big\}\bigg] \geq 0 \nonumber
\end{equation}
In order to prove the Theorem we need to show that:
\begin{equation}
\displaystyle P_{e}\big(\underbrace{1,...,1}_{k+1}\big) \geq P_{e}\big(\underbrace{2,1,...,1,0}_{k+1}\big)\nonumber
\end{equation}
\par
\textbf{Calculation of} $\bf \displaystyle P_{e}\big(\underbrace{1,...,1}_{k+1}\big)$:
\par
Using equations (\ref{pj}), (\ref{Pewf}) and (\ref{Pek}) we get the following expression:
\begin{align}
\label{Pek+1}
\displaystyle & P_{e}\big(\underbrace{1,...,1}_{k+1}\big) = \frac{1}{k+1}\bigg[ \sum_{\bar{y}\in\{0,1\}^{k}}\bigg\{ \min  \big\{ (1-P_{_{F}})A_{1}(\bar{y}) + c(\bar{y}),.....,(1-P_{_{F}})A_{k}(\bar{y}) + c(\bar{y}),    \nonumber \\
\displaystyle & (1-P_{_{F}})A_{k}(\bar{y})+ c^{*}(\bar{y})\big\}+\min \big\{ (P_{_{F}}) A_{1}(\bar{y}) +d(\bar{y}),.....,(P_{_{F}})A_{k}(\bar{y}) + d(\bar{y}), (P_{_{F}})A_{k}(\bar{y}) + d^{*}(\bar{y})\big\}\bigg\}\bigg]
\end{align}
where $c(\bar{y}),c^{*}(\bar{y}),d(\bar{y}),d^{*}(\bar{y})$ are provided below:
\begin{align}
\label{cc*dd*}
\displaystyle & c(\bar{y}) = (1-P_{_{D}})(P_{_{F}})^{(y_{_{1}}+...+y_{_{k}})}(1-P_{_{F}})^{k-(y_{_{1}}+...+y_{_{k}})} \nonumber \\
\displaystyle & c^{*}(\bar{y}) = (P_{_{D}})^{y_{_{k}}}(1-P_{_{D}})^{1-y_{_{k}}}(P_{_{F}})^{(y_{_{1}}+...+y_{_{k-1}})} (1-P_{_{F}})^{k-(y_{_{1}}+...+y_{_{k-1}})} \nonumber \\
\displaystyle & d(\bar{y}) = P_{_{D}}(P_{_{F}})^{(y_{_{1}}+...+y_{_{k}})}(1-P_{_{F}})^{k-(y_{_{1}}+...+y_{_{k}})} \nonumber \\
\displaystyle & d^{*}(\bar{y}) = (P_{_{D}})^{y_{_{k}}}(1-P_{_{D}})^{1-y_{_{k}}}(P_{_{F}})^{y_{_{1}}+..+y_{_{k-1}}+1} (1-P_{_{F}})^{k-1-(y_{_{1}}+..+y_{_{k-1}})}
\end{align}
\par
\textbf{Calculation of} $\bf \displaystyle P_{e}\big(\underbrace{2,1,...,1,0}_{k+1}\big)$:
\par
Using equations (\ref{pj}), (\ref{Pewf}) and (\ref{Pe2k}) we get the following expression:
\begin{align}
\label{Pe2k+1}
\displaystyle & P_{e}\big(\underbrace{2,1,...,1,0}_{k+1}\big) = \frac{1}{k+1}\bigg[ \sum_{\bar{y}\in\{0,1\}^{k}}\bigg\{ \min  \big\{ (1-P_{_{F}})B_{1}(\bar{y}) + c(\bar{y}),.....,(1-P_{_{F}})B_{k-1}(\bar{y}) + c(\bar{y}),  \nonumber\\ 
\displaystyle &(1-P_{_{F}})B_{k-1}(\bar{y}) + c^{*}(\bar{y}),(1-P_{_{F}})B_{k}(\bar{y})+ c(\bar{y})\big\} + \min\big\{(P_{_{F}})B_{1}(\bar{y})+   d(\bar{y}),.....,(P_{_{F}})B_{k-1}(\bar{y})+d(\bar{y}), \nonumber \\
\displaystyle & (P_{_{F}})B_{k-1}(\bar{y})+d^{*}(\bar{y}),(P_{_{F}})B_{k}(\bar{y})+d(\bar{y})\big\}\bigg\}\bigg]
\end{align}
If $P_{_{D}}=1$ or $P_{_{F}}=0$ then $c^{*}(\bar{y}) \geq c(\bar{y})$. If $P_{_{D}}<1$ and $P_{_{F}}>0$ then $c^{*}(\bar{y})$ can be expressed in terms of $c(\bar{y})$ as follows:
\begin{equation}
\label{c*}
\displaystyle c^{*}(\bar{y})= c(\bar{y})\left(\frac{P_{_{D}}(1-P_{_{F}})}{P_{_{F}}(1-P_{_{D}})}\right)^{y_{_{k}}}
\end{equation}
First consider the case where we have good sensors i-e sensors for which $\displaystyle \bf P_{_{D}}\geq P_{_{F}}$. From equation (\ref{c*}) we have that $c^{*}(\bar{y}) \geq c(\bar{y})$. Therefore, we can eliminate the terms involving $c^{*}(\bar{y})$ from equations (\ref{Pek+1}) and (\ref{Pe2k+1}). We will first consider the situations where we can eliminate the terms involving $d^{*}(\bar{y})$ from equation (\ref{Pek+1}) and then proceed by doing a similar analysis for equation (\ref{Pe2k+1}). If $y_{_{k}}=1$, then $d^{*}(\bar{y})=d(\bar{y})$. Using equations (\ref{Pek}), (\ref{cc*dd*}), and $(y_{_{k-1}},y_{_{k}})=(0,0)$ we obtain:
\begin{align}
\label{e36}
\displaystyle &\big(P_{_{F}}A_{k}(\bar{y}) + d^{*}(y)\big) - \big(P_{_{F}}A_{k-1}(\bar{y})-d(\bar{y})\big)= (P_{_{F}})^{(y_{_{1}}+...+y_{_{k-2}})}(1-P_{_{F}})^{k-1-(y_{_{1}}+...+y_{_{k-2}})}(P_{_{F}}-P_{_{D}}) \leq 0
\end{align}
From equations (\ref{Pek}), (\ref{cc*dd*}), and $(y_{_{k-1}},y_{_{k}})=(1,0)$ we obtain:
\begin{equation}
\label{e37}
\displaystyle (P_{_{F}})A_{k}(\bar{y}) + d^{*}(\bar{y}) = (P_{_{F}})A_{k-1}(\bar{y}) + d(\bar{y})
\end{equation}
Similarly it can be shown for $(y_{_{j-1}},y_{_{k}}) = (1,0)$:
\begin{equation}
\label{e38}
\displaystyle (P_{_{F}})A_{k}(\bar{y}) + d^{*}(\bar{y}) = (P_{_{F}})A_{j-1}(\bar{y}) + d(\bar{y}), j = 2,...,k
\end{equation}
So for $\bar{y} \in \{0,1\}^{k}/(0,...,0)$ we can eliminate the term $(P_{_{F}})A_{k}(\bar{y}) + d^{*}(\bar{y})$ from equation (\ref{Pek+1}). For the observation $\bar{y}=(0,...,0), \min\big\{(P_{_{F}})A_{1}(\bar{y})+d(\bar{y}),.....,(P_{_{F}})A_{k}(\bar{y})+d(\bar{y}),(P_{_{F}})A_{k}(\bar{y})+d^{*}(\bar{y})\big\} = (P_{_{F}})A_{k}(\bar{y})+d^{*}(\bar{y})$. Let $a_{_{1}},...,a_{_{2^{k}}}$ be given as follows:
\begin{equation}
\label{sau}
\displaystyle \sum_{u=1}^{2^{k}} a_{_{u}}=\sum_{\bar{y}\in\{0,1\}^{k}}\min_{i=1,..,k}\{P_{_{F}}A_{i}(\bar{y})+d(\bar{y})\}
\end{equation}
where $\displaystyle a_{_{u}} = \min_{i=1,...,k}\{P_{_{F}}A_{i}(y_{_{1}},...,y_{_{k}})+d(y_{_{1}},...,y_{_{k}})\}$ and $u = dec(y_{_{1}}...y_{_{k}})+1$. Here $\displaystyle dec(y_{_{1}}...y_{_{k}})$ is the decimal value of the binary number $y_{_{1}}...y_{_{k}}$. Equations (\ref{e36}), (\ref{e37}), and (\ref{e38}) imply that:
\begin{align}
\displaystyle \sum_{\bar{y}\in\{0,1\}^{k}}&\min \{P_{_{F}}A_{1}(\bar{y})+d(\bar{y}),...,P_{_{F}}A_{k}(\bar{y})+d(\bar{y}),P_{_{F}}A_{k}(\bar{y})+d^{*}(\bar{y})\} = (a_{_{1}}^{'}-a_{_{1}}) + \sum_{u=1}^{2^{k}} a_{_{u}} \nonumber
\end{align}
where $\displaystyle a_{_{1}}^{'} = (k-1)(1-P_{_{D}})(1-P_{_{F}})^{k-1} + P_{_{F}}(1-P_{_{D}})(1-P_{_{F}})^{k-1}, a_{_{1}}= (k-1)(1-P_{_{D}})(1-P_{_{F}})^{k-1} + P_{_{D}}(1-P_{_{F}})^{k}$. Using the same notation as in (\ref{sau}), $\displaystyle b_{_{1}},...,b_{_{2^{k}}}$ can be given as follows: 
\begin{equation}
\displaystyle \sum_{u=1}^{2^{k}} b_{u}=\sum_{\bar{y}\in\{0,1\}^{k}}\min_{i=1,..,k}\{P_{_{F}}B_{i}(\bar{y})+d(\bar{y})\} \nonumber
\end{equation}    
Using equation (\ref{Bis}) for $\bar{y}=(0,...,0)$ we get:
\begin{equation}
\displaystyle B_{k}(\bar{y}) \leq B_{i}(\bar{y}) \leq B_{1}(\bar{y}), i=2,..,k-1 \nonumber
\end{equation}
\begin{equation}
\displaystyle \Rightarrow b_{_{1}} = P_{_{F}}B_{k}(0,...,0)+d(0,...,0) \nonumber
\end{equation}
Let $\displaystyle b_{_{1}}^{'} = P_{_{F}}B_{k-1}(0,...,0)+d^{*}(0,...,0)$. Using the expressions for $\displaystyle a_{_{1}},a_{_{1}}^{'},b_{_{1}},b_{_{1}}^{'}$ we get:
\begin{equation}
\label{e44}
\displaystyle (a_{_{1}}^{'} - a_{_{1}}) - (b_{_{1}}^{'}-b_{_{1}}) = -(P_{_{F}})(P_{_{D}}-P_{_{F}})(1-P_{_{F}})^{k-1} 
\end{equation} 
We have two cases here $\displaystyle P_{_{D}}\geq P_{_{F}}, P_{_{D}}(1-P_{_{D}})\geq P_{_{F}}(1-P_{_{F}})$ and $\displaystyle P_{_{D}}\geq P_{_{F}}, P_{_{D}}(1-P_{_{D}})<P_{_{F}}(1-P_{_{F}})$. 
\par
\textbf{Case A:} $\displaystyle \bf P_{_{D}}\geq P_{_{F}}, P_{_{D}}(1-P_{_{D}})<P_{_{F}}(1-P_{_{F}})$
\\
Using equation (\ref{Bis}) and $\displaystyle P_{_{D}}(1-P_{_{D}})<P_{_{F}}(1-P_{_{F}})$ we get:
\begin{equation}
\displaystyle B_{k}(\bar{y}) < B_{1}(\bar{y})\leq B_{i}(\bar{y}),\bar{y}=(1,0,...,0) \nonumber
\end{equation}  
\begin{equation}
\displaystyle \Rightarrow b_{_{2^{k-1}+1}} = P_{_{F}}B_{k}(1,0,...,0)+d(1,0,...,0) \nonumber
\end{equation}
Let $\displaystyle b_{_{2^{k-1}+1}}^{'} = P_{_{F}}B_{k-1}(1,0,...,0)+d^{*}(1,0,...,0)$. Then using $b_{_{2^{k-1}+1}},b_{_{2^{k-1}+1}}^{'}$ we get:
\begin{equation}
\label{e47}
\displaystyle b_{_{2^{k-1}+1}} - b_{_{2^{k-1}+1}}^{'} = P_{_{F}}(P_{_{D}}-P_{_{F}})(1-P_{_{F}})^{k-1}
\end{equation} 
Using (\ref{e44}) and (\ref{e47}) we get:
\begin{equation}
\displaystyle (a_{_{1}}^{'} - a_{_{1}}) - (b_{_{1}}^{'}-b_{_{1}}) - (b_{_{2^{k-1}+1}}^{'}-b_{_{2^{k-1}+1}})=0 \nonumber
\end{equation}
These equations imply that:
\begin{equation}
\displaystyle \sum_{\bar{y}\in\{0,1\}^{k}} \min \big\{ P_{_{F}}B_{1}(\bar{y})+d(\bar{y}),...,P_{_{F}}B_{k-1}(\bar{y})+d(\bar{y}),P_{_{F}}B_{k-1}(\bar{y})+d^{*}(\bar{y}),P_{_{F}}B_{k}(\bar{y})+d(\bar{y}) \big\} \nonumber 
\end{equation}
\begin{equation}
\displaystyle \leq b_{1}' + b_{2^{k-1}+1}'+\sum_{i=2,i\neq 2^{k-1}+1 }^{2^{k}}b_{i} \nonumber
\end{equation}
Therefore we can conclude that:
\begin{align}
\displaystyle &\sum_{\bar{y}\in\{0,1\}^{k}} \min \big\{ P_{_{F}}A_{1}(\bar{y})+d(\bar{y}),...,P_{_{F}}A_{k}(\bar{y})+d(\bar{y}), P_{_{F}}A_{k}(\bar{y}) +d^{*}(\bar{y}) \big\} - \sum_{\bar{y}\in\{0,1\}^{k}} \min \big\{ P_{_{F}}B_{1}(\bar{y})+d(\bar{y}),...,  
\nonumber \\
\displaystyle & P_{_{F}}B_{k-1}(\bar{y})+d(\bar{y}),P_{_{F}}B_{k-1}(\bar{y})+d^{*}(\bar{y}),P_{_{F}}B_{k}(\bar{y})+ d(\bar{y}) \big\} \nonumber \\
\displaystyle &  ~\geq~ 
(a_{_{1}}'-a_{_{1}}) - (b_{_{1}}' - b_{_{1}}) - (b_{2^{k-1}+1}' - b_{2^{k-1}+1})+\sum_{i=1}^{2^{k}}a_{_{i}} - \sum_{i=1}^{2^{k}}b_{_{i}} ~\geq~ 0 \nonumber
\end{align}
\begin{equation}
\displaystyle \bf \Rightarrow (k+1)\bigg[P_{e}(\underbrace{1,...,1}_{k+1}) - P_{e}(\underbrace{2,1,..,1,0}_{k+1})\bigg] \geq \bigg[(1-P_{_{F}})\big\{\sum_{\bar{y}\in\{0,1\}^{k}} \min_{i=1,..,k}A_{i}(\bar{y}) - \sum_{\bar{y}\in\{0,1\}^{k}}\min_{i=1,..,k}B_{i}(\bar{y})\big\}   \nonumber \end{equation}
\begin{equation}
\displaystyle +~P_{_{F}}\big\{\sum_{\bar{y}\in\{0,1\}^{k}} \min_{i=1,..,k}A_{i}(\bar{y}) - \sum_{\bar{y}\in\{0,1\}^{k}}\min_{i=1,..,k}B_{i}(\bar{y})\big\}\bigg] \geq 0 \nonumber
\end{equation}
which proves the result for Case A. 
\par
\textbf{Case B:} $\displaystyle \bf P_{_{D}}\geq P_{_{F}}, P_{_{D}}(1-P_{_{D}})\geq P_{_{F}}(1-P_{_{F}})$
\\
\newline
For this case we will modify the induction steps slightly and use the above results and discussions to complete the proof. We will show that:
\begin{equation}
\displaystyle (k+1)\big[P_{e}(\underbrace{1,...,1}_{k+1}) - P_{e}(\underbrace{2,1,...,1,0}_{k+1})\big] \geq (P_{_{D}}-P_{_{F}})(1-P_{_{D}}-P_{_{F}})(1-P_{_{F}})^{k-1} \nonumber
\end{equation}
From (\ref{Pe11Pe20}) it is clear that this results holds for M=N=2:
\begin{equation}
\displaystyle 2\big(P_{e}(1,1) - P_{e}(2,0)\big) = (P_{_{D}}-P_{_{F}})(1-P_{_{D}}-P_{_{F}}) \nonumber
\end{equation}
Assume that it holds for M=N=k:
\begin{equation}
\label{Istep2}
\displaystyle (k)\big[P_{e}(\underbrace{1,...,1}_{k}) - P_{e}(\underbrace{2,1,...,1,0}_{k})\big] \geq(P_{_{D}}-P_{_{F}})(1-P_{_{D}}-P_{_{F}})(1-P_{_{F}})^{k-2}
\end{equation}
Using equation (\ref{Bis}) and $P_{_{D}}(1-P_{_{D}}) \geq P_{_{F}}(1-P_{_{F}})$:
\begin{equation}
\displaystyle B_{1}(\bar{y}) \leq B_{k}(\bar{y})\leq B_{i}(\bar{y}),\bar{y}=(1,0,...,0) \nonumber
\end{equation}
\begin{equation}
\displaystyle \Rightarrow b_{_{2^{k-1}+1}} = P_{_{F}}B_{1}(1,0,...,0) + d(1,0,...,0) \nonumber
\end{equation}
Let $\displaystyle b_{_{2^{k-1}+1}}^{'} = P_{_{F}}B_{k-1}(1,0,...,0) + d^{*}(1,0,...,0)$. Using the values of $\displaystyle b_{_{2^{k-1}+1}}, b_{_{2^{k-1}+1}}^{'}$ we get:
\begin{equation}
\displaystyle b_{_{2^{k-1}+1}} - b_{_{2^{k-1}+1}}^{'} = P_{_{D}}P_{_{F}}(P_{_{D}}-P_{_{F}})(1-P_{_{F}})^{k-2} \nonumber
\end{equation}
\begin{equation}
\label{e58}
\displaystyle \Rightarrow ~(a_{_{1}}^{'} - a_{_{1}}) - (b_{_{1}}^{'} - b_{_{1}}) - (b_{_{2^{k-1}+1}}^{'} - b_{_{2^{k-1}+1}}) = P_{_{F}}(P_{_{D}}-P_{_{F}})(P_{_{D}}+P_{_{F}}-1)(1-P_{_{F}})^{k-2}  
\end{equation}
Consider the observation $(y_{_{1}},...,y_{_{k}}) = (0,1,0,..,0)$:
\begin{equation}
\displaystyle B_{1}(\bar{y}) \leq B_{k}(\bar{y}) \leq B_{i}(\bar{y}),\bar{y}=(0,1,0,...,0) \nonumber
\end{equation}
\begin{equation}
\label{e60}
\displaystyle \Rightarrow b_{_{2^{k-2}+1}} = P_{_{F}}B_{1}(0,1,0,...,0) + d(0,1,0,...,0)
\end{equation} 
Let $\displaystyle b_{_{2^{k-2}+1}}^{'} = P_{_{F}}B_{k-1}(0,1,0,...,0) + d^{*}(0,1,0,...,0)$.\\ 
Using (\ref{e58} and (\ref{e60}) we get:
\begin{align}
\displaystyle & (a_{_{1}}^{'} - a_{_{1}}) - (b_{_{1}}^{'} - b_{_{1}}) - (b_{_{2^{k-1}+1}}^{'} - b_{_{2^{k-1}+1}}) ~- (b_{_{2^{k-2}+1}}^{'} - b_{_{2^{k-2}+1}}) =P_{_{F}}(P_{_{D}}-P_{_{F}})(P_{_{D}}+P_{_{F}}-1)(1-P_{_{F}})^{k-2} \nonumber \\
\displaystyle &  \hspace{3cm}+ P_{_{D}}P_{_{F}}(P_{_{D}}-P_{_{F}})(1-P_{_{F}})^{k-2} \nonumber
\end{align}
This implies that:
\begin{equation}
\displaystyle (k+1)\big[P_{e}(\underbrace{1,...,1}_{k+1}) - P_{e}(\underbrace{2,1,...,1,0}_{k+1})\big] \geq (1-P_{_{F}})\big\{\sum_{\bar{y}\in\{0,1\}^{k}} \big(\min_{i=1,..,k}(A_{i}(\bar{y})+c(\bar{y})) - \min_{i=1,..,k}(B_{i}(\bar{y})+c(\bar{y}))\big)\big\}  \nonumber
\end{equation}
\begin{equation}
\displaystyle +(P_{_{F}})\big\{\sum_{\bar{y}\in\{0,1\}^{k}} \big(\min_{i=1,..,k}(A_{i}(\bar{y})+d(\bar{y})) - \min_{i=1,..,k}(B_{i}(\bar{y})+d(\bar{y}))\big)\big\}  - P_{_{F}}(P_{_{D}}-P_{_{F}})(1-P_{_{D}}-P_{_{F}})(1-P_{_{F}})^{k-2} \nonumber
\end{equation}
\begin{equation}
\label{e62}
\displaystyle + P_{_{D}}P_{_{F}}(P_{_{D}}-P_{_{F}})(1-P_{_{F}})^{k-2} \hspace{3cm}
\end{equation}
Using (\ref{Istep2}), (\ref{e62}), and the fact that $\displaystyle P_{_{D}}P_{_{F}}(P_{_{D}}-P_{_{F}})(1-P_{_{F}})^{k-2} \geq 0$ we conclude that:
\begin{equation}
\displaystyle (k+1)\big[P_{e}(\underbrace{1,...,1}_{k+1}) - P_{e}(\underbrace{2,1,...,1,0}_{k+1})\big] \geq (P_{_{D}}-P_{_{F}})(1-P_{_{D}}-P_{_{F}})(1-P_{_{F}})^{k-1} \geq 0  \nonumber
\end{equation}
This proves the claim for the case $\bf P_{_{D}} \geq P_{_{F}}$. 
\par
A proof for the aforementioned case is sufficient to conclude that the result holds for all values of $P_{_{D}},P_{_{F}}$. This is due to the fact that for the case $P_{_{D}}<P_{_{F}}$ the detector will simply flip the observation bits and the same optimal strategy, which was employed for the case $P_{_{D}}>P_{_{F}}$, will be employed by the fusion center.  
\end{proof}
\par
Theorem 4.1 is an extremely useful result since it holds for any values of $P_{_{D}}$ and $P_{_{F}}$. It is quite natural to assume that uniform placement of the sensors should be strictly optimal for sensors which possess a high detection probability and a low probability of false alarm. This results proves that this natural assumption fails for the case when the number of sensors equal the number of placement points. 
\par
Next we will prove some properties of the optimal placement on the $(P_{_{F}},P_{_{D}})$ plane by varying the number of placement points $N$. We will prove that the optimal solution structure on the $(P_{_{F}},P_{_{D}})$ plane for $M < N$ remains the same as we increase $N$.
\\
\par
\textbf{\textit{Theorem 4.2:}} The sensor-placement point pairs $(M,N_{1})$ and $(M,N_{2})$, where $M < N_{1} < N_{2}$, have the same optimal placement structure on the $(P_{_{F}},P_{_{D}})$ plane.          
\par
\begin{proof} We will use the notation $P_{e}(v_{_{1}},...,v_{_{M}})\big|_{N_{i}},i=1,2$ to distinguish between the two situations. 
\par
First consider the case $\displaystyle (M,N_{1})$ where we have $M$ agents and $N_{1}$ placement points. For a specific placement $\displaystyle \bar{v}=(v_{_{1}},...,v_{_{M}}) \in \Lambda^{M}$,
\begin{equation}
\label{PeN1}
\displaystyle P_{e}(v_{_{1}},...,v_{_{M}})\big|_{N_{1}}= \frac{1}{N_{1}} \sum_{\bar{y}\in \{0,1\}^{M}}\min_{i=1,..,N_{1}} \{\sum_{j=1,j\neq i}^{N_{1}} p_{j}(\bar{y},\bar{v})\}
\end{equation}  
where $\displaystyle p_{j}(\bar{y},\bar{v})$ is given by equation (\ref{pj}). Now $v_{_{j}}=0,j=(M+1),...,N_{1}$, therefore we can take $p_{_{j}}(\bar{y},\bar{v})= p_{_{M+1}}(\bar{y}), j=(M+1),...,N_{1}$. Note that $p_{_{M+1}}(\bar{y})$ does not depend on the placement vector $\bar{v}$. Using this fact (\ref{PeN1}) can be updated as follows: 
\begin{equation}
\displaystyle P_{e}(v_{_{1}},...,v_{_{M}})\big|_{N_{1}}= \frac{1}{N_{1}} \sum_{\bar{y}\in \{0,1\}^{M}}\min_{i=1,..,M+1}\bigg[\sum_{j=1,j\neq i}^{M} p_{_{j}}(\bar{y},\bar{v}) +(N_{1}-M)p_{_{M+1}}(\bar{y})\chi_{(i\neq M+1)}\nonumber 
\end{equation}
\begin{equation}
\label{PeN1updated} 
\displaystyle  + (N_{1}-M-1)p_{_{M+1}}(\bar{y})\chi_{(i=M+1)}\bigg]
\end{equation}
Consider another placement vector $\hat{\bar{{v}}} = (\hat{v}_{_{1}},...,\hat{v}_{_{M}}) \in \Lambda^{M}$. 
\begin{equation}
\displaystyle  P_{e}(\hat{v}_{_{1}},...,\hat{v}_{_{M}})\big|_{N_{1}}= \frac{1}{N_{1}} \sum_{\bar{y}\in \{0,1\}^{M}}\min_{i=1,..,M+1}\bigg[\sum_{j=1,j\neq i}^{M} p_{_{j}}(\bar{y},\hat{\bar{v}})+ (N_{1}-M)p_{_{M+1}}(\bar{y})\chi_{(i\neq M+1)}  \nonumber 
\end{equation}
\begin{equation}
\label{PeN1hat} 
\displaystyle + (N_{1}-M-1)p_{_{M+1}}(\bar{y})\chi_{(i=M+1)}\bigg]
\end{equation}
Both $(N_{1},M)$ and $(N_{2},M)$ have the same set of admissible placements, $\Lambda^{M}$. $\displaystyle P_{e}(v_{_{1}},...,v_{_{M}})\big|_{N_{2}}, P_{e}(\hat{v}_{_{1}},...,\hat{v}_{_{M}})\big|_{N_{2}}$ can be written as: 
\begin{equation}
\displaystyle  P_{e}(v_{_{1}},...,v_{_{M}})\big|_{N_{2}}= \frac{1}{N_{2}} \sum_{\bar{y}\in \{0,1\}^{M}}\min_{i=1,..,M+1}\bigg[\sum_{j=1,j\neq i}^{M} p_{_{j}}(\bar{y},\bar{v}) + (N_{1}-M)p_{_{M+1}}(\bar{y})\chi_{(i\neq M+1)} +  \nonumber
\end{equation}
\begin{equation}
\label{PeN2}
\displaystyle (N_{1}-M-1)p_{_{M+1}}(\bar{y})\chi_{(i=M+1)}+ (N_{2}-N_{1})p_{_{M+1}}(\bar{y})\chi_{(i\neq M+1)} + (N_{2}-N_{1})p_{_{M+1}}(\bar{y})\chi_{(i=M+1)}\bigg]
\end{equation}
\begin{equation}
\displaystyle  P_{e}(\hat{v}_{_{1}},...,\hat{v}_{_{M}})\big|_{N_{2}}= \frac{1}{N_{2}} \sum_{\bar{y}\in \{0,1\}^{M}}\min_{i=1,..,M+1}\bigg[\sum_{j=1,j\neq i}^{M} p_{_{j}}(\bar{y},\hat{\bar{v}}) + (N_{1}-M)p_{_{M+1}}(\bar{y})\chi_{(i\neq M+1)} +  \nonumber
\end{equation}
\begin{equation}
\label{PeN2hat}
\displaystyle (N_{1}-M-1)p_{_{M+1}}(\bar{y})\chi_{(i=M+1)}+ (N_{2}-N_{1})p_{_{M+1}}(\bar{y})\chi_{(i\neq M+1)} + (N_{2}-N_{1})p_{_{M+1}}(\bar{y})\chi_{(i=M+1)}\bigg]
\end{equation}
Equations (\ref{PeN1updated}), (\ref{PeN1hat}), (\ref{PeN2}), and (\ref{PeN2hat}) imply that:
\begin{equation}
\label{comparison}
\displaystyle N_{2}\times P_{e}(v_{_{1}},...,v_{_{M}})\big|_{N_{2}} - N_{2}\times P_{e}(\hat{v}_{_{1}},...,\hat{v}_{_{M}})\big|_{N_{2}} = 
N_{1}\times P_{e}(v_{_{1}},...,v_{_{M}})\big|_{N_{1}} - N_{1}\times P_{e}(\hat{v}_{_{1}},...,\hat{v}_{_{M}})\big|_{N_{1}}
\end{equation}
Since $\bar{v}$ and $\hat{\bar{v}}$ are two arbitrarily chosen placements from the set $\Lambda^{M}$, (\ref{comparison}) implies that as we transition from $(M,N_{1})$ to $(M,N_{2})$ the comparison equations which determine the optimal placements do not change. Therefore, we conclude that both $(M,N_{1})$ and $(M,N_{2})$ have the same optimal placement structure on the $(P_{_{F}},P_{_{D}})$ plane. 
\end{proof}  
\par 
Theorem 4.2 has important practical implications. It implies that if we are given a sensor-placement pair $(M,N)$ where $M<N$ we only need to analyze the sensor-placement pair $(M,M+1)$ and its optimal placements will be valid for all $N > M$. This results in a significant reduction in computational complexity. 
\\   
\par
\textbf{\textit{Corollary 4.1}}: Let $\bf \Delta^{^{M,N_{1}}}$ be the set of strictly optimal placements for the sensor-placement pair $\displaystyle (M,N_{1}), N_{1}=M$. Then $\displaystyle \bf \Delta^{^{M,N_{2}}} = \Delta^{^{M,N_{1}}}$ $\displaystyle \cup \big\{(v_{_{1}},...,v_{_{M}})=(1,...,1)\big\}$ is the set of strictly optimal placements for the sensor-placement pair $\displaystyle (M,N_{2}), N_{2}> M$. 
\par
Corollary 4.1 follows directly from Theorem 4.1, Theorem 4.2, and some basic facts regarding the uniform placement of sensors. One can easily recognize the fact that uniform placement of sensors will always belong to the set of strictly optimal placements $\bf \Delta^{M,N}$, for a sensor-placement pair $(M,N),M < N$. This can be established by considering extremely high values of $P_{_{D}}$ and extremely low values of $P_{_{F}}$.

\section{A Majorization Approach}
\par
In this section, we will utilize concepts from Majorization Theory to characterize some properties of the optimal solution. We will utilize the notion of a majorization based placement scale defined in Section II. 
\par
\par
It should be noted that majorization is a partial ordering and therefore not all elements of $\Lambda^{M}$ can be placed on a majorization based scale. There can be many sets of optimal placements dependent on what we choose as the optimal placement at points where no placement is strictly optimal. For a given $(M,N)$, an important question is the existence of a set of optimal placements which can be placed on a majorization based scale. 
\\
\par
\textbf{\textit{Conjecture 5.1:}} For any $(M,N)$, there exists a set of optimal placements $O^{M} \subseteq \Lambda^{M}$ which can be placed on a majorization based placement scale.  
\\
\par
We explain the intution behind this conjecture by considering the sensor-placement pair, $(M,N)=(6,6)$, which has $\Lambda^{M}$ of size $|\Lambda^{M}|=11$. Note that the placements $(4,1,1)$ and $(3,3)$ are not comparable with respect to majorization. Similarly $(3,1,1,1)$ and $(2,2,2)$ are not comparable. The conjecture would be false if every set of optimal placements contained both $(4,1,1)$ and $(3,3)$ or contained both $(3,1,1,1)$ and $(2,2,2)$. $\{(6,0),(5,1),(4,2),(3,2,1),(2,2,1,1),(2,1,1,1,1)\}$ is a set of optimal placements for this case. It should be noted that for high values of $P_{_{F}}$, $P_{_{D}}\geq P_{_{F}}$, the optimal placements tend to have large number of sensors at a small number of points (concentrated placement) and for low values of $P_{_{F}}$ the optimal placements tend to have small number of sensors placed at a higher number of points (spread out placement). In this case the placements $(4,2),(3,2,1),$ and $(2,2,1,1)$ are optimal in the regions where one would have naturally expected the placements $(3,1,1,1),(2,2,2),(4,1,1),$ and $(3,3)$ to be optimal. The aforementioned example indicates existence of the following properties of non-comparable placements which form the basis of this conjecture: 
\begin{itemize}
\item Placements containing sensors concentrated at few points but not enough to outperform other placements, for higher values of $P_{_{F}}$, which are highly concentrated and can be placed on a majorization scale.
\item Placements containing sensors that are relatively spread out but not enough to outperform other placements, for lower values of $P_{_{F}}$, which have a higher spread of and can be placed on a majorization based scale. 
\item Placements not possesing a fine balance between concentration and spread of sensors to be optimal for values of $P_{_{F}}$ that are neither high nor low.
\end{itemize}
The main result of this section can be stated as follows:
\\
\par
\textbf{\textit{Proposition 5.1:}} By fixing $P_{_{D}}$ and increasing $P_{_{F}}$, for $M\leq5$ and $P_{_{D}}\geq P_{_{F}}$, the optimal placement occurs non decreasingly on a majorization based placement scale. 

\begin{proof} We will only consider the sensor-placement pair $\displaystyle (M,N)=(4,4)$. Other cases where $M \leq 5$ can be proved similarly.  
\par
The admissible placement set is given as $\Lambda^{M} = \{(1,1,1,1),(2,1,1,0),(2,2,0,0),(3,1,0,0),(4,0,0,0)\}$. Since $M=N$ we do not need to consider the placement $(1,1,1,1)$. Using majorization theory the following majorization based placement scale can be defined:
\begin{equation}
\label{mscale}
\displaystyle (4,0,0,0) \succ (3,1,0,0) \succ (2,2,0,0) \succ (2,1,1,0) 
\end{equation}
Using (\ref{Pewf}) we will derive the regions on the $(P_{b},P_{a})$ plane on which these placements are optimal, respectively.
First we provide the conditional probabilties $\displaystyle p_{_{1}}(y_{_{1}},y_{_{2}},y_{_{3}},y_{_{4}})\big|_{(v_{_{1}},v_{_{2}},v_{_{3}},v_{_{4}})}$ for the sensor placement $\displaystyle (v_{_{1}},v_{_{2}},v_{_{3}},v_{_{4}}) = (2,1,1,0)$ using equation (5) and (6). 
\begin{equation}
\displaystyle p_{_{1}}(1,1,1,1)\big|_{(2,1,1,0)} = P_{_{D}}^{2}P_{_{F}}^{2}~,~ p_{_{1}}(1,1,1,0)\big|_{(2,1,1,0)} = P_{_{D}}^{2}P_{_{F}}(1-P_{_{F}}) \nonumber
\end{equation}
\begin{equation}
\displaystyle p_{_{1}}(1,1,0,1)\big|_{(2,1,1,0)} = P_{_{D}}^{2}P_{_{F}}(1-P_{_{F}})~,~ p_{_{1}}(1,1,0,0)\big|_{(2,1,1,0)} = P_{_{D}}^{2}(1-P_{_{F}})^{2} \nonumber
\end{equation}
\begin{equation}
\displaystyle p_{_{1}}(1,0,1,1)\big|_{(2,1,1,0)} = P_{_{D}}P_{_{F}}^{2}(1-P_{_{D}})~,~ p_{_{1}}(1,0,1,0)\big|_{(2,1,1,0)} = P_{_{D}}P_{_{F}}(1-P_{_{D}})(1-P_{_{F}}) \nonumber
\end{equation}
\begin{equation}
\displaystyle p_{_{1}}(1,0,0,1)\big|_{(2,1,1,0)} = P_{_{D}}P_{_{F}}(1-P_{_{D}})(1-P_{_{F}})~,~ p_{_{1}}(1,0,0,0)\big|_{(2,1,1,0)} = P_{_{D}}(1-P_{_{D}})(1-P_{_{F}})^{2} \nonumber
\end{equation}
\begin{equation}
\displaystyle p_{_{1}}(0,1,1,1)\big|_{(2,1,1,0)} = P_{_{D}}P_{_{F}}^{2}(1-P_{_{D}})~,~ p_{_{1}}(0,1,1,0)\big|_{(2,1,1,0)} = P_{_{D}}P_{_{F}}(1-P_{_{D}})(1-P_{_{F}}) \nonumber
\end{equation}
\begin{equation}
\displaystyle p_{_{1}}(0,1,0,1)\big|_{(2,1,1,0)} = P_{_{D}}P_{_{F}}(1-P_{_{D}})(1-P_{_{F}})~,~ p_{_{1}}(0,1,0,0)\big|_{(2,1,1,0)} = P_{_{D}}(1-P_{_{D}})(1-P_{_{F}})^{2} \nonumber
\end{equation}
\begin{equation}
\displaystyle p_{_{1}}(0,0,1,1)\big|_{(2,1,1,0)} = P_{_{F}}^{2}(1-P_{_{D}})^{2}~,~ p_{_{1}}(0,0,1,0)\big|_{(2,1,1,0)} = P_{_{F}}(1-P_{_{F}})(1-P_{_{D}})^{2} \nonumber
\end{equation}
\begin{equation}
\displaystyle p_{_{1}}(0,0,0,1)\big|_{(2,1,1,0)} = P_{F}(1-P_{F})(1-P_{D})^{2}~,~ p_{_{1}}(0,0,0,0)\big|_{(2,1,1,0)} = (1-P_{D})^{2}(1-P_{F})^{2} \nonumber
\end{equation}
The probabilities $\displaystyle p_{_{2}}(y_{_{1}},y_{_{2}},y_{_{3}},y_{_{4}})\big|_{(2,1,1,0)}$ are given as follows:
\begin{equation}
\displaystyle p_{_{2}}(1,1,1,1)\big|_{(2,1,1,0)} = P_{D}P_{F}^{3}~,~ p_{_{2}}(1,1,1,0)\big|_{(2,1,1,0)} = P_{D}P_{F}^{2}(1-P_{F}) \nonumber
\end{equation}
\begin{equation}
\displaystyle p_{_{2}}(1,1,0,1)\big|_{(2,1,1,0)} = P_{_{F}}^{3}(1-P_{_{D}})~,~ p_{_{2}}(1,1,0,0)\big|_{(2,1,1,0)} = P_{_{F}}^{2}(1-P_{_{D}})(1-P_{_{F}}) \nonumber
\end{equation}
\begin{equation}
\displaystyle p_{_{2}}(1,0,1,1)\big|_{(2,1,1,0)} = P_{_{D}}P_{_{F}}^{2}(1-P_{_{F}})~,~ p_{_{2}}(1,0,1,0)\big|_{(2,1,1,0)} = P_{_{D}}P_{_{F}}(1-P_{_{F}})^{2} \nonumber
\end{equation}
\begin{equation}
\displaystyle p_{_{2}}(1,0,0,1)\big|_{(2,1,1,0)} = P_{_{F}}^{2}(1-P_{_{D}})(1-P_{_{F}})~,~ p_{_{2}}(1,0,0,0)\big|_{(2,1,1,0)} = P_{_{F}}(1-P_{_{D}})(1-P_{_{F}})^{2} \nonumber
\end{equation}
\begin{equation}
\displaystyle p_{_{2}}(0,1,1,1)\big|_{(2,1,1,0)} = P_{_{D}}P_{_{F}}^{2}(1-P_{_{F}})~,~ p_{_{2}}(0,1,1,0)\big|_{(2,1,1,0)} = P_{_{D}}P_{_{F}}(1-P_{_{F}})^{2} \nonumber
\end{equation}
\begin{equation}
\displaystyle p_{_{2}}(0,1,0,1)\big|_{(2,1,1,0)} = P_{_{F}}^{2}(1-P_{_{D}})(1-P_{_{F}})~,~ p_{_{2}}(0,1,0,0)\big|_{(2,1,1,0)} = P_{_{F}}(1-P_{_{D}})(1-P_{_{F}})^{2}\nonumber 
\end{equation}
\begin{equation}
\displaystyle p_{_{2}}(0,0,1,1)\big|_{(2,1,1,0)} = P_{_{D}}P_{_{F}}(1-P_{_{F}})^{2}~,~ p_{_{2}}(0,0,1,0)\big|_{(2,1,1,0)} = P_{_{D}}(1-P_{_{F}})^{3} \nonumber
\end{equation}
\begin{equation}
\displaystyle p_{_{2}}(0,0,0,1)\big|_{(2,1,1,0)} = P_{_{F}}(1-P_{_{D}})(1-P_{_{F}})^{2}~,~ p_{_{2}}(0,0,0,0)\big|_{(2,1,1,0)} = (1-P_{_{D}})(1-P_{_{F}})^{3}\nonumber 
\end{equation}
The probabilities $\displaystyle p_{_{3}}(y_{_{1}},y_{_{2}},y_{_{3}},y_{_{4}})\big|_{(2,1,1,0)}$ are given as follows:
\begin{equation}
\displaystyle p_{_{3}}(1,1,1,1)\big|_{(2,1,1,0)} = P_{_{D}}P_{_{F}}^{3}~,~ p_{_{3}}(1,1,1,0)\big|_{(2,1,1,0)} = P_{_{F}}^{3}(1-P_{_{D}}) \nonumber
\end{equation}
\begin{equation}
\displaystyle p_{_{3}}(1,1,0,1)\big|_{(2,1,1,0)} = P_{_{D}}P_{_{F}}^{2}(1-P_{_{F}})~,~ p_{_{3}}(1,1,0,0)\big|_{(2,1,1,0)} = P_{_{F}}^{2}(1-P_{_{D}})(1-P_{_{F}}) \nonumber
\end{equation}
\begin{equation}
\displaystyle p_{_{3}}(1,0,1,1)\big|_{(2,1,1,0)} = P_{_{D}}P_{_{F}}^{2}(1-P_{_{F}})~,~ p_{_{3}}(1,0,1,0)\big|_{(2,1,1,0)} = P_{_{F}}^{2}(1-P_{_{D}})(1-P_{_{F}}) \nonumber
\end{equation}
\begin{equation}
\displaystyle p_{_{3}}(1,0,0,1)\big|_{(2,1,1,0)} = P_{_{D}}P_{_{F}}(1-P_{_{F}})^{2}~,~ p_{_{3}}(1,0,0,0)\big|_{(2,1,1,0)} = P_{_{F}}(1-P_{_{D}})(1-P_{_{F}})^{2} \nonumber
\end{equation}
\begin{equation}
\displaystyle p_{_{3}}(0,1,1,1)\big|_{(2,1,1,0)} = P_{_{D}}P_{_{F}}^{2}(1-P_{_{F}})~,~ p_{_{3}}(0,1,1,0)\big|_{(2,1,1,0)} = P_{_{F}}^{2}(1-P_{_{D}})(1-P_{_{F}}) \nonumber
\end{equation}
\begin{equation}
\displaystyle p_{_{3}}(0,1,0,1)\big|_{(2,1,1,0)} = P_{_{D}}P_{_{F}}(1-P_{_{F}})^{2}~,~ p_{_{3}}(0,1,0,0)\big|_{(2,1,1,0)} = P_{_{F}}(1-P_{_{D}})(1-P_{_{F}})^{2} \nonumber
\end{equation}
\begin{equation}
\displaystyle p_{_{3}}(0,0,1,1)\big|_{(2,1,1,0)} = P_{_{D}}P_{_{F}}(1-P_{_{F}})^{2}~,~ p_{_{3}}(0,0,1,0)\big|_{(2,1,1,0)} = P_{_{F}}(1-P_{_{D}})(1-P_{_{F}})^{2} \nonumber
\end{equation}
\begin{equation}
\displaystyle p_{_{3}}(0,0,0,1)\big|_{(2,1,1,0)} = P_{_{D}}(1-P_{_{F}})^{3}~,~ p_{_{3}}(0,0,0,0)\big|_{(2,1,1,0)} = (1-P_{_{D}})(1-P_{_{F}})^{3} \nonumber
\end{equation}
The probabilities $\displaystyle p_{_{4}}(y_{_{1}},y_{_{2}},y_{_{3}},y_{_{4}})\big|_{(2,1,1,0)}$ are given as follows:
\begin{equation}
\displaystyle p_{_{4}}(1,1,1,1)\big|_{(2,1,1,0)} = P_{_{F}}^{4}~,~ p_{_{4}}(1,1,1,0)\big|_{(2,1,1,0)} = P_{_{F}}^{3}(1-P_{_{F}}) \nonumber
\end{equation}
\begin{equation}
\displaystyle p_{_{4}}(1,1,0,1)\big|_{(2,1,1,0)} = P_{_{F}}^{3}(1-P_{_{F}})~,~ p_{_{4}}(1,1,0,0)\big|_{(2,1,1,0)} = P_{_{F}}^{2}(1-P_{_{F}})^{2} \nonumber
\end{equation}
\begin{equation}
\displaystyle p_{_{4}}(1,0,1,1)\big|_{(2,1,1,0)} = P_{_{F}}^{3}(1-P_{_{F}})~,~ p_{_{4}}(1,0,1,0)\big|_{(2,1,1,0)} = P_{_{F}}^{2}(1-P_{_{F}})^{2} \nonumber
\end{equation}
\begin{equation}
\displaystyle p_{_{4}}(1,0,0,1)\big|_{(2,1,1,0)} = P_{_{F}}^{2}(1-P_{_{F}})^{2}~,~ p_{_{4}}(1,0,0,0)\big|_{(2,1,1,0)} = P_{_{F}}(1-P_{_{F}})^{3} \nonumber
\end{equation}
\begin{equation}
\displaystyle p_{_{4}}(0,1,1,1)\big|_{(2,1,1,0)} = P_{_{F}}^{3}(1-P_{_{F}})~,~ p_{_{4}}(0,1,1,0)\big|_{(2,1,1,0)} = P_{_{F}}^{2}(1-P_{_{F}})^{2} \nonumber
\end{equation}
\begin{equation}
\displaystyle p_{_{4}}(0,1,0,1)\big|_{(2,1,1,0)} = P_{_{F}}^{2}(1-P_{_{F}})^{2}~,~ p_{_{4}}(0,1,0,0)\big|_{(2,1,1,0)} = P_{_{F}}(1-P_{_{F}})^{3} \nonumber
\end{equation}
\begin{equation}
\displaystyle p_{_{4}}(0,0,1,1)\big|_{(2,1,1,0)} = P_{_{F}}^{2}(1-P_{_{F}})^{2}~,~ p_{_{4}}(0,0,1,0)\big|_{(2,1,1,0)} = P_{_{F}}(1-P_{_{F}})^{3}\nonumber 
\end{equation}
\begin{equation}
\displaystyle p_{_{4}}(0,0,0,1)\big|_{(2,1,1,0)} = P_{_{F}}(1-P_{_{F}})^{3}~,~ p_{_{4}}(0,0,0,0)\big|_{(2,1,1,0)} = (1-P_{_{F}})^{4} \nonumber
\end{equation}
Let us denote $I_{i}(\bar{y})$ by:
\begin{equation}
\label{I(y)}
\displaystyle I_{i}(y_{_{1}}, y_{_{2}}, y_{_{3}}, y_{_{4}})\big|_{(v_{1},v_{2},v_{3},v_{4})}  = \sum_{j=1,j\neq i} p_{j}(y_{_{1}},y_{_{2}},y_{_{3}},y_{_{4}})\big|_{(v_{1},v_{2},v_{3},v_{4})} 
\end{equation}
Using equation (\ref{I(y)}) and the conditional probabilities we get the following values for $I_{i}(y_{_{1}}, y_{_{2}}, y_{_{3}}, y_{_{4}})\big|_{(2,1,1,0)}$:
\begin{align}
\displaystyle &I_{1}(1,1,1,1)\big|_{(2,1,1,0)} = 2P_{_{D}}P_{_{F}}^{3} + P_{_{F}}^{4} \nonumber\\
\displaystyle &I_{2}(1,1,1,1)\big|_{(2,1,1,0)} = P_{_{D}}^{2}P_{_{F}}^{2} + P_{_{D}}P_{_{F}}^{3} + P_{_{F}}^{4} \nonumber\\
\displaystyle &I_{3}(1,1,1,1)\big|_{(2,1,1,0)} = P_{_{D}}^{2}P_{_{F}}^{2} + P_{_{D}}P_{_{F}}^{3} + P_{_{F}}^{4} \nonumber\\
\displaystyle &I_{4}(1,1,1,1)\big|_{(2,1,1,0)} = P_{_{D}}^{2}P_{_{F}}^{2} + 2P_{_{D}}P_{_{F}}^{3} \nonumber \\
\displaystyle &I_{1}(1,1,1,0)\big|_{(2,1,1,0)} = P_{_{D}}P_{_{F}}^{2}(1-P_{_{F}}) + P_{_{F}}^{3}(1-P_{_{D}}) + P_{_{F}}^{3}(1-P_{_{F}})\nonumber \\
\displaystyle &I_{2}(1,1,1,0)\big|_{(2,1,1,0)} = P_{_{D}}^{2}P_{_{F}}(1-P_{_{F}}) + P_{_{F}}^{3}(1-P_{_{D}}) + P_{_{F}}^{3}(1-P_{_{F}})\nonumber \\
\displaystyle &I_{3}(1,1,1,0)\big|_{(2,1,1,0)} = P_{_{D}}^{2}P_{_{F}}(1-P_{_{F}}) + P_{_{D}}P_{_{F}}^{2}(1-P_{_{F}}) + P_{_{F}}^{3}(1-P_{_{F}})\nonumber 
\end{align}
\begin{align}
\displaystyle &I_{4}(1,1,1,0)\big|_{(2,1,1,0)} = P_{_{D}}^{2}(1-P_{_{F}}) + P_{_{D}}P_{_{F}}^{2}(1-P_{_{F}}) + P_{_{F}}^{3}(1-P_{_{D}}) \nonumber \\
\displaystyle &I_{1}(1,1,0,1)\big|_{(2,1,1,0)} = P_{_{D}}^{3}(1-P_{_{D}}) + P_{_{D}}P_{_{F}}^{2}(1-P_{_{F}}) + P_{_{F}}^{3}(1-P_{_{F}})\nonumber \\
\displaystyle &I_{2}(1,1,0,1)\big|_{(2,1,1,0)} = P_{_{D}}^{2}P_{_{F}}(1-P_{_{F}}) + P_{_{D}}P_{_{F}}^{2}(1-P_{_{F}}) + P_{_{F}}^{3}(1-P_{_{F}})\nonumber\\
\displaystyle &I_{3}(1,1,0,1)\big|_{(2,1,1,0)} = P_{_{D}}^{2}P_{_{F}}(1-P_{_{F}}) + P_{_{F}}^{3}(1-P_{_{D}}) + P_{_{F}}^{3}(1-P_{_{F}})\nonumber\\
\displaystyle &I_{4}(1,1,0,1)\big|_{(2,1,1,0)} = P_{_{D}}^{2}P_{_{F}}(1-P_{_{F}}) + P_{_{F}}^{3}(1-P_{_{D}}) + P_{_{D}}P_{_{F}}^{2}(1-P_{_{F}})\nonumber \\
\displaystyle &I_{1}(1,1,0,0)\big|_{(2,1,1,0)} = 2P_{_{F}}^{2}(1-P_{_{D}})(1-P_{_{F}}) +  P_{_{F}}^{2}(1-P_{_{F}})^{2}\nonumber\\
\displaystyle &I_{2}(1,1,0,0)\big|_{(2,1,1,0)} = P_{_{D}}^{2}(1-P_{_{F}})^{2} + P_{_{F}}^{2}(1-P_{_{D}})(1-P_{_{F}}) + P_{_{F}}^{2}(1-P_{_{F}})^{2}\nonumber\\
\displaystyle &I_{3}(1,1,0,0)\big|_{(2,1,1,0)} = P_{_{D}}^{2}(1-P_{_{F}})^{2} + P_{_{F}}^{2}(1-P_{_{D}})(1-P_{_{F}}) + P_{_{F}}^{2}(1-P_{_{F}})^{2}\nonumber\\
\displaystyle &I_{4}(1,1,0,0)\big|_{(2,1,1,0)} = P_{_{D}}^{2}(1-P_{_{F}})^{2} + 2P_{_{F}}^{2}(1-P_{_{D}})(1-P_{_{F}})\nonumber \\
\displaystyle &I_{1}(1,0,1,1)\big|_{(2,1,1,0)} = 2P_{_{D}}P_{_{F}}^{2}(1-P_{_{F}}) + P_{_{F}}^{3}(1-P_{_{F}})\nonumber\\
\displaystyle &I_{2}(1,0,1,1)\big|_{(2,1,1,0)} = P_{_{D}}P_{_{F}}^{2}(1-P_{_{D}}) + P_{_{D}}P_{_{F}}^{2}(1-P_{_{F}}) + P_{_{F}}^{3}(1-P_{_{F}})\nonumber\\
\displaystyle &I_{3}(1,0,1,1)\big|_{(2,1,1,0)} = P_{_{D}}P_{_{F}}^{2}(1-P_{_{D}}) + P_{_{D}}P_{_{F}}^{2}(1-P_{_{F}}) + P_{_{F}}^{3}(1-P_{_{F}})\nonumber\\
\displaystyle &I_{4}(1,0,1,1)\big|_{(2,1,1,0)} = P_{_{D}}P_{_{F}}^{2}(1-P_{_{D}}) + 2P_{_{D}}P_{_{F}}^{2}(1-P_{_{F}})\nonumber\\
\displaystyle &I_{1}(1,0,1,0)\big|_{(2,1,1,0)} = P_{_{D}}P_{_{F}}(1-P_{_{F}})^{2} + P_{_{F}}^{2}(1-P_{_{D}})(1-P_{_{F}}) + P_{_{F}}^{2}(1-P_{_{F}})^{2}\nonumber\\
\displaystyle &I_{2}(1,0,1,0)\big|_{(2,1,1,0)} = P_{_{D}}P_{_{F}}(1-P_{_{D}})(1-P_{_{F}}) + P_{_{F}}^{2}(1-P_{_{D}})(1-P_{_{F}}) + P_{_{F}}^{2}(1-P_{_{F}})^{2}\nonumber\\
\displaystyle &I_{3}(1,0,1,0)\big|_{(2,1,1,0)} = P_{_{D}}P_{_{F}}(1-P_{_{D}})(1-P_{_{F}}) + P_{_{D}}P_{_{F}}(1-P_{_{F}})^{2} + P_{_{F}}^{2}(1-P_{_{F}})^{2}\nonumber\\
\displaystyle &I_{4}(1,0,1,0)\big|_{(2,1,1,0)} = P_{_{D}}P_{_{F}}(1-P_{_{D}})(1-P_{_{F}}) + P_{_{D}}P_{_{F}}(1-P_{_{F}})^{2} + P_{_{F}}^{2}(1-P_{_{D}})(1-P_{_{F}})\nonumber\\
\displaystyle &I_{1}(1,0,0,1)\big|_{(2,1,1,0)} = P_{_{F}}^{2}(1-P_{_{D}})(1-P_{_{F}}) + P_{_{D}}P_{_{F}}(1-P_{_{F}})^{2} + P_{_{F}}^{2}(1-P_{_{F}})^{2}\nonumber\\
\displaystyle &I_{2}(1,0,0,1)\big|_{(2,1,1,0)} = P_{_{D}}P_{_{F}}(1-P_{_{D}})(1-P_{_{F}}) + P_{_{D}}P_{_{F}}(1-P_{_{F}})^{2} + P_{_{F}}^{2}(1-P_{_{F}})^{2}\nonumber\\
\displaystyle &I_{3}(1,0,0,1)\big|_{(2,1,1,0)} = P_{_{D}}P_{_{F}}(1-P_{_{D}})(1-P_{_{F}}) + P_{_{F}}^{2}(1-P_{_{D}})(1-P_{_{F}}) + P_{_{F}}^{2}(1-P_{_{F}})^{2}\nonumber\\
\displaystyle &I_{4}(1,0,0,1)\big|_{(2,1,1,0)} = P_{_{D}}P_{_{F}}(1-P_{_{D}})(1-P_{_{F}}) + P_{_{F}}^{2}(1-P_{_{D}})(1-P_{_{F}}) + P_{_{D}}P_{_{F}}(1-P_{_{F}})^{2}\nonumber\\
\displaystyle &I_{1}(1,0,0,0)\big|_{(2,1,1,0)} = 2P_{_{F}}(1-P_{_{D}})(1-P_{_{F}})^{2} + P_{_{F}}(1-P_{_{F}})^{3}\nonumber\\
\displaystyle &I_{2}(1,0,0,0)\big|_{(2,1,1,0)} = P_{_{D}}(1-P_{_{D}})(1-P_{_{F}})^{2} + P_{_{F}}(1-P_{_{D}})(1-P_{_{F}})^{2} + P_{_{F}}(1-P_{_{F}})^{3}\nonumber\\
\displaystyle &I_{3}(1,0,0,0)\big|_{(2,1,1,0)} = P_{_{D}}(1-P_{_{D}})(1-P_{_{F}})^{2} + P_{_{F}}(1-P_{_{D}})(1-P_{_{F}})^{2} + P_{_{F}}(1-P_{_{F}})^{3}\nonumber
\end{align}
\begin{align}
\displaystyle &I_{4}(1,0,0,0)\big|_{(2,1,1,0)} = P_{_{D}}(1-P_{_{D}})(1-P_{_{F}})^{2} + 2P_{_{F}}(1-P_{_{D}})(1-P_{_{F}})^{2}\nonumber\\
\displaystyle &I_{1}(0,1,1,1)\big|_{(2,1,1,0)} = 2P_{_{D}}P_{_{F}}^{2}(1-P_{_{F}}) + P_{_{F}}^{3}(1-P_{_{F}})\nonumber\\
\displaystyle &I_{2}(0,1,1,1)\big|_{(2,1,1,0)} = P_{_{D}}P_{_{F}}^{2}(1-P_{_{D}}) + P_{_{D}}P_{_{F}}^{2}(1-P_{_{F}}) + P_{_{F}}^{3}(1-P_{_{F}})\nonumber\\
\displaystyle &I_{3}(0,1,1,1)\big|_{(2,1,1,0)} = P_{_{D}}P_{_{F}}^{2}(1-P_{_{D}}) + P_{_{D}}P_{_{F}}^{2}(1-P_{_{F}}) + P_{_{F}}^{3}(1-P_{_{F}})\nonumber\\
\displaystyle &I_{4}(0,1,1,1)\big|_{(2,1,1,0)} = P_{_{D}}P_{_{F}}^{2}(1-P_{_{D}}) + 2P_{_{D}}P_{_{F}}^{2}(1-P_{_{F}})\nonumber\\
\displaystyle &I_{1}(0,1,1,0)\big|_{(2,1,1,0)} = P_{_{D}}P_{_{F}}(1-P_{_{F}})^{2} + P_{_{F}}^{2}(1-P_{_{D}})(1-P_{_{F}}) + P_{_{F}}^{2}(1-P_{_{F}})^{2}\nonumber\\
\displaystyle &I_{2}(0,1,1,0)\big|_{(2,1,1,0)} = P_{_{D}}P_{_{F}}(1-P_{_{D}})(1-P_{_{F}}) + P_{_{F}}^{2}(1-P_{_{D}})(1-P_{_{F}}) + P_{_{F}}^{2}(1-P_{_{F}})^{2}\nonumber\\
\displaystyle &I_{3}(0,1,1,0)\big|_{(2,1,1,0)} = P_{_{D}}P_{_{F}}(1-P_{_{D}})(1-P_{_{F}}) + P_{_{D}}P_{_{F}}(1-P_{_{F}})^{2} + P_{_{F}}^{2}(1-P_{_{F}})^{2}\nonumber\\
\displaystyle &I_{4}(0,1,1,0)\big|_{(2,1,1,0)} = P_{_{D}}P_{_{F}}(1-P_{_{D}})(1-P_{_{F}}) + P_{_{D}}P_{_{F}}(1-P_{_{F}})^{2} + P_{_{F}}^{2}(1-P_{_{D}})(1-P_{_{F}})\nonumber\\
\displaystyle &I_{1}(0,1,0,1)\big|_{(2,1,1,0)} = P_{_{F}}^{2}(1-P_{_{D}})(1-P_{_{F}}) + P_{_{D}}P_{_{F}}(1-P_{_{F}})^{2} + P_{_{F}}^{2}(1-P_{_{F}})^{2}\nonumber\\
\displaystyle &I_{2}(0,1,0,1)\big|_{(2,1,1,0)} = P_{_{D}}P_{_{F}}(1-P_{_{D}})(1-P_{_{F}})+ P_{_{D}}P_{_{F}}(1-P_{_{F}})^{2} + P_{_{F}}^{2}(1-P_{_{F}})^{2}\nonumber\\
\displaystyle &I_{3}(0,1,0,1)\big|_{(2,1,1,0)} = P_{_{D}}P_{_{F}}(1-P_{_{D}})(1-P_{_{F}})+ P_{_{F}}^{2}(1-P_{_{D}})(1-P_{_{F}}) + P_{_{F}}^{2}(1-P_{_{F}})^{2}\nonumber\\
\displaystyle &I_{4}(0,1,0,1)\big|_{(2,1,1,0)} = P_{_{D}}P_{_{F}}(1-P_{_{D}})(1-P_{_{F}})+ P_{_{F}}^{2}(1-P_{_{D}})(1-P_{_{F}}) + P_{_{D}}P_{_{F}}(1-P_{_{F}})^{2}\nonumber\\
\displaystyle &I_{1}(0,1,0,0)\big|_{(2,1,1,0)} = 2P_{_{F}}(1-P_{_{D}})(1-P_{_{F}})^{2} + P_{_{F}}(1-P_{_{F}})^{3}\nonumber\\
\displaystyle &I_{2}(0,1,0,0)\big|_{(2,1,1,0)} = P_{_{D}}(1-P_{_{D}})(1-P_{_{F}})^{2} + P_{_{F}}(1-P_{_{D}})(1-P_{_{F}})^{2} + P_{_{F}}(1-P_{_{F}})^{3}\nonumber \\
\displaystyle &I_{3}(0,1,0,0)\big|_{(2,1,1,0)} = P_{_{D}}(1-P_{_{D}})(1-P_{_{F}})^{2} + P_{_{F}}(1-P_{_{D}})(1-P_{_{F}})^{2} + P_{_{F}}(1-P_{_{F}})^{3}\nonumber\\
\displaystyle &I_{4}(0,1,0,0)\big|_{(2,1,1,0)} = P_{_{D}}(1-P_{_{D}})(1-P_{_{F}})^{2} + 2P_{_{F}}(1-P_{_{D}})(1-P_{_{F}})^{2}\nonumber\\
\displaystyle &I_{1}(0,0,1,1)\big|_{(2,1,1,0)} = 2P_{_{D}}P_{_{F}}(1-P_{_{F}})^{2} + P_{_{F}}^{2}(1-P_{_{F}})^{2}\nonumber\\
\displaystyle &I_{2}(0,0,1,1)\big|_{(2,1,1,0)} = P_{_{F}}^{2}(1-P_{_{D}})^{2} + P_{_{D}}P_{_{F}}(1-P_{_{F}})^{2} + P_{_{F}}^{2}(1-P_{_{F}})^{2}\nonumber\\
\displaystyle &I_{3}(0,0,1,1)\big|_{(2,1,1,0)} = P_{_{F}}^{2}(1-P_{_{D}})^{2} + P_{_{D}}P_{_{F}}(1-P_{_{F}})^{2} + P_{_{F}}^{2}(1-P_{_{F}})^{2}\nonumber\\
\displaystyle &I_{4}(0,0,1,1)\big|_{(2,1,1,0)} = P_{_{F}}^{2}(1-P_{_{D}})^{2} + 2P_{_{D}}P_{_{F}}(1-P_{_{F}})^{2}\nonumber\\
\displaystyle &I_{1}(0,0,1,0)\big|_{(2,1,1,0)} = P_{_{D}}(1-P_{_{F}})^{3} + P_{_{F}}(1-P_{_{D}})(1-P_{_{F}})^{2} + P_{_{F}}(1-P_{_{F}})^{3}\nonumber\\
\displaystyle &I_{2}(0,0,1,0)\big|_{(2,1,1,0)} = P_{_{F}}(1-P_{_{F}})(1-P_{_{D}})^{2} + P_{_{F}}(1-P_{_{D}})(1-P_{_{F}})^{2} + P_{_{F}}(1-P_{_{F}})^{3}\nonumber
\end{align}
\begin{align}
\label{I2110}
\displaystyle &I_{3}(0,0,1,0)\big|_{(2,1,1,0)} = P_{_{F}}(1-P_{_{F}})(1-P_{_{D}})^{2} + P_{_{D}}(1-P_{_{F}})^{3} + P_{_{F}}(1-P_{_{F}})^{3}\nonumber\\
\displaystyle &I_{4}(0,0,1,0)\big|_{(2,1,1,0)} = P_{_{F}}(1-P_{_{F}})(1-P_{_{D}})^{2} + P_{_{D}}(1-P_{_{F}})^{3} + P_{_{F}}(1-P_{_{D}})(1-P_{_{F}})^{2}\nonumber\\
\displaystyle &I_{1}(0,0,0,1)\big|_{(2,1,1,0)} = P_{_{F}}(1-P_{_{D}})(1-P_{_{F}})^{2} + P_{_{D}}(1-P_{_{F}})^{3} + P_{_{F}}(1-P_{_{F}})^{3}\nonumber\\
\displaystyle &I_{2}(0,0,0,1)\big|_{(2,1,1,0)} = P_{_{F}}(1-P_{_{F}})(1-P_{_{D}})^{2} + P_{_{D}}(1-P_{_{F}})^{3} + P_{_{F}}(1-P_{_{F}})^{3}\nonumber\\
\displaystyle &I_{3}(0,0,0,1)\big|_{(2,1,1,0)} = P_{_{F}}(1-P_{_{F}})(1-P_{_{D}})^{2} + P_{_{F}}(1-P_{_{D}})(1-P_{_{F}})^{2} + P_{_{F}}(1-P_{_{F}})^{3}\nonumber\\
\displaystyle &I_{4}(0,0,0,1)\big|_{(2,1,1,0)} = P_{_{F}}(1-P_{_{F}})(1-P_{_{D}})^{2} + P_{_{F}}(1-P_{_{D}})(1-P_{_{F}})^{2} + P_{_{D}}(1-P_{_{F}})^{3}\nonumber\\
\displaystyle &I_{1}(0,0,0,0)\big|_{(2,1,1,0)} = 2(1-P_{_{D}})(1-P_{_{F}})^{3} + (1-P_{_{F}})^{4}\nonumber\\
\displaystyle &I_{2}(0,0,0,0)\big|_{(2,1,1,0)} = (1-P_{_{D}})^{2}(1-P_{_{F}})^{2} + (1-P_{_{D}})(1-P_{_{F}})^{3} + (1-P_{_{F}})^{4}\nonumber\\
\displaystyle &I_{3}(0,0,0,0)\big|_{(2,1,1,0)} = (1-P_{_{D}})^{2}(1-P_{_{F}})^{2} + (1-P_{_{D}})(1-P_{_{F}})^{3} + (1-P_{_{F}})^{4}\nonumber\\
\displaystyle &I_{4}(0,0,0,0)\big|_{(2,1,1,0)} = (1-P_{_{D}})^{2}(1-P_{_{F}})^{2} + 2(1-P_{_{D}})(1-P_{_{F}})^{3}
\end{align}

Now we compute the conditional probabilties $\displaystyle p_{_{1}}(y_{_{1}},y_{_{2}},y_{_{3}},y_{_{4}})\big|_{(v_{_{1}},v_{_{2}},v_{_{3}},v_{_{4}})}$ for the sensor placement $\displaystyle (v_{_{1}},v_{_{2}},v_{_{3}},v_{_{4}}) = (2,2,0,0)$ using equation (5) and (6). 
\begin{equation}
\displaystyle p_{_{1}}(1,1,1,1)\big|_{(2,2,0,0)} = P_{_{D}}^{2}P_{_{F}}^{2}~,~ p_{_{1}}(1,1,1,0)\big|_{(2,2,0,0)} = P_{_{D}}^{2}P_{_{F}}(1-P_{_{F}})\nonumber  
\end{equation}
\begin{equation}
\displaystyle p_{_{1}}(1,1,0,1)\big|_{(2,2,0,0)} = P_{_{D}}^{2}P_{_{F}}(1-P_{_{F}}) ~,~ p_{_{1}}(1,1,0,0)\big|_{(2,2,0,0)} = P_{_{D}}^{2}(1-P_{_{F}})^{2} \nonumber
\end{equation}
\begin{equation}
\displaystyle p_{_{1}}(1,0,1,1)\big|_{(2,2,0,0)} = P_{_{D}}(P_{_{F}}^{2})(1-P_{_{F}})~,~ p_{_{1}}(1,0,1,0)\big|_{(2,2,0,0)} = P_{_{D}}P_{_{F}}(1-P_{_{D}})(1-P_{_{F}})\nonumber 
\end{equation}
\begin{equation}
\displaystyle p_{_{1}}(1,0,0,1)\big|_{(2,2,0,0)} = P_{_{D}}P_{_{F}}(1-P_{_{D}})(1-P_{_{F}})~,~ p_{_{1}}(1,0,0,0)\big|_{(2,2,0,0)} = P_{_{D}}(1-P_{_{D}})(1-P_{_{F}})^{2} \nonumber
\end{equation}
\begin{equation}
\displaystyle p_{_{1}}(0,1,1,1)\big|_{(2,2,0,0)} = P_{_{D}}(P_{_{F}}^{2})(1-P_{_{D}})~,~ p_{_{1}}(0,1,1,0)\big|_{(2,2,0,0)} = P_{_{D}}P_{_{F}}(1-P_{_{D}})(1-P_{_{F}}) \nonumber
\end{equation}
\begin{equation}
\displaystyle p_{_{1}}(0,1,0,1)\big|_{(2,2,0,0)} = P_{_{D}}P_{_{F}}(1-P_{_{D}})(1-P_{_{F}})~,~ p_{_{1}}(0,1,0,0)\big|_{(2,2,0,0)} = P_{_{D}}(1-P_{_{D}})(1-P_{_{F}})^{2} \nonumber
\end{equation}
\begin{equation}
\displaystyle p_{_{1}}(0,0,1,1)\big|_{(2,2,0,0)} = P_{_{F}}^{2}(1-P_{_{D}})^{2}~,~ p_{_{1}}(0,0,1,0)\big|_{(2,2,0,0)} = P_{_{F}}(1-P_{_{F}})(1-P_{_{D}})^{2} \nonumber
\end{equation}
\begin{equation}
\displaystyle p_{_{1}}(0,0,0,1)\big|_{(2,2,0,0)} = P_{_{F}}(1-P_{_{F}})(1-P_{_{D}})^{2}~,~ p_{_{1}}(0,0,0,0)\big|_{(2,2,0,0)} = (1-P_{_{D}})^{2}(1-P_{_{F}})^{2} \nonumber
\end{equation}
The conditional probabilities $\displaystyle p_{_{2}}(y_{_{1}},y_{_{2}},y_{_{3}},y_{_{4}})\big|_{(2,2,0,0)}$ are given as follows:
\begin{equation}
\displaystyle p_{_{2}}(1,1,1,1)\big|_{(2,2,0,0)} = P_{_{D}}^{2}P_{_{F}}^{2}~,~ p_{_{2}}(1,1,1,0)\big|_{(2,2,0,0)} = P_{_{D}}P_{_{F}}^{2}(1-P_{_{D}})  \nonumber
\end{equation}
\begin{equation}
\displaystyle p_{_{2}}(1,1,0,1)\big|_{(2,2,0,0)} = P_{_{D}}P_{_{F}}^{2}(1-P_{_{D}}) ~,~ p_{_{2}}(1,1,0,0)\big|_{(2,2,0,0)} = P_{_{F}}^{2}(1-P_{_{D}})^{2} \nonumber
\end{equation}
\begin{equation}
\displaystyle p_{_{2}}(1,0,1,1)\big|_{(2,2,0,0)} = P_{_{D}}^{2}P_{_{F}}(1-P_{_{F}})~,~ p_{_{2}}(1,0,1,0)\big|_{(2,2,0,0)} = P_{_{D}}P_{_{F}}(1-P_{_{D}})(1-P_{_{F}}) \nonumber
\end{equation}
\begin{equation}
\displaystyle p_{_{2}}(1,0,0,1)\big|_{(2,2,0,0)} = P_{D}P_{F}(1-P_{D})(1-P_{F})~,~ p_{_{2}}(1,0,0,0)\big|_{(2,2,0,0)} = P_{F}(1-P_{F})(1-P_{D})^{2}\nonumber
\end{equation}
\begin{equation}
\displaystyle p_{_{2}}(0,1,1,1)\big|_{(2,2,0,0)} = P_{_{D}}^{2}P_{_{F}}(1-P_{_{F}})~,~ p_{_{2}}(0,1,1,0)\big|_{(2,2,0,0)} = P_{_{D}}P_{_{F}}(1-P_{_{D}})(1-P_{_{F}}) \nonumber 
\end{equation}
\begin{equation}
\displaystyle p_{_{2}}(0,1,0,1)\big|_{(2,2,0,0)} = P_{_{D}}P_{_{F}}(1-P_{_{D}})(1-P_{_{F}})~,~ p_{_{2}}(0,1,0,0)\big|_{(2,2,0,0)} = P_{_{F}}(1-P_{_{F}})(1-P_{_{D}})^{2} \nonumber
\end{equation}
\begin{equation}
\displaystyle p_{_{2}}(0,0,1,1)\big|_{(2,2,0,0)} = P_{_{D}}^{2}(1-P_{_{F}})^{2}~,~ p_{_{2}}(0,0,1,0)\big|_{(2,2,0,0)} = P_{_{D}}(1-P_{_{D}})(1-P_{_{F}})^{2} \nonumber
\end{equation}
\begin{equation}
\displaystyle p_{_{2}}(0,0,0,1)\big|_{(2,2,0,0)} = P_{_{D}}(1-P_{_{D}})(1-P_{_{F}})^{2}~,~ p_{_{2}}(0,0,0,0)\big|_{(2,2,0,0)} = (1-P_{_{D}})^{2}(1-P_{_{F}})^{2} \nonumber
\end{equation}
The conditional probabilities $\displaystyle p_{_{3}}(y_{_{1}},y_{_{2}},y_{_{3}},y_{_{4}})\big|_{(2,2,0,0)}$ are given as follows:
\begin{equation}
\displaystyle p_{_{3}}(1,1,1,1)\big|_{(2,2,0,0)} = P_{_{F}}^{4}~,~ p_{_{3}}(1,1,1,0)\big|_{(2,2,0,0)} = P_{_{F}}^{3}(1-P_{_{F}})  \nonumber
\end{equation}
\begin{equation}
\displaystyle p_{_{3}}(1,1,0,1)\big|_{(2,2,0,0)} = P_{_{F}}^{3}(1-P_{_{F}})~,~ p_{_{3}}(1,1,0,0)\big|_{(2,2,0,0)} = P_{_{F}}^{2}(1-P_{_{F}})^{2} \nonumber
\end{equation}
\begin{equation}
\displaystyle p_{_{3}}(1,0,1,1)\big|_{(2,2,0,0)} = P_{_{F}}^{3}(1-P_{_{F}})~,~ p_{_{3}}(1,0,1,0)\big|_{(2,2,0,0)} = P_{_{F}}^{2}(1-P_{_{F}})^{2} \nonumber
\end{equation}
\begin{equation}
\displaystyle p_{_{3}}(1,0,0,1)\big|_{(2,2,0,0)} = P_{_{F}}^{2}(1-P_{_{F}})^{2}~,~ p_{_{3}}(1,0,0,0)\big|_{(2,2,0,0)} = P_{_{F}}(1-P_{_{F}})^{3} \nonumber
\end{equation}
\begin{equation}
\displaystyle p_{_{3}}(0,1,1,1)\big|_{(2,2,0,0)} = P_{_{F}}^{3}(1-P_{_{F}})~,~ p_{_{3}}(0,1,1,0)\big|_{(2,2,0,0)} = P_{_{F}}^{2}(1-P_{_{F}})^{2} \nonumber
\end{equation}
\begin{equation}
\displaystyle p_{_{3}}(0,1,0,1)\big|_{(2,2,0,0)} = P_{_{F}}^{2}(1-P_{_{F}})^{2} ~,~ p_{_{3}}(0,1,0,0)\big|_{(2,2,0,0)} = P_{_{F}}(1-P_{_{F}})^{3} \nonumber
\end{equation}
\begin{equation}
\displaystyle p_{_{3}}(0,0,1,1)\big|_{(2,2,0,0)} = P_{_{F}}^{2}(1-P_{_{F}})^{2}~,~ p_{_{3}}(0,0,1,0)\big|_{(2,2,0,0)} = P_{_{F}}(1-P_{_{F}})^{3} \nonumber  
\end{equation}
\begin{equation}
\displaystyle p_{_{3}}(0,0,0,1)\big|_{(2,2,0,0)} = P_{_{F}}(1-P_{_{F}})^{3} ~,~ p_{_{3}}(0,0,0,0)\big|_{(2,2,0,0)} = (1-P_{_{F}})^{4}\nonumber 
\end{equation}

It should be noted that in this case $\displaystyle p_{_{3}}(y_{_{1}},y_{_{2}},y_{_{3}},y_{_{4}})\big|_{(2,2,0,0)} = \displaystyle p_{_{4}}(y_{_{1}},y_{_{2}},y_{_{3}},y_{_{4}})\big|_{(2,2,0,0)}$. Using equation (\ref{I(y)}) and the conditional probabilities we get the following values for $I_{i}(y_{_{1}}, y_{_{2}}, y_{_{3}}, y_{_{4}})\big|_{(2,2,0,0)}$:
\begin{align}
\displaystyle &I_{1}(1,1,1,1)\big|_{(2,2,0,0)} = P_{_{D}}^{2}P_{_{F}}^{2} + 2P_{_{F}}^{4}\nonumber
\end{align}
\begin{align}
\displaystyle &I_{2}(1,1,1,1)\big|_{(2,2,0,0)} = P_{_{D}}^{2}P_{_{F}}^{2} + 2P_{_{F}}^{4}\nonumber\\
\displaystyle &I_{3}(1,1,1,1)\big|_{(2,2,0,0)} = 2P_{_{D}}^{2}P_{_{F}}^{2} + P_{_{F}}^{4}\nonumber\\
\displaystyle &I_{1}(1,1,1,0)\big|_{(2,2,0,0)} = P_{_{D}}P_{_{F}}^{2}(1-P_{_{D}}) + 2P_{_{F}}^{3}(1-P_{_{F}})\nonumber\\
\displaystyle &I_{2}(1,1,1,0)\big|_{(2,2,0,0)} = P_{_{D}}^{2}P_{_{F}}(1-P_{_{F}}) + 2P_{_{F}}^{3}(1-P_{_{F}})\nonumber\\
\displaystyle &I_{3}(1,1,1,0)\big|_{(2,2,0,0)} = P_{_{D}}^{2}P_{_{F}}(1-P_{_{F}}) + P_{_{D}}P_{_{F}}^{2}(1-P_{_{D}}) + P_{_{F}}^{3}(1-P_{_{F}})\nonumber\\
\displaystyle &I_{1}(1,1,0,1)\big|_{(2,2,0,0)} = P_{D}P_{_{F}}^{2}(1-P_{_{D}}) + 2P_{_{F}}^{3}(1-P_{_{F}}) \nonumber\\
\displaystyle &I_{2}(1,1,0,1)\big|_{(2,2,0,0)} = P_{_{D}}^{2}P_{_{F}}(1-P_{_{F}}) + 2P_{_{F}}^{3}(1-P_{_{F}})\nonumber\\
\displaystyle &I_{3}(1,1,0,1)\big|_{(2,2,0,0)} = P_{_{D}}^{2}P_{_{F}}(1-P_{_{F}}) +P_{_{D}}P_{_{F}}^{2}(1-P_{_{D}}) + P_{_{F}}^{3}(1-P_{_{F}})\nonumber\\
\displaystyle &I_{1}(1,1,0,0)\big|_{(2,2,0,0)} = P_{_{F}}^{2}(1-P_{_{D}})^{2} + 2P_{_{F}}^{2}(1-P_{_{F}})^{2}\nonumber\\
\displaystyle &I_{2}(1,1,0,0)\big|_{(2,2,0,0)} = P_{_{D}}^{2}(1-P_{_{F}})^{2} + 2P_{_{F}}^{2}(1-P_{_{F}})^{2}\nonumber\\
\displaystyle &I_{3}(1,1,0,0)\big|_{(2,2,0,0)} = P_{_{D}}^{2}(1-P_{_{F}})^{2} +P_{_{F}}^{2}(1-P_{_{D}})^{2} + P_{_{F}}^{2}(1-P_{_{F}})^{2}\nonumber\\
\displaystyle &I_{1}(1,0,1,1)\big|_{(2,2,0,0)} = P_{_{D}}^{2}P_{_{F}}(1-P_{_{F}}) + 2P_{_{F}}^{3}(1-P_{_{F}})\nonumber\\
\displaystyle &I_{2}(1,0,1,1)\big|_{(2,2,0,0)} = P_{_{D}}P_{_{F}}^{2}(1-P_{_{D}}) + 2P_{_{F}}^{3}(1-P_{_{F}})\nonumber\\
\displaystyle &I_{3}(1,0,1,1)\big|_{(2,2,0,0)} = P_{_{D}}P_{_{F}}^{2}(1-P_{_{F}}) +P_{_{F}}^{2}P_{_{F}}(1-P_{_{F}}) + P_{_{F}}^{3}(1-P_{_{F}})\nonumber\\
\displaystyle &I_{1}(1,0,1,0)\big|_{(2,2,0,0)} = P_{_{D}}P_{_{F}}(1-P_{_{D}})(1-P_{_{F}}) + 2P_{_{F}}^{2}(1-P_{_{F}})^{2}\nonumber\\
\displaystyle &I_{2}(1,0,1,0)\big|_{(2,2,0,0)} = P_{_{D}}P_{_{F}}(1-P_{_{D}})(1-P_{_{F}}) + 2P_{_{F}}^{2}(1-P_{_{F}})^{2}\nonumber\\
\displaystyle &I_{3}(1,0,1,0)\big|_{(2,2,0,0)} = 2P_{_{D}}P_{_{F}}(1-P_{_{D}})(1-P_{_{F}}) + P_{_{F}}^{2}(1-P_{_{F}})^{2}\nonumber\\
\displaystyle &I_{1}(1,0,0,1)\big|_{(2,2,0,0)} = P_{_{D}}P_{_{F}}(1-P_{_{D}})(1-P_{_{F}}) + 2P_{_{F}}^{2}(1-P_{_{F}})^{2}\nonumber\\
\displaystyle &I_{2}(1,0,0,1)\big|_{(2,2,0,0)} = P_{_{D}}P_{_{F}}(1-P_{_{D}})(1-P_{_{F}}) + 2P_{_{F}}^{2}(1-P_{_{F}})^{2}\nonumber\\
\displaystyle &I_{3}(1,0,0,1)\big|_{(2,2,0,0)} = 2P_{_{D}}P_{_{F}}(1-P_{_{D}})(1-P_{_{F}}) + P_{_{F}}^{2}(1-P_{_{F}})^{2}\nonumber\\
\displaystyle &I_{1}(1,0,0,0)\big|_{(2,2,0,0)} = P_{_{F}}(1-P_{_{F}})(1-P_{_{D}})^{2} + 2P_{_{F}}(1-P_{_{F}})^{3}\nonumber\\
\displaystyle &I_{2}(1,0,0,0)\big|_{(2,2,0,0)} = P_{_{D}}(1-P_{_{D}})(1-P_{_{F}})^{2} + 2P_{_{F}}(1-P_{_{F}})^{3}\nonumber\\
\displaystyle &I_{3}(1,0,0,0)\big|_{(2,2,0,0)} = P_{_{D}}(1-P_{_{D}})(1-P_{_{F}})^{2} + P_{_{F}}(1-P_{_{F}})(1-P_{_{D}})^{2} + P_{_{F}}(1-P_{_{F}})^{3}\nonumber\\
\displaystyle &I_{1}(0,1,1,1)\big|_{(2,2,0,0)} = P_{_{D}}^{2}P_{_{F}}(1-P_{_{F}}) + 2P_{_{F}}^{3}(1-P_{_{F}})\nonumber\\
\displaystyle &I_{2}(0,1,1,1)\big|_{(2,2,0,0)} = P_{_{D}}P_{_{F}}^{2}(1-P_{_{D}}) + 2P_{_{F}}^{3}(1-P_{_{F}})\nonumber
\end{align}
\begin{align}
\label{I2200}
\displaystyle &I_{3}(0,1,1,1)\big|_{(2,2,0,0)} = P_{_{D}}P_{_{F}}^{2}(1-P_{_{D}}) +P_{_{D}}^{2}P_{_{F}}(1-P_{_{F}}) + P_{_{F}}^{3}(1-P_{_{F}})\nonumber \\
\displaystyle &I_{1}(0,1,1,0)\big|_{(2,2,0,0)} = P_{_{D}}P_{_{F}}(1-P_{_{D}})(1-P_{_{F}}) + 2P_{_{F}}^{2}(1-P_{_{F}})^{2}\nonumber\\
\displaystyle &I_{2}(0,1,1,0)\big|_{(2,2,0,0)} = P_{_{D}}P_{_{F}}(1-P_{_{D}})(1-P_{_{F}}) + 2P_{_{F}}^{2}(1-P_{_{F}})^{2}\nonumber\\
\displaystyle &I_{3}(0,1,1,0)\big|_{(2,2,0,0)} = 2P_{_{D}}P_{_{F}}(1-P_{_{D}})(1-P_{_{F}}) + P_{_{F}}^{2}(1-P_{_{F}})^{2}\nonumber\\
\displaystyle &I_{1}(0,1,0,1)\big|_{(2,2,0,0)} = P_{_{D}}P_{_{F}}(1-P_{_{D}})(1-P_{_{F}}) + 2P_{_{F}}^{2}(1-P_{_{F}})^{2}\nonumber\\
\displaystyle &I_{2}(0,1,0,1)\big|_{(2,2,0,0)} = P_{_{D}}P_{_{F}}(1-P_{_{D}})(1-P_{_{F}}) + 2P_{_{F}}^{2}(1-P_{_{F}})^{2}\nonumber\\
\displaystyle &I_{3}(0,1,0,1)\big|_{(2,2,0,0)} = 2P_{_{D}}P_{_{F}}(1-P_{_{D}})(1-P_{_{F}}) + P_{_{F}}^{2}(1-P_{_{F}})^{2}\nonumber\\
\displaystyle &I_{1}(0,1,0,0)\big|_{(2,2,0,0)} = P_{_{F}}(1-P_{_{F}})(1-P_{_{D}})^{2} + 2P_{_{F}}(1-P_{_{F}})^{3}\nonumber\\
\displaystyle &I_{2}(0,1,0,0)\big|_{(2,2,0,0)} = P_{_{D}}(1-P_{_{D}})(1-P_{_{F}})^{2} + 2P_{_{F}}(1-P_{_{F}})^{3}\nonumber\\
\displaystyle &I_{3}(0,1,0,0)\big|_{(2,2,0,0)} = P_{_{D}}(1-P_{_{D}})(1-P_{_{F}})^{2} + P_{_{F}}(1-P_{_{F}})(1-P_{_{D}})^{2} + P_{_{F}}(1-P_{_{F}})^{3}\nonumber\\
\displaystyle &I_{1}(0,0,1,1)\big|_{(2,2,0,0)} = P_{_{D}}^{2}(1-P_{_{F}})^{2} + 2P_{_{F}}^{2}(1-P_{_{F}})^{2}\nonumber\\
\displaystyle &I_{2}(0,0,1,1)\big|_{(2,2,0,0)} = P_{_{F}}^{2}(1-P_{_{D}})^{2} + 2P_{_{F}}^{2}(1-P_{_{F}})^{2}\nonumber\\
\displaystyle &I_{3}(0,0,1,1)\big|_{(2,2,0,0)} = P_{_{F}}^{2}(1-P_{_{D}})^{2} +P_{_{D}}^{2}(1-P_{_{F}})^{2} + P_{_{F}}^{2}(1-P_{_{F}})^{2}\nonumber\\
\displaystyle &I_{1}(0,0,1,0)\big|_{(2,2,0,0)} = P_{_{D}}(1-P_{_{D}})(1-P_{_{F}})^{2} + 2P_{_{F}}(1-P_{_{F}})^{3}\nonumber\\
\displaystyle &I_{2}(0,0,1,0)\big|_{(2,2,0,0)} = P_{_{F}}(1-P_{_{F}})(1-P_{_{D}})^{2} + 2P_{_{F}}(1-P_{_{F}})^{3}\nonumber\\
\displaystyle &I_{3}(0,0,1,0)\big|_{(2,2,0,0)} = P_{_{F}}(1-P_{_{F}})(1-P_{_{D}})^{2} +P_{_{D}}(1-P_{_{D}})(1-P_{_{F}})^{2} + P_{_{F}}(1-P_{_{F}})^{3}\nonumber\\
\displaystyle &I_{1}(0,0,0,1)\big|_{(2,2,0,0)} = P_{_{D}}(1-P_{_{D}})(1-P_{_{F}})^{2} + 2P_{_{F}}(1-P_{_{F}})^{3}\nonumber\\
\displaystyle &I_{2}(0,0,0,1)\big|_{(2,2,0,0)} = P_{_{F}}(1-P_{_{F}})(1-P_{_{D}})^{2} + 2P_{_{F}}(1-P_{_{F}})^{3}\nonumber\\
\displaystyle &I_{3}(0,0,0,1)\big|_{(2,2,0,0)} = P_{_{F}}(1-P_{_{F}})(1-P_{_{D}})^{2} +P_{_{D}}(1-P_{_{D}})(1-P_{_{F}})^{2} + P_{_{F}}(1-P_{_{F}})^{3}\nonumber\\
\displaystyle &I_{1}(0,0,0,0)\big|_{(2,2,0,0)} = (1-P_{_{D}})^{2}(1-P_{_{F}})^{2} + 2(1-P_{_{F}})^{4}\nonumber\\
\displaystyle &I_{2}(0,0,0,0)\big|_{(2,2,0,0)} = (1-P_{_{D}})^{2}(1-P_{_{F}})^{2} + 2(1-P_{_{F}})^{4}\nonumber\\
\displaystyle &I_{3}(0,0,0,0)\big|_{(2,2,0,0)} = 2(1-P_{_{D}})^{2}(1-P_{_{F}})^{2} + (1-P_{_{F}})^{4}
\end{align}
The conditional probabilties $\displaystyle p_{_{1}}(y_{_{1}},y_{_{2}},y_{_{3}},y_{_{4}})\big|_{(v_{_{1}},v_{_{2}},v_{_{3}},v_{_{4}})}$ for the sensor placement $\displaystyle (v_{_{1}},v_{_{2}},v_{_{3}},v_{_{4}}) = (3,1,0,0)$ are given by: 
\begin{equation}
\displaystyle p_{_{1}}(1,1,1,1)\big|_{(3,1,0,0)} = P_{_{D}}^{3}P_{_{F}}~,~ p_{_{1}}(1,1,1,0)\big|_{(3,1,0,0)} = P_{_{D}}^{3}(1-P_{_{F}})  \nonumber
\end{equation}
\begin{equation}
\displaystyle p_{_{1}}(1,1,0,1)\big|_{(3,1,0,0)} = P_{_{D}}^{2}P_{_{F}}(1-P_{_{D}}) ~,~ p_{_{1}}(1,1,0,0)\big|_{(3,1,0,0)} = P_{_{D}}^{2}(1-P_{_{D}})(1-P_{_{F}})  \nonumber
\end{equation}
\begin{equation}
\displaystyle p_{_{1}}(1,0,1,1)\big|_{(3,1,0,0)} = P_{_{D}}^{2}P_{_{F}}(1-P_{_{D}})~,~ p_{_{1}}(1,0,1,0)\big|_{(3,1,0,0)} = P_{_{D}}^{2}(1-P_{_{D}})(1-P_{_{F}}) \nonumber
\end{equation}
\begin{equation}
\displaystyle p_{_{1}}(1,0,0,1)\big|_{(3,1,0,0)} = P_{_{D}}P_{_{F}}(1-P_{_{D}})^{2}~,~ p_{_{1}}(1,0,0,0)\big|_{(3,1,0,0)} = P_{_{D}}(1-P_{_{F}})(1-P_{_{D}})^{2} \nonumber 
\end{equation}
\begin{equation}
\displaystyle p_{_{1}}(0,1,1,1)\big|_{(3,1,0,0)} = P_{_{D}}^{2}P_{_{F}}(1-P_{_{D}})~,~ p_{_{1}}(0,1,1,0)\big|_{(3,1,0,0)} = P_{_{D}}^{2}(1-P_{_{D}})(1-P_{_{F}}) \nonumber
\end{equation}
\begin{equation}
\displaystyle p_{_{1}}(0,1,0,1)\big|_{(3,1,0,0)} = P_{_{D}}P_{_{F}}(1-P_{_{D}})^{2}~,~ p_{_{1}}(0,1,0,0)\big|_{(3,1,0,0)} = P_{_{D}}(1-P_{_{F}})(1-P_{_{D}})^{2} \nonumber 
\end{equation}
\begin{equation}
\displaystyle p_{_{1}}(0,0,1,1)\big|_{(3,1,0,0)} = P_{_{D}}P_{_{F}}(1-P_{_{D}})^{2}~,~ p_{_{1}}(0,0,1,0)\big|_{(3,1,0,0)} = P_{_{D}}(1-P_{_{F}})(1-P_{_{D}})^{2} \nonumber 
\end{equation}
\begin{equation}
\displaystyle p_{_{1}}(0,0,0,1)\big|_{(3,1,0,0)} = P_{_{F}}(1-P_{_{D}})^{3}~,~ p_{_{1}}(0,0,0,0)\big|_{(3,1,0,0)} = (1-P_{_{D}})^{3}(1-P_{_{F}}) \nonumber
\end{equation}

The conditional probabilties $\displaystyle p_{_{2}}(y_{_{1}},y_{_{2}},y_{_{3}},y_{_{4}})\big|_{(v_{_{1}},v_{_{2}},v_{_{3}},v_{_{4}})}$ for the sensor placement $\displaystyle (v_{_{1}},v_{_{2}},v_{_{3}},v_{_{4}}) = (3,1,0,0)$ are given by: 
\begin{equation}
\displaystyle p_{_{2}}(1,1,1,1)\big|_{(3,1,0,0)} = P_{_{D}}P_{_{F}}^{3}~,~ p_{_{2}}(1,1,1,0)\big|_{(3,1,0,0)} =  P_{_{F}}^{3}(1-P_{_{D}}) \nonumber
\end{equation}
\begin{equation}
\displaystyle p_{_{2}}(1,1,0,1)\big|_{(3,1,0,0)} = P_{_{D}}P_{_{F}}^{2}(1-P_{_{F}})~,~ p_{_{2}}(1,1,0,0)\big|_{(3,1,0,0)} = P_{_{F}}^{2}(1-P_{_{D}})(1-P_{_{F}})  \nonumber
\end{equation}
\begin{equation}
\displaystyle p_{_{2}}(1,0,1,1)\big|_{(3,1,0,0)} = P_{_{D}}P_{_{F}}^{2}(1-P_{_{F}})~,~ p_{_{2}}(1,0,1,0)\big|_{(3,1,0,0)} = P_{_{F}}^{2}(1-P_{_{D}})(1-P_{_{F}})  \nonumber
\end{equation}
\begin{equation}
\displaystyle p_{_{2}}(1,0,0,1)\big|_{(3,1,0,0)} = P_{_{D}}P_{_{F}}(1-P_{_{F}})^{2}~,~ p_{_{2}}(1,0,0,0)\big|_{(3,1,0,0)} = P_{_{F}}(1-P_{_{D}})(1-P_{_{F}})^{2} \nonumber 
\end{equation}
\begin{equation}
\displaystyle p_{_{2}}(0,1,1,1)\big|_{(3,1,0,0)} = P_{_{D}}P_{_{F}}^{2}(1-P_{_{F}})~,~ p_{_{2}}(0,1,1,0)\big|_{(3,1,0,0)} = P_{_{F}}^{2}(1-P_{_{D}})(1-P_{_{F}})  \nonumber
\end{equation}
\begin{equation}
\displaystyle p_{_{2}}(0,1,0,1)\big|_{(3,1,0,0)} = P_{_{D}}P_{_{F}}(1-P_{_{F}})^{2}~,~ p_{_{2}}(0,1,0,0)\big|_{(3,1,0,0)} = P_{_{F}}(1-P_{_{D}})(1-P_{_{D}})^{2}\nonumber 
\end{equation}
\begin{equation}
\displaystyle p_{_{2}}(0,0,1,1)\big|_{(3,1,0,0)} = P_{_{D}}P_{_{F}}(1-P_{_{F}})^{2}~,~ p_{_{2}}(0,0,1,0)\big|_{(3,1,0,0)} = P_{_{F}}(1-P_{_{D}})(1-P_{_{F}})^{2} \nonumber 
\end{equation}
\begin{equation}
\displaystyle p_{_{2}}(0,0,0,1)\big|_{(3,1,0,0)} = P_{_{D}}(1-P_{_{F}})^{3}~,~ p_{_{2}}(0,0,0,0)\big|_{(3,1,0,0)} = (1-P_{_{D}})(1-P_{_{F}})^{3}\nonumber
\end{equation}
The conditional probabilties $\displaystyle p_{_{3}}(y_{_{1}},y_{_{2}},y_{_{3}},y_{_{4}})\big|_{(v_{_{1}},v_{_{2}},v_{_{3}},v_{_{4}})}$ for the sensor placement $\displaystyle (v_{_{1}},v_{_{2}},v_{_{3}},v_{_{4}}) = (3,1,0,0)$ are given by: 
\begin{equation}
\displaystyle p_{_{3}}(1,1,1,1)\big|_{(3,1,0,0)} = P_{_{F}}^{4}~,~ p_{_{3}}(1,1,1,0)\big|_{(3,1,0,0)} = P_{_{F}}^{3}(1-P_{_{F}})  \nonumber
\end{equation}
\begin{equation}
\displaystyle p_{_{3}}(1,1,0,1)\big|_{(3,1,0,0)} = P_{_{F}}^{3}(1-P_{_{F}})~,~ p_{_{3}}(1,1,0,0)\big|_{(3,1,0,0)} = P_{_{F}}^{2}(1-P_{_{F}})^{2}  \nonumber
\end{equation}
\begin{equation}
\displaystyle p_{_{3}}(1,0,1,1)\big|_{(3,1,0,0)} = P_{_{F}}^{3}(1-P_{_{F}})~,~ p_{_{3}}(1,0,1,0)\big|_{(3,1,0,0)} =P_{_{F}}^{2}(1-P_{_{F}})^{2}   \nonumber
\end{equation}
\begin{equation}
\displaystyle p_{_{3}}(1,0,0,1)\big|_{(3,1,0,0)} = P_{_{F}}^{2}(1-P_{_{F}})^{2} ~,~ p_{_{3}}(1,0,0,0)\big|_{(3,1,0,0)} = P_{_{F}}(1-P_{_{F}})^{3} \nonumber 
\end{equation}
\begin{equation}
\displaystyle p_{_{3}}(0,1,1,1)\big|_{(3,1,0,0)} = P_{_{F}}^{3}(1-P_{_{F}})~,~ p_{_{3}}(0,1,1,0)\big|_{(3,1,0,0)} = P_{_{F}}^{2}(1-P_{_{F}})^{2} \nonumber
\end{equation}
\begin{equation}
\displaystyle p_{_{3}}(0,1,0,1)\big|_{(3,1,0,0)} = P_{_{F}}^{2}(1-P_{_{F}})^{2}~,~ p_{_{3}}(0,1,0,0)\big|_{(3,1,0,0)} = P_{_{F}}(1-P_{_{F}})^{3}\nonumber 
\end{equation}
\begin{equation}
\displaystyle p_{_{3}}(0,0,1,1)\big|_{(3,1,0,0)} = P_{_{F}}^{2}(1-P_{_{F}})^{2}~,~ p_{_{3}}(0,0,1,0)\big|_{(3,1,0,0)} = P_{_{F}}(1-P_{_{F}})^{3}\nonumber 
\end{equation}
\begin{equation}
\displaystyle p_{_{3}}(0,0,0,1)\big|_{(3,1,0,0)} = P_{_{F}}(1-P_{_{F}})^{3}~,~ p_{_{3}}(0,0,0,0)\big|_{(3,1,0,0)} = (1-P_{_{F}})^{4}\nonumber
\end{equation}
Using equation (\ref{I(y)}) and the conditional probabilities we get the following values for $I_{i}(y_{_{1}}, y_{_{2}}, y_{_{3}}, y_{_{4}})\big|_{(3,1,0,0)}$:
\begin{align}
\displaystyle &I_{1}(1,1,1,1)\big|_{(3,1,0,0)} = P_{_{D}}P_{_{F}}^{3} + 2P_{_{F}}^{4}\nonumber\\
\displaystyle &I_{2}(1,1,1,1)\big|_{(3,1,0,0)} = P_{_{D}}^{3}P_{_{F}} + 2P_{_{F}}^{4}\nonumber\\
\displaystyle &I_{3}(1,1,1,1)\big|_{(3,1,0,0)} = P_{_{D}}^{3}P_{_{F}} + P_{_{D}}P_{_{F}}^{3} + P_{_{F}}^{4}\nonumber\\
\displaystyle &I_{1}(1,1,1,0)\big|_{(3,1,0,0)} = P_{_{F}}^{3}(1-P_{_{D}}) + 2P_{_{F}}^{3}(1-P_{_{F}})\nonumber\\
\displaystyle &I_{2}(1,1,1,0)\big|_{(3,1,0,0)} = P_{_{D}}^{3}(1-P_{_{F}}) + 2P_{_{F}}^{3}(1-P_{_{F}})\nonumber\\
\displaystyle &I_{3}(1,1,1,0)\big|_{(3,1,0,0)} = P_{_{D}}^{3}(1-P_{_{F}}) + P_{_{F}}^{3}(1-P_{_{D}}) + P_{_{F}}^{3}(1-P_{_{F}})\nonumber\\
\displaystyle &I_{1}(1,1,0,1)\big|_{(3,1,0,0)} = P_{_{D}}P_{_{F}}^{2}(1-P_{_{F}}) + 2P_{_{F}}^{3}(1-P_{_{F}})\nonumber\\
\displaystyle &I_{2}(1,1,0,1)\big|_{(3,1,0,0)} = P_{_{D}}^{2}P_{_{F}}(1-P_{_{D}}) + 2P_{_{F}}^{3}(1-P_{_{F}})\nonumber\\
\displaystyle &I_{3}(1,1,0,1)\big|_{(3,1,0,0)} = P_{_{D}}^{2}P_{_{F}}(1-P_{_{D}}) + P_{_{D}}P_{_{F}}^{2}(1-P_{_{F}}) + P_{_{F}}^{3}(1-P_{_{F}})\nonumber \\
\displaystyle &I_{1}(1,1,0,0)\big|_{(3,1,0,0)} = P_{_{F}}^{2}(1-P_{_{D}})(1-P_{_{F}}) + 2P_{_{F}}^{2}(1-P_{_{F}})^{2}\nonumber \\
\displaystyle &I_{2}(1,1,0,0)\big|_{(3,1,0,0)} = P_{_{D}}^{2}(1-P_{_{D}})(1-P_{_{F}}) + 2P_{_{F}}^{2}(1-P_{_{F}})^{2}\nonumber
\end{align}
\begin{align}
\displaystyle &I_{3}(1,1,0,0)\big|_{(3,1,0,0)} = P_{_{D}}^{2}(1-P_{_{D}})(1-P_{_{F}}) +P_{_{F}}^{2}(1-P_{_{D}})(1-P_{_{F}}) + P_{_{F}}^{2}(1-P_{_{F}})^{2}\nonumber\\
\displaystyle &I_{1}(1,0,1,1)\big|_{(3,1,0,0)} = P_{D}P_{F}^{2}(1-P_{F}) + 2P_{F}^{3}(1-P_{F})\nonumber\\
\displaystyle &I_{2}(1,0,1,1)\big|_{(3,1,0,0)} = P_{_{D}}^{2}P_{_{F}}(1-P_{_{D}}) + 2P_{_{F}}^{3}(1-P_{_{F}})\nonumber\\
\displaystyle &I_{3}(1,0,1,1)\big|_{(3,1,0,0)} = P_{_{D}}^{2}P_{_{F}}(1-P_{_{D}}) + P_{_{D}}P_{_{F}}^{2}(1-P_{_{F}}) + P_{_{F}}^{3}(1-P_{_{F}})\nonumber\\
\displaystyle &I_{1}(1,0,1,0)\big|_{(3,1,0,0)} = P_{_{F}}^{2}(1-P_{_{D}})(1-P_{_{F}}) + 2P_{_{F}}^{2}(1-P_{_{F}})^{2}\nonumber\\
\displaystyle &I_{2}(1,0,1,0)\big|_{(3,1,0,0)} = P_{_{D}}^{2}(1-P_{_{D}})(1-P_{_{F}}) + 2P_{_{F}}^{2}(1-P_{_{F}})^{2}\nonumber\\
\displaystyle &I_{3}(1,0,1,0)\big|_{(3,1,0,0)} = P_{_{D}}^{2}(1-P_{_{D}})(1-P_{_{F}}) + P_{_{F}}^{2}(1-P_{_{D}})(1-P_{_{F}}) + P_{_{F}}^{2}(1-P_{_{F}})^{2}\nonumber\\
\displaystyle &I_{1}(1,0,0,1)\big|_{(3,1,0,0)} = P_{_{D}}P_{_{F}}(1-P_{_{F}})^{2} + 2P_{_{F}}^{2}(1-P_{_{F}})^{2}\nonumber \\
\displaystyle &I_{2}(1,0,0,1)\big|_{(3,1,0,0)} = P_{_{D}}P_{_{F}}(1-P_{_{D}})^{2} + 2P_{_{F}}^{2}(1-P_{_{F}})^{2}\nonumber\\
\displaystyle &I_{3}(1,0,0,1)\big|_{(3,1,0,0)} = P_{_{D}}P_{_{F}}(1-P_{_{D}})^{2} +P_{_{D}}P_{_{F}}(1-P_{_{F}})^{2} + P_{_{F}}^{2}(1-P_{_{F}})^{2}\nonumber \\
\displaystyle &I_{1}(1,0,0,0)\big|_{(3,1,0,0)} = P_{_{F}}(1-P_{_{D}})(1-P_{_{F}})^{2} + 2P_{_{F}}(1-P_{_{F}})^{3}\nonumber\\
\displaystyle &I_{2}(1,0,0,0)\big|_{(3,1,0,0)} = P_{_{D}}(1-P_{_{F}})(1-P_{_{D}})^{2} + 2P_{_{F}}(1-P_{_{F}})^{3}\nonumber\\
\displaystyle &I_{3}(1,0,0,0)\big|_{(3,1,0,0)} = P_{_{D}}(1-P_{_{F}})(1-P_{_{D}})^{2} +P_{_{F}}(1-P_{_{D}})(1-P_{_{F}})^{2} + P_{_{F}}(1-P_{_{F}})^{3}\nonumber\\
\displaystyle &I_{1}(0,1,1,1)\big|_{(3,1,0,0)} = P_{_{D}}P_{_{F}}^{2}(1-P_{_{F}}) + 2P_{_{F}}^{3}(1-P_{_{F}})\nonumber\\
\displaystyle &I_{2}(0,1,1,1)\big|_{(3,1,0,0)} = P_{_{D}}^{2}P_{_{F}}(1-P_{_{D}}) + 2P_{_{F}}^{3}(1-P_{_{F}})\nonumber\\
\displaystyle &I_{3}(0,1,1,1)\big|_{(3,1,0,0)} = P_{_{D}}^{2}P_{_{F}}(1-P_{_{D}}) + P_{D}P_{_{F}}^{2}(1-P_{_{F}}) + P_{_{F}}^{3}(1-P_{_{F}})\nonumber\\
\displaystyle &I_{1}(0,1,1,0)\big|_{(3,1,0,0)} = P_{_{F}}^{2}(1-P_{_{D}})(1-P_{_{F}}) + 2P_{_{F}}^{2}(1-P_{_{F}})^{2}\nonumber\\
\displaystyle &I_{2}(0,1,1,0)\big|_{(3,1,0,0)} = P_{_{D}}^{2}(1-P_{_{D}})(1-P_{_{F}}) + 2P_{_{F}}^{2}(1-P_{_{F}})^{2}\nonumber\\
\displaystyle &I_{3}(0,1,1,0)\big|_{(3,1,0,0)} = P_{_{D}}^{2}(1-P_{_{D}})(1-P_{_{F}}) + P_{_{F}}^{2}(1-P_{_{D}})(1-P_{_{F}}) + P_{_{F}}^{2}(1-P_{_{F}})^{2}\nonumber \\
\displaystyle &I_{1}(0,1,0,1)\big|_{(3,1,0,0)} = P_{_{D}}P_{_{F}}(1-P_{_{F}})^{2} + 2P_{_{F}}^{2}(1-P_{_{F}})^{2}\nonumber\\
\displaystyle &I_{2}(0,1,0,1)\big|_{(3,1,0,0)} = P_{_{D}}P_{_{F}}(1-P_{_{D}})^{2} + 2P_{_{F}}^{2}(1-P_{_{F}})^{2}\nonumber\\
\displaystyle &I_{3}(0,1,0,1)\big|_{(3,1,0,0)} = P_{_{D}}P_{_{F}}(1-P_{_{D}})^{2} + P_{_{D}}P_{_{F}}(1-P_{_{F}})^{2} + P_{_{F}}^{2}(1-P_{_{F}})^{2}\nonumber \\
\displaystyle &I_{1}(0,1,0,0)\big|_{(3,1,0,0)} = P_{_{F}}(1-P_{_{D}})(1-P_{_{F}})^{2} + 2P_{_{F}}(1-P_{_{F}})^{3}\nonumber\\
\displaystyle &I_{2}(0,1,0,0)\big|_{(3,1,0,0)} = P_{_{D}}(1-P_{_{F}})(1-P_{_{D}})^{2} + 2P_{_{F}}(1-P_{_{F}})^{3}\nonumber\\
\displaystyle &I_{3}(0,1,0,0)\big|_{(3,1,0,0)} = P_{_{D}}(1-P_{_{F}})(1-P_{_{D}})^{2} + P_{_{F}}(1-P_{_{D}})(1-P_{_{F}}^{2}) + P_{_{F}}(1-P_{_{F}})^{3}\nonumber
\end{align}
\begin{align}
\label{I3100}
\displaystyle &I_{1}(0,0,1,1)\big|_{(3,1,0,0)} = P_{_{D}}P_{_{F}}(1-P_{_{F}})^{2} + 2P_{_{F}}^{2}(1-P_{_{F}})^{2}\nonumber\\
\displaystyle &I_{2}(0,0,1,1)\big|_{(3,1,0,0)} = P_{_{D}}P_{_{F}}(1-P_{_{D}})^{2} + 2P_{_{F}}^{2}(1-P_{_{F}})^{2}\nonumber\\
\displaystyle &I_{3}(0,0,1,1)\big|_{(3,1,0,0)} = P_{_{D}}P_{_{F}}(1-P_{_{D}})^{2} +P_{_{D}}P_{_{F}}(1-P_{_{F}})^{2} + P_{_{F}}^{2}(1-P_{_{F}})^{2}\nonumber\\
\displaystyle &I_{1}(0,0,1,0)\big|_{(3,1,0,0)} = P_{_{F}}(1-P_{_{D}})(1-P_{_{F}})^{2} + 2P_{_{F}}(1-P_{_{F}})^{3}\nonumber\\
\displaystyle &I_{2}(0,0,1,0)\big|_{(3,1,0,0)} = P_{_{D}}(1-P_{_{F}})(1-P_{_{D}})^{2} + 2P_{_{F}}(1-P_{_{F}})^{3}\nonumber\\
\displaystyle &I_{3}(0,0,1,0)\big|_{(3,1,0,0)} = P_{_{D}}(1-P_{_{F}})(1-P_{_{D}})^{2} + P_{_{F}}(1-P_{_{D}})(1-P_{_{F}})^{2} + P_{_{F}}(1-P_{_{F}})^{3}\nonumber\\
\displaystyle &I_{1}(0,0,0,1)\big|_{(3,1,0,0)} = P_{_{D}}(1-P_{_{F}})^{3} + 2P_{_{F}}(1-P_{_{F}})^{3}\nonumber \\
\displaystyle &I_{2}(0,0,0,1)\big|_{(3,1,0,0)} = P_{_{F}}(1-P_{_{D}})^{3} + 2P_{_{F}}(1-P_{_{F}})^{3}\nonumber\\
\displaystyle &I_{3}(0,0,0,1)\big|_{(3,1,0,0)} = P_{_{D}}(1-P_{_{F}})^{3} + P_{_{F}}(1-P_{_{D}})^{3} + P_{_{F}}(1-P_{_{F}})^{3}\nonumber \\
\displaystyle &I_{1}(0,0,0,0)\big|_{(3,1,0,0)} = (1-P_{_{D}})(1-P_{_{F}})^{3} + 2(1-P_{_{F}})^{4}\nonumber\\
\displaystyle &I_{2}(0,0,0,0)\big|_{(3,1,0,0)} = (1-P_{_{D}})^{3}(1-P_{_{F}}) + 2(1-P_{_{F}})^{4}\nonumber\\
\displaystyle &I_{3}(0,0,0,0)\big|_{(3,1,0,0)} = (1-P_{_{D}})^{3}(1-P_{_{F}}) + (1-P_{_{D}})(1-P_{_{F}})^{3} + (1-P_{_{F}})^{4}
\end{align}

The conditional probabilties $\displaystyle p_{_{1}}(y_{_{1}},y_{_{2}},y_{_{3}},y_{_{4}})\big|_{(v_{_{1}},v_{_{2}},v_{_{3}},v_{_{4}})}$ for the sensor placement $\displaystyle (v_{_{1}},v_{_{2}},v_{_{3}},v_{_{4}}) = (4,0,0,0)$ are given by: 
\begin{equation}
\displaystyle p_{_{1}}(1,1,1,1)\big|_{(4,0,0,0)} = P_{_{D}}^{4}~,~ p_{_{1}}(1,1,1,0)\big|_{(4,0,0,0)} = P_{_{D}}^{3}(1-P_{_{D}})  \nonumber
\end{equation}
\begin{equation}
\displaystyle p_{_{1}}(1,1,0,1)\big|_{(4,0,0,0)} = P_{_{D}}^{3}(1-P_{_{D}}) ~,~ p_{_{1}}(1,1,0,0)\big|_{(4,0,0,0)} = P_{_{D}}^{2}(1-P_{_{D}})^{2} \nonumber
\end{equation}
\begin{equation}
\displaystyle p_{_{1}}(1,0,1,1)\big|_{(4,0,0,0)} = P_{_{D}}^{3}(1-P_{_{D}})~,~ p_{_{1}}(1,0,1,0)\big|_{(4,0,0,0)} = P_{_{D}}^{2}(1-P_{_{D}})^{2} \nonumber
\end{equation}
\begin{equation}
\displaystyle p_{_{1}}(1,0,0,1)\big|_{(4,0,0,0)} = P_{_{D}}^{2}(1-P_{_{D}})^{2}~,~ p_{_{1}}(1,0,0,0)\big|_{(4,0,0,0)} = P_{_{D}}(1-P_{_{D}})^{3}\nonumber 
\end{equation}
\begin{equation}
\displaystyle p_{_{1}}(0,1,1,1)\big|_{(4,0,0,0)} = P_{_{D}}^{3}(1-P_{_{D}})~,~ p_{_{1}}(0,1,1,0)\big|_{(4,0,0,0)} = P_{_{D}}^{2}(1-P_{_{D}})^{2} \nonumber
\end{equation}
\begin{equation}
\displaystyle p_{_{1}}(0,1,0,1)\big|_{(4,0,0,0)} = P_{_{D}}^{2}(1-P_{_{D}})^{2}~,~ p_{_{1}}(0,1,0,0)\big|_{(4,0,0,0)} = P_{_{D}}(1-P_{_{D}})^{3} \nonumber 
\end{equation}
\begin{equation}
\displaystyle p_{_{1}}(0,0,1,1)\big|_{(4,0,0,0)} = P_{_{D}}^{2}(1-P_{_{D}})^{2}~,~ p_{_{1}}(0,0,1,0)\big|_{(4,0,0,0)} = P_{_{D}}(1-P_{_{D}})^{3} \nonumber 
\end{equation}
\begin{equation}
\displaystyle p_{_{1}}(0,0,0,1)\big|_{(4,0,0,0)} = P_{_{D}}(1-P_{_{D}})^{3}~,~ p_{_{1}}(0,0,0,0)\big|_{(4,0,0,0)} = (1-P_{_{D}})^{4} \nonumber
\end{equation}

The conditional probabilties $\displaystyle p_{_{2}}(y_{_{1}},y_{_{2}},y_{_{3}},y_{_{4}})\big|_{(v_{_{1}},v_{_{2}},v_{_{3}},v_{_{4}})}$ for the sensor placement $\displaystyle (v_{_{1}},v_{_{2}},v_{_{3}},v_{_{4}}) = (4,0,0,0)$ are given by: 
\begin{equation}
\displaystyle p_{_{2}}(1,1,1,1)\big|_{(4,0,0,0)} = P_{_{F}}^{4}~,~ p_{_{2}}(1,1,1,0)\big|_{(4,0,0,0)} = P_{_{F}}^{3}(1-P_{_{F}}) \nonumber
\end{equation}
\begin{equation}
\displaystyle p_{_{2}}(1,1,0,1)\big|_{(4,0,0,0)} = P_{_{F}}^{3}(1-P_{_{F}})~,~ p_{_{2}}(1,1,0,0)\big|_{(4,0,0,0)} = P_{_{F}}^{2}(1-P_{_{F}})^{2} \nonumber
\end{equation}
\begin{equation}
\displaystyle p_{_{2}}(1,0,1,1)\big|_{(4,0,0,0)} = P_{_{F}}^{3}(1-P_{_{F}})~,~ p_{_{2}}(1,0,1,0)\big|_{(4,0,0,0)} = P_{_{F}}^{2}(1-P_{_{F}})^{2} \nonumber
\end{equation}
\begin{equation}
\displaystyle p_{_{2}}(1,0,0,1)\big|_{(4,0,0,0)} =P_{_{F}}^{2}(1-P_{_{F}})^{2} ~,~ p_{_{2}}(1,0,0,0)\big|_{(4,0,0,0)} = P_{_{F}}(1-P_{_{F}})^{3} \nonumber 
\end{equation}
\begin{equation}
\displaystyle p_{_{2}}(0,1,1,1)\big|_{(4,0,0,0)} = P_{_{F}}^{3}(1-P_{_{F}})~,~ p_{_{2}}(0,1,1,0)\big|_{(4,0,0,0)} = P_{_{F}}^{2}(1-P_{_{F}})^{2}  \nonumber
\end{equation}
\begin{equation}
\displaystyle p_{_{2}}(0,1,0,1)\big|_{(4,0,0,0)} = P_{_{F}}^{2}(1-P_{_{F}})^{2}~,~ p_{_{2}}(0,1,0,0)\big|_{(4,0,0,0)} = P_{_{F}}(1-P_{_{F}})^{3}\nonumber 
\end{equation}
\begin{equation}
\displaystyle p_{_{2}}(0,0,1,1)\big|_{(4,0,0,0)} = P_{_{F}}^{2}(1-P_{_{F}})^{2}~,~ p_{_{2}}(0,0,1,0)\big|_{(4,0,0,0)} = P_{_{F}}(1-P_{_{F}})^{3} \nonumber 
\end{equation}
\begin{equation}
\displaystyle p_{_{2}}(0,0,0,1)\big|_{(4,0,0,0)} = P_{_{F}}(1-P_{_{F}})^{3}~,~ p_{_{2}}(0,0,0,0)\big|_{(4,0,0,0)} = (1-P_{_{F}})^{4}\nonumber
\end{equation}
It should be noted that for this placement: $\displaystyle p_{_{2}}(y_{_{1}},y_{_{2}},y_{_{3}},y_{_{4}})\big|_{(4,0,0,0)} = p_{_{3}}(y_{_{1}},y_{_{2}},y_{_{3}},y_{_{4}})\big|_{(4,0,0,0)} = p_{_{4}}(y_{_{1}},y_{_{2}},y_{_{3}},y_{_{4}})\big|_{(4,0,0,0)}$.  
Using equation (\ref{I(y)}) and the conditional probabilities we get the following values for $I_{i}(y_{_{1}}, y_{_{2}}, y_{_{3}}, y_{_{4}})\big|_{(4,0,0,0)}$:
\begin{align}
\displaystyle &I_{1}(1,1,1,1)\big|_{(4,0,0,0)} = 3P_{_{F}}^{4}~,~I_{2}(1,1,1,1)\big|_{(4,0,0,0)} = P_{_{D}}^{4} + 2P_{_{F}}^{4}\nonumber\\
\displaystyle &I_{1}(1,1,1,0)\big|_{(4,0,0,0)} = 3P_{_{F}}^{3}(1-P_{_{F}})~,~I_{2}(1,1,1,0)\big|_{(4,0,0,0)} = P_{_{D}}^{3}(1-P_{_{D}}) + 2P_{_{F}}^{3}(1-P_{_{F}})\nonumber\\
\displaystyle &I_{1}(1,1,0,1)\big|_{(4,0,0,0)} = 3P_{_{F}}^{3}(1-P_{_{F}})~,~I_{2}(1,1,0,1)\big|_{(4,0,0,0)} = P_{_{D}}^{3}(1-P_{_{D}}) + 2P_{_{F}}^{3}(1-P_{_{F}})\nonumber\\
\displaystyle &I_{1}(1,1,0,0)\big|_{(4,0,0,0)} = 3P_{_{F}}^{2}(1-P_{_{F}})^{2}~,~I_{2}(1,1,0,0)\big|_{(4,0,0,0)} = P_{_{D}}^{2}(1-P_{_{D}})^{2} + 2P_{_{F}}^{2}(1-P_{_{F}})^{2}\nonumber\\
\displaystyle &I_{1}(1,0,1,1)\big|_{(4,0,0,0)} = 3P_{_{F}}^{3}(1-P_{_{F}})~,~I_{2}(1,0,1,1)\big|_{(4,0,0,0)} = P_{_{D}}^{3}(1-P_{_{D}}) + 2P_{_{F}}^{3}(1-P_{_{F}})\nonumber\\
\displaystyle &I_{1}(1,0,1,0)\big|_{(4,0,0,0)} = 3P_{_{F}}^{2}(1-P_{_{F}})^{2}~,~I_{2}(1,0,1,0)\big|_{(4,0,0,0)} = P_{_{D}}^{2}(1-P_{_{D}})^{2} + 2P_{_{F}}^{2}(1-P_{_{F}})^{2}\nonumber\\
\displaystyle &I_{1}(1,0,0,1)\big|_{(4,0,0,0)} = 3P_{_{F}}^{2}(1-P_{_{F}})^{2}~,~I_{2}(1,0,0,1)\big|_{(4,0,0,0)} = P_{_{D}}^{2}(1-P_{_{D}})^{2} + 2P_{_{F}}^{2}(1-P_{_{F}})^{2}\nonumber\\
\displaystyle &I_{1}(1,0,0,0)\big|_{(4,0,0,0)} = 3P_{_{F}}(1-P_{_{F}})^{3}~,~I_{2}(1,0,0,0)\big|_{(4,0,0,0)} = P_{_{D}}(1-P_{_{D}})^{3} + 2P_{_{F}}(1-P_{_{F}})^{3}\nonumber\\
\displaystyle &I_{1}(0,1,1,1)\big|_{(4,0,0,0)} = 3P_{_{F}}^{3}(1-P_{_{F}})~,~I_{2}(0,1,1,1)\big|_{(4,0,0,0)} = P_{_{D}}^{3}(1-P_{_{D}}) + 2P_{_{F}}^{3}(1-P_{_{F}})\nonumber
\end{align}
\begin{align}
\label{I4000}
\displaystyle &I_{1}(0,1,1,0)\big|_{(4,0,0,0)} = 3P_{_{F}}^{2}(1-P_{_{F}})^{2}~,~I_{2}(0,1,1,0)\big|_{(4,0,0,0)} = P_{_{D}}^{2}(1-P_{_{D}})^{2} + 2P_{_{F}}^{2}(1-P_{_{F}})^{2}\nonumber\\
\displaystyle &I_{1}(0,1,0,1)\big|_{(4,0,0,0)} = 3P_{_{F}}^{2}(1-P_{_{F}})^{2}~,~I_{2}(0,1,0,1)\big|_{(4,0,0,0)} = P_{_{D}}^{2}(1-P_{_{D}})^{2} + 2P_{_{F}}^{2}(1-P_{_{F}})^{2}\nonumber\\
\displaystyle &I_{1}(0,1,0,0)\big|_{(4,0,0,0)} = 3P_{_{F}}(1-P_{_{F}})^{3}~,~I_{2}(0,1,0,0)\big|_{(4,0,0,0)} = P_{_{D}}(1-P_{_{D}})^{3} + 2P_{_{F}}(1-P_{_{F}})^{3}\nonumber\\
\displaystyle &I_{1}(0,0,1,1)\big|_{(4,0,0,0)} = 3P_{_{F}}^{2}(1-P_{_{F}})^{2}~,~I_{2}(0,0,1,1)\big|_{(4,0,0,0)} = P_{_{D}}^{2}(1-P_{_{D}})^{2} + 2P_{_{F}}^{2}(1-P_{_{F}})^{2}\nonumber\\
\displaystyle &I_{1}(0,0,1,0)\big|_{(4,0,0,0)} = 3P_{_{F}}(1-P_{_{F}})^{3}~,~I_{2}(0,0,1,0)\big|_{(4,0,0,0)} = P_{_{D}}(1-P_{_{D}})^{3} + 2P_{_{F}}(1-P_{_{F}})^{3}\nonumber\\
\displaystyle &I_{1}(0,0,0,1)\big|_{(4,0,0,0)} = 3P_{_{F}}(1-P_{_{F}})^{3}~,~I_{2}(0,0,0,1)\big|_{(4,0,0,0)} = P_{_{D}}(1-P_{_{D}})^{3} + 2P_{_{F}}(1-P_{_{F}})^{3}\nonumber \\
\displaystyle &I_{1}(0,0,0,0)\big|_{(4,0,0,0)} = 3(1-P_{_{F}})^{4}~,~I_{2}(0,0,0,0)\big|_{(4,0,0,0)} = (1-P_{_{D}})^{4} + 2(1-P_{_{F}})^{4}
\end{align}
Using equations (\ref{I2110}), (\ref{I2200}), (\ref{I3100}), and (\ref{I4000}) we can obtain the probability of error of each placement respectively:
\begin{equation}
\displaystyle 4P_{e}(2,1,1,0)=  \min\bigg\{2P_{_{D}}P_{_{F}}^{3} + P_{_{F}}^{4}, P_{_{D}}^{2}P_{_{F}}^{2} + P_{D}P_{_{F}}^{3} + P_{_{F}}^{4}, P_{_{D}}^{2}P_{_{F}}^{2} + 2P_{_{D}}P_{_{F}}^{3}\bigg\} + 2\min\bigg\{ P_{_{D}}P_{_{F}}^{2}(1-P_{_{F}}) \nonumber 
\end{equation}
\begin{equation}
\displaystyle  + P_{_{F}}^{3}(1-P_{_{D}}) + P_{_{F}}^{3}(1-P_{_{F}}), P_{_{D}}^{2}P_{_{F}}(1-P_{_{F}}) + P_{_{F}}^{3}(1-P_{_{D}}) + P_{_{F}}^{3}(1-P_{_{F}}), P_{_{D}}^{2}P_{_{F}}(1-P_{_{F}}) + P_{_{D}}P_{_{F}}^{2}(1-P_{_{F}})  \nonumber 
\end{equation}
\begin{equation}
\displaystyle  + P_{_{F}}^{3}(1-P_{_{F}}), P_{_{D}}^{2}P_{_{F}}(1-P_{_{F}}) + P_{_{D}}P_{_{F}}^{2}(1-P_{_{F}}) + P_{_{F}}^{3}(1-P_{_{D}})\bigg\} + \min\bigg\{2P_{_{F}}^{2}(1-P_{_{D}})(1-P_{_{F}}) +  \nonumber 
\end{equation}
\begin{equation} 
\displaystyle P_{_{F}}^{2}(1-P_{_{F}})^{2},P_{_{D}}^{2}(1-P_{_{F}})^{2} + P_{_{F}}^{2}(1-P_{_{D}})(1-P_{_{F}}) + P_{_{F}}^{2}(1-P_{_{F}})^{2}, P_{_{D}}^{2}(1-P_{_{F}})^{2} + 2P_{_{F}}^{2}(1-P_{_{D}})(1-P_{_{F}})\bigg\} + \nonumber 
\end{equation}
\begin{equation}
\displaystyle 2\min\bigg\{ 2P_{_{D}}P_{_{F}}^{2}(1-P_{_{F}}) + P_{_{F}}^{3}(1-P_{_{F}}), P_{_{D}}P_{_{F}}^{2}(1-P_{_{D}}) + P_{_{D}}P_{_{F}}^{2}(1-P_{_{F}}) + P_{_{F}}^{3}(1-P_{_{F}}), P_{_{D}}P_{_{F}}^{2}(1-P_{_{D}}) \nonumber 
\end{equation}
\begin{equation}
\displaystyle + 2P_{_{D}}P_{_{F}}^{2}(1-P_{_{F}})\bigg\}+ 4\min\bigg\{P_{_{D}}P_{_{F}}(1-P_{_{F}})^{2} + P_{_{F}}^{2}(1-P_{_{D}})(1-P_{_{F}}) + P_{_{F}}^{2}(1-P_{_{F}})^{2}, P_{_{D}}P_{_{F}}(1-P_{_{D}})(1-P_{_{F}})\nonumber 
\end{equation}
\begin{equation}
\displaystyle  + P_{_{F}}^{2}(1-P_{_{D}})(1-P_{_{F}}) + P_{_{F}}^{2}(1-P_{_{F}})^{2}, P_{_{D}}P_{_{F}}(1-P_{_{D}})(1-P_{_{F}}) + P_{_{D}}P_{_{F}}(1-P_{_{F}})^{2} + P_{_{F}}^{2}(1-P_{_{F}})^{2},\nonumber 
\end{equation}
\begin{equation}
\displaystyle  P_{_{D}}P_{_{F}}(1-P_{_{D}})(1-P_{_{F}}) + P_{_{D}}P_{_{F}}(1-P_{_{F}})^{2} + P_{_{F}}^{2}(1-P_{_{D}})(1-P_{_{F}})\bigg\} + 2\min\bigg\{ 2P_{_{F}}(1-P_{_{D}})(1-P_{_{F}})^{2}  \nonumber 
\end{equation}
\begin{equation}
\displaystyle + P_{_{F}}(1-P_{_{F}})^{3},P_{_{D}}(1-P_{_{D}})(1-P_{_{F}})^{2} + P_{_{F}}(1-P_{_{D}})(1-P_{_{F}})^{2} + P_{_{F}}(1-P_{_{F}})^{3}, P_{_{D}}(1-P_{_{D}})(1-P_{_{F}})^{2} + \nonumber 
\end{equation}
\begin{equation}
\displaystyle  2P_{_{F}}(1-P_{_{D}})(1-P_{_{F}})^{2} \bigg\} + \min\bigg\{ 2P_{_{D}}P_{_{F}}(1-P_{_{F}})^{2} + P_{_{F}}^{2}(1-P_{_{F}})^{2}, P_{_{F}}^{2}(1-P_{_{D}})^{2} + P_{_{D}}P_{_{F}}(1-P_{_{F}})^{2} \nonumber 
\end{equation}
\begin{equation}
\displaystyle  + P_{_{F}}^{2}(1-P_{_{F}})^{2}, P_{_{F}}^{2}(1-P_{_{D}})^{2} + 2P_{_{D}}P_{_{F}}(1-P_{_{F}})^{2}\bigg\} + 2\min\bigg\{ P_{_{D}}(1-P_{_{F}})^{3} + P_{_{F}}(1-P_{_{D}})(1-P_{_{F}})^{2} \nonumber 
\end{equation}
\begin{equation}
\displaystyle  P_{_{F}}(1-P_{_{F}})^{3}, P_{_{F}}(1-P_{_{F}})(1-P_{_{D}})^{2} + P_{_{F}}(1-P_{_{D}})(1-P_{_{F}})^{2} + P_{_{F}}(1-P_{_{F}})^{3}, P_{_{F}}(1-P_{_{F}})(1-P_{_{D}})^{2} \nonumber 
\end{equation}
\begin{equation}
\displaystyle + P_{_{D}}(1-P_{_{F}})^{3} + P_{_{F}}(1-P_{_{F}})^{3}, P_{_{F}}(1-P_{_{F}})(1-P_{_{D}})^{2} + P_{_{D}}(1-P_{_{F}})^{3} + P_{_{F}}(1-P_{_{D}})(1-P_{_{F}})^{2}\bigg\}+ \nonumber 
\end{equation}
\begin{equation}
\displaystyle \min\bigg\{2(1-P_{_{D}})(1-P_{_{F}})^{3}+ (1-P_{_{F}})^{4}, (1-P_{_{D}})^{2}(1-P_{_{F}})^{2} + (1-P_{_{D}})(1-P_{_{F}})^{3} + (1-P_{_{F}})^{4},  \nonumber 
\end{equation}
\begin{equation}
\label{Pe2110}
\displaystyle  (1-P_{_{D}})^{2}(1-P_{_{F}})^{2} +2(1-P_{_{D}})(1-P_{_{F}})^{3} \bigg\}
\end{equation}
Using the fact $P_{_{D}} \geq P_{_{F}}$, the above expression can be simplified to give:
\begin{equation}
\displaystyle 4P_{e}(2,1,1,0) = 2P_{_{D}}P_{_{F}}^{3} + P_{_{F}}^{4} + 2P_{_{D}}P_{_{F}}^{2}(1-P_{_{F}}) + 2P_{_{F}}^{3}(1-P_{_{D}}) + 2P_{_{F}}^{3}(1-P_{_{F}}) + 2P_{_{F}}^{2}(1-P_{_{D}})(1-P_{_{F}})\nonumber
\end{equation}
\begin{equation}
\displaystyle  + P_{_{F}}^{2}(1-P_{_{F}})^{2} + 2P_{_{D}}P_{_{F}}^{2}(1-P_{_{D}}) + 2P_{_{D}}P_{_{F}}^{2}(1-P_{_{F}})+ 2P_{_{F}}^{3}(1-P_{_{F}}) + 4P_{_{D}}P_{_{F}}(1-P_{_{D}})(1-P_{_{F}}) +  \nonumber
\end{equation}  
\begin{equation}
\displaystyle 4P_{_{F}}^{2}(1-P_{_{D}})(1-P_{_{F}})+ 4P_{_{F}}^{2}(1-P_{_{F}})^{2} + 2\min\{2P_{_{F}}(1-P_{_{D}})(1-P_{_{F}})^{2} + P_{_{F}}(1-P_{_{F}})^{3}, P_{_{D}}(1-P_{_{D}})(1-P_{_{F}})^{2} \nonumber
\end{equation}
\begin{equation}
\displaystyle + 2P_{_{F}}(1-P_{_{D}})(1-P_{_{F}})^{2} \}+ P_{_{F}}^{2}(1-P_{_{D}})^{2} + P_{_{D}}P_{_{F}}(1-P_{_{F}})^{2} + P_{_{F}}^{2}(1-P_{_{F}})^{2} + 2P_{_{F}}(1-P_{_{F}})(1-P_{_{D}})^{2}  \nonumber
\end{equation}
\begin{equation}
\label{Pe2110Sim}
\displaystyle + 2P_{_{F}}(1-P_{_{D}})(1-P_{_{F}})^{2}+ 2P_{_{F}}(1-P_{_{F}})^{3} + (1-P_{_{D}})^{2}(1-P_{_{F}})^{2} + 2(1-P_{_{D}})(1-P_{_{F}})^{3}
\end{equation}
There are two cases here, $P_{_{D}}(1-P_{_{D}}) \geq P_{_{F}}(1-P_{_{F}})$ and $P_{_{D}}(1-P_{_{D}}) < P_{_{F}}(1-P_{_{F}})$:
\par
\textbf{Case 1:} $\bf P_{_{D}}(1-P_{_{D}}) \geq P_{_{F}}(1-P_{_{F}})$
\begin{equation}
\displaystyle 4P_{e}(2,1,1,0) = 2P_{_{D}}P_{_{F}}^{3} + P_{_{F}}^{4} + 2P_{_{D}}P_{_{F}}^{2}(1-P_{_{F}}) + 2P_{_{F}}^{3}(1-P_{_{D}}) + 2P_{_{F}}^{3}(1-P_{_{F}}) + 2P_{_{F}}^{2}(1-P_{_{D}})(1-P_{_{F}})\nonumber
\end{equation}
\begin{equation}
\displaystyle  + P_{_{F}}^{2}(1-P_{_{F}})^{2} + 2P_{_{D}}P_{_{F}}^{2}(1-P_{_{D}}) + 2P_{_{D}}P_{_{F}}^{2}(1-P_{_{F}})+ 2P_{_{F}}^{3}(1-P_{_{F}}) + 4P_{_{D}}P_{_{F}}(1-P_{_{D}})(1-P_{_{F}}) +  \nonumber
\end{equation}  
\begin{equation}
\displaystyle 4P_{_{F}}^{2}(1-P_{_{D}})(1-P_{_{F}})+ 4P_{_{F}}^{2}(1-P_{_{F}})^{2} + 4P_{_{F}}(1-P_{_{D}})(1-P_{_{F}})^{2} + 2P_{_{F}}(1-P_{_{F}})^{3} + P_{_{F}}^{2}(1-P_{_{D}})^{2} \nonumber
\end{equation}
\begin{equation}
\displaystyle + P_{_{D}}P_{_{F}}(1-P_{_{F}})^{2} + P_{_{F}}^{2}(1-P_{_{F}})^{2} + 2P_{_{F}}(1-P_{_{F}})(1-P_{_{D}})^{2} + 2P_{_{F}}(1-P_{_{D}})(1-P_{_{F}})^{2}+ 2P_{_{F}}(1-P_{_{F}})^{3} \nonumber
\end{equation}
\begin{equation}
\label{Pe2110C1}
\displaystyle +~ (1-P_{_{D}})^{2}(1-P_{_{F}})^{2} + 2(1-P_{_{D}})(1-P_{_{F}})^{3}
\end{equation}
\textbf{Case 2:} $\bf P_{_{D}}(1-P_{_{D}}) < P_{_{F}}(1-P_{_{F}})$
\begin{equation}
\displaystyle 4P_{e}(2,1,1,0) = 2P_{_{D}}P_{_{F}}^{3} + P_{_{F}}^{4} + 2P_{_{D}}P_{_{F}}^{2}(1-P_{_{F}}) + 2P_{_{F}}^{3}(1-P_{_{D}}) + 2P_{_{F}}^{3}(1-P_{_{F}}) + 2P_{_{F}}^{2}(1-P_{_{D}})(1-P_{_{F}})\nonumber
\end{equation}
\begin{equation}
\displaystyle + P_{_{F}}^{2}(1-P_{_{F}})^{2} + 2P_{_{D}}P_{_{F}}^{2}(1-P_{_{D}}) + 2P_{_{D}}P_{_{F}}^{2}(1-P_{_{F}})+ 2P_{_{F}}^{3}(1-P_{_{F}}) + 4P_{_{D}}P_{_{F}}(1-P_{_{D}})(1-P_{_{F}}) +  \nonumber
\end{equation}  
\begin{equation}
\displaystyle 4P_{_{F}}^{2}(1-P_{_{D}})(1-P_{_{F}})+ 4P_{_{F}}^{2}(1-P_{_{F}})^{2} + 4P_{_{F}}(1-P_{_{D}})(1-P_{_{F}})^{2} + 2P_{_{D}}(1-P_{_{D}})(1-P_{_{F}})^{2} + P_{_{F}}^{2}(1-P_{_{D}})^{2} \nonumber
\end{equation}
\begin{equation}
\displaystyle + P_{_{D}}P_{_{F}}(1-P_{_{F}})^{2} + P_{_{F}}^{2}(1-P_{_{F}})^{2} + 2P_{_{F}}(1-P_{_{F}})(1-P_{_{D}})^{2} + 2P_{_{F}}(1-P_{_{D}})(1-P_{_{F}})^{2}+ 2P_{_{F}}(1-P_{_{F}})^{3} \nonumber
\end{equation}
\begin{equation}
\label{Pe2110C2}
\displaystyle +~ (1-P_{_{D}})^{2}(1-P_{_{F}})^{2} + 2(1-P_{_{D}})(1-P_{_{F}})^{3}
\end{equation}
Now we provide an expression for $P_{e}(2,2,0,0)$ using (\ref{Pewf}), (\ref{I(y)}), and (\ref{I2200}):
\begin{equation}
\displaystyle 4P_{e}(2,2,0,0) =  \min\{ P_{_{D}}^{2}P_{_{F}}^{2} + 2P_{_{F}}^{4}, 2P_{_{D}}^{2}P_{_{F}}^{2} + P_{_{F}}^{4}\} + 4\min\{P_{_{D}}P_{_{F}}^{2}(1-P_{_{D}}) + 2P_{_{F}}^{3}(1-P_{_{F}}), P_{_{D}}^{2}P_{_{F}}(1-P_{_{F}}) +  \nonumber
\end{equation}
\begin{equation}
\displaystyle 2P_{_{F}}^{3}(1-P_{_{F}}),P_{_{D}}^{2}P_{_{F}}(1-P_{_{F}}) + P_{_{D}}P_{_{F}}^{2}(1-P_{_{D}}) + P_{_{F}}^{3}(1-P_{_{F}})\} + 2\min\{P_{_{F}}^{2}(1-P_{_{D}})^{2} + 2P_{_{F}}^{2}(1-P_{_{F}})^{2}, P_{_{D}}^{2}(1-P_{_{F}})^{2} +   \nonumber
\end{equation}
\begin{equation}
\displaystyle 2P_{_{F}}^{2}(1-P_{_{F}})^{2},P_{_{D}}^{2}(1-P_{_{F}})^{2} + P_{_{F}}^{2}(1-P_{_{D}})^{2} + P_{_{F}}^{2}(1-P_{_{F}})^{2}\} +4\min\{P_{_{D}}P_{_{F}}(1-P_{_{D}})(1-P_{_{F}}) + 2P_{_{F}}^{2}(1-P_{_{F}})^{2},   \nonumber
\end{equation}
\begin{equation}
\displaystyle 2P_{_{D}}P_{_{F}}(1-P_{_{D}})(1-P_{_{F}})+ P_{_{F}}^{2}(1-P_{_{F}})^{2}\} + 4\min\{P_{_{D}}(1-P_{_{D}})(1-P_{_{F}})^{2} + 2P_{_{F}}(1-P_{_{F}})^{3}, P_{_{F}}(1-P_{_{F}})(1-P_{_{D}})^{2} +   \nonumber
\end{equation}
\begin{equation}
\displaystyle 2P_{_{F}}(1-P_{_{F}})^{3},P_{_{F}}(1-P_{_{F}})(1-P_{_{D}})^{2}+ P_{_{D}}(1-P_{_{D}})(1-P_{_{F}})^{2} + P_{_{F}}(1-P_{_{F}})^{3}\}+ \min\{(1-P_{_{D}})^{2}(1-P_{_{F}})^{2} + 2(1-P_{_{F}})^{4}, \nonumber
\end{equation}
\begin{equation}
\label{Pe2200}
\displaystyle  2(1-P_{_{D}})^{2}(1-P_{_{F}})^{2} + (1-P_{_{F}})^{4}\}
\end{equation}
Using the fact $P_{_{D}} \geq P_{_{F}}$, the above expression can be simplified to give:
\begin{equation}
\displaystyle 4P_{e}(2,2,0,0) = P_{_{D}}^{2}P_{_{F}}^{2} + 2P_{_{F}}^{4} + 4P_{_{D}}P_{_{F}}^{2}(1-P_{_{D}}) + 8P_{_{F}}^{3}(1-P_{_{F}}) + 2P_{_{F}}^{2}(1-P_{_{D}})^{2} + 4P_{_{F}}^{2}(1-P_{_{F}})^{2} +  \nonumber
\end{equation}
\begin{equation}
\displaystyle 4\min\{P_{_{D}}P_{_{F}}(1-P_{_{D}})(1-P_{_{F}}) +2P_{_{F}}^{2}(1-P_{_{F}})^{2}, 2P_{_{D}}P_{_{F}}(1-P_{_{D}})(1-P_{_{F}}) + P_{_{F}}^{2}(1-P_{_{F}})^{2}\} +   \nonumber
\end{equation}
\begin{equation}
\displaystyle 4\min\{P_{_{F}}(1-P_{_{F}})(1-P_{_{D}})^{2} + 2P_{_{F}}(1-P_{_{F}})^{3}, P_{_{F}}(1-P_{_{F}})(1-P_{_{D}})^{2}+ P_{_{D}}(1-P_{_{D}})(1-P_{_{F}})^{2}+P_{_{F}}(1-P_{_{F}})^{3}\} \nonumber
\end{equation}
\begin{equation}
\label{Pe2200Sim}
\displaystyle +2(1-P_{_{D}})^{2}(1-P_{_{F}})^{2} + (1-P_{_{F}})^{4}
\end{equation}
Now there are two cases here: $P_{_{D}}(1-P_{_{D}}) \geq P_{_{F}}(1-P_{_{F}})$ and $P_{D}(1-P_{D}) < P_{_{F}}(1-P_{_{F}})$ 
\par
\textbf{Case 1:} $\bf P_{_{D}}(1-P_{_{D}}) \geq P_{_{F}}(1-P_{_{F}})$
\begin{equation}
\displaystyle 4P_{e}(2,2,0,0) = P_{_{D}}^{2}P_{_{F}}^{2} + 2P_{_{F}}^{4} + 4P_{_{D}}P_{_{F}}^{2}(1-P_{_{D}}) + 8P_{_{F}}^{3}(1-P_{_{F}}) + 2P_{_{F}}^{2}(1-P_{_{D}})^{2}+ 4P_{_{F}}^{2}(1-P_{_{F}})^{2}+  \nonumber 
\end{equation}
\begin{equation}
\displaystyle 4P_{_{D}}P_{_{F}}(1-P_{_{D}})(1-P_{_{F}}) + 8P_{_{F}}^{2}(1-P_{_{F}})^{2} + 4P_{_{F}}(1-P_{_{F}})(1-P_{_{D}})^{2} + 8P_{_{F}}(1-P_{_{F}})^{3} + 2(1-P_{_{D}})(1-P_{_{F}})^{2} \nonumber
\end{equation}
\begin{equation}
\label{Pe2200C1}
\displaystyle  ~+~ (1-P_{_{F}})^{4}
\end{equation}
\par
\textbf{Case 2:} $\bf P_{_{D}}(1-P_{_{D}}) < P_{_{F}}(1-P_{_{F}})$
\begin{equation}
\displaystyle 4P_{e}(2,2,0,0) = P_{_{D}}^{2}P_{_{F}}^{2} + 2P_{_{F}}^{4} + 4P_{_{D}}P_{_{F}}^{2}(1-P_{_{D}}) + 8P_{_{F}}^{3}(1-P_{_{F}}) + 2P_{_{F}}^{2}(1-P_{_{D}})^{2} + 4P_{_{F}}^{2}(1-P_{_{F}})^{2}+  \nonumber
\end{equation}
\begin{equation}
\displaystyle  8P_{_{D}}P_{_{F}}(1-P_{_{D}})(1-P_{_{F}}) + 4P_{_{F}}^{2}(1-P_{_{F}})^{2} + 4P_{_{F}}(1-P_{_{F}})(1-P_{_{D}})^{2} + 4P_{_{D}}(1-P_{_{D}})(1-P_{_{F}})^{2} + 4P_{_{F}}(1-P_{_{F}})^{3}  \nonumber
\end{equation}
\begin{equation}
\label{Pe2200C2} 
\displaystyle +2(1-P_{_{D}})^{2}(1-P_{_{F}})^{2} + (1-P_{_{F}})^{4}
\end{equation}
Now we provide an expression for $P_{e}(3,1,0,0)$ using (\ref{Pewf}), (\ref{I(y)}), and (\ref{I3100}):
\begin{equation}
\displaystyle 4P_{e}(3,1,0,0) = \min\{P_{_{D}}P_{_{F}}^{3} + 2P_{_{F}}^{4}, P_{_{D}}^{3}P_{_{F}} + 2P_{_{F}}^{4}, P_{_{D}}^{3}P_{_{F}}+ P_{_{D}}P_{_{F}}^{3} + P_{_{F}}^{4}\} + \min\{P_{_{F}}^{3}(1-P_{_{D}}) + 2P_{_{F}}^{3}(1-P_{_{F}}), \nonumber
\end{equation}
\begin{equation}
\displaystyle P_{_{D}}^{3}(1-P_{_{F}}) + 2P_{_{F}}^{3}(1-P_{_{F}}), P_{_{D}}^{3}(1-P_{_{F}}) + P_{_{F}}^{3}(1-P_{_{D}}) + P_{_{F}}^{3}(1-P_{_{F}})\} + 3\min\{P_{_{D}}P_{_{F}}^{2}(1-P_{_{F}}) + 2P_{_{F}}^{3}(1-P_{_{F}}),  \nonumber
\end{equation}
\begin{equation}
\displaystyle P_{_{D}}^{2}P_{_{F}}(1-P_{_{D}})+2P_{_{F}}^{3}(1-P_{_{F}}), P_{_{D}}^{2}P_{_{F}}(1-P_{_{D}}) + P_{_{D}}P_{_{F}}^{2}(1-P_{_{F}}) + P_{_{F}}^{3}(1-P_{_{F}})\} + 3\min\{P_{_{F}}^{2}(1-P_{_{D}})(1-P_{_{F}})  \nonumber
\end{equation}
\begin{equation}
\displaystyle + 2P_{_{F}}^{2}(1-P_{_{F}})^{2},P_{_{D}}^{2}(1-P_{_{D}})(1-P_{_{F}}) + 2P_{_{F}}^{2}(1-P_{_{F}})^{2}, P_{_{D}}^{2}(1-P_{_{D}})(1-P_{_{F}}) +P_{_{F}}^{2}(1-P_{_{D}})(1-P_{_{F}})+ P_{_{F}}^{2}(1-P_{_{F}})^{2} \} \nonumber
\end{equation}
\begin{equation}
\displaystyle +3\min\{P_{_{D}}P_{_{F}}(1-P_{_{F}})^{2} + 2P_{_{F}}^{2}(1-P_{_{F}})^{2}, P_{_{D}}P_{_{F}}(1-P_{_{D}})^{2} + 2P_{_{F}}^{2}(1-P_{_{F}})^{2}, P_{_{D}}P_{_{F}}(1-P_{_{D}})^{2} + P_{_{D}}P_{_{F}}(1-P_{_{F}})^{2} +  \nonumber
\end{equation}
\begin{equation}
\displaystyle P_{_{F}}^{2}(1-P_{_{F}})^{2}\} + 3\min\{P_{_{F}}(1-P_{_{D}})(1-P_{_{F}})^{2} + 2P_{_{F}}(1-P_{_{F}})^{3}, P_{_{D}}(1-P_{_{F}})(1-P_{_{D}})^{2} + 2P_{_{F}}(1-P_{_{F}})^{3},   \nonumber
\end{equation}
\begin{equation}
\displaystyle P_{_{D}}(1-P_{_{F}})(1-P_{_{D}})^{2}+P_{_{F}}(1-P_{_{D}})(1-P_{_{F}})^{2} + P_{_{F}}(1-P_{_{F}})^{3}\} + \min\{P_{_{D}}(1-P_{_{F}})^{3} + 2P_{_{F}}(1-P_{_{F}})^{3}, P_{_{F}}(1-P_{_{D}})^{3} +   \nonumber
\end{equation}
\begin{equation}
\displaystyle 2P_{_{F}}(1-P_{_{F}})^{3},P_{_{F}}(1-P_{_{D}})^{3}+P_{_{D}}(1-P_{_{F}})^{3} + P_{_{F}}(1-P_{_{F}})^{3}\} + \min\{(1-P_{_{D}})(1-P_{_{F}})^{3} + 2(1-P_{_{F}})^{4}, (1-P_{_{D}})^{3}(1-P_{_{F}})   \nonumber
\end{equation}
\begin{equation}
\label{Pe3100}
\displaystyle + 2(1-P_{_{F}})^{4},(1-P_{_{D}})^{3}(1-P_{_{F}})+ (1-P_{_{D}})(1-P_{_{F}})^{3} + (1-P_{_{F}})^{4}\} 
\end{equation}
Using the fact $P_{_{D}} \geq P_{_{F}}$, the above expression can be simplified to give:
\begin{equation}
\displaystyle 4P_{e}(3,1,0,0) = P_{_{D}}P_{_{F}}^{3} + 2P_{_{F}}^{4} + P_{_{F}}^{3}(1-P_{_{D}}) + 2P_{_{F}}^{3}(1-P_{_{F}}) + 3P_{_{D}}P_{_{F}}\min\{P_{_{D}}(1-P_{_{D}}),P_{_{F}}(1-P_{_{F}})\} \nonumber
\end{equation}
\begin{equation}
\displaystyle +6P_{_{F}}^{3}(1-P_{_{F}}) + 3\min\{P_{_{F}}^{2}(1-P_{_{D}})(1-P_{_{F}}) + 2P_{_{F}}^{2}(1-P_{_{F}})^{2}, P_{_{D}}^{2}(1-P_{_{D}})(1-P_{_{F}}) + P_{_{F}}^{2}(1-P_{_{D}})(1-P_{_{F}})\nonumber
\end{equation}
\begin{equation}
\displaystyle P_{_{F}}^{2}(1-P_{_{F}})^{2}\} + 3P_{_{D}}P_{_{F}}(1-P_{_{D}})^{2} + 6P_{_{F}}^{2}(1-P_{_{F}})^{2} + 3\min\{P_{_{F}}(1-P_{_{D}})(1-P_{_{F}})^{2} + 2P_{_{F}}(1-P_{_{F}})^{3},  \nonumber
\end{equation}
\begin{equation}
\displaystyle P_{_{D}}(1-P_{_{F}})(1-P_{_{D}})^{2}+P_{_{F}}(1-P_{_{D}})(1-P_{_{F}})^{2} + P_{_{F}}(1-P_{_{F}})^{3}\} + P_{_{F}}(1-P_{_{D}})^{3} + 2P_{_{F}}(1-P_{_{F}})^{3} \nonumber
\end{equation}
\begin{equation}
\label{Pe3100Sim}
\displaystyle + (1-P_{_{D}})^{3}(1-P_{_{F}}) + (1-P_{_{F}})^{4}
\end{equation}
\par
We have four cases here which are given below:
\par
\textbf{Case W:} $\displaystyle \bf P_{_{D}}(1-P_{_{D}}) \geq P_{_{F}}(1-P_{_{F}}), P_{_{D}}^{2}(1-P_{_{D}})\geq P_{_{F}}^{2}(1-P_{_{F}}), P_{_{D}}(1-P_{_{D}})^{2}\geq P_{_{F}}(1-P_{_{F}})^{2}$
\begin{equation}
\displaystyle 4P_{e}(3,1,0,0) = P_{_{D}}P_{_{F}}^{3} + 2P_{_{F}}^{4} + P_{_{F}}^{3}(1-P_{_{D}}) + 2P_{_{F}}^{3}(1-P_{_{F}}) + 3P_{_{D}}P_{_{F}}^{2}(1-P_{_{F}}) + 6P_{_{F}}^{3}(1-P_{_{F}}) \nonumber
\end{equation}
\begin{equation}
\displaystyle 3P_{_{F}}^{2}(1-P_{_{D}})(1-P_{_{F}}) + 6P_{_{F}}^{2}(1-P_{_{F}})^{2} + 3P_{_{D}}P_{_{F}}(1-P_{_{D}})^{2} + 6P_{_{F}}^{2}(1-P_{_{F}})^{2} + 3P_{_{F}}(1-P_{_{D}})(1-P_{_{F}})^{2} \nonumber
\end{equation}
\begin{equation}
\label{Pe3100CW}
\displaystyle 6P_{_{F}}(1-P_{_{F}})^{3} + P_{_{F}}(1-P_{_{D}})^{3} + 2P_{_{F}}(1-P_{_{F}})^{3} + (1-P_{_{D}})^{3}(1-P_{_{F}}) + (1-P_{_{D}})(1-P_{_{F}})^{3} + (1-P_{_{F}})^{4}
\end{equation}
\par
\textbf{Case X:} $\displaystyle \bf P_{_{D}}(1-P_{_{D}}) \geq P_{_{F}}(1-P_{_{F}}), P_{_{D}}^{2}(1-P_{_{D}})\geq P_{_{F}}^{2}(1-P_{_{F}}), P_{_{D}}(1-P_{_{D}})^{2}< P_{_{F}}(1-P_{_{F}})^{2}$
\begin{equation}
\displaystyle 4P_{e}(3,1,0,0) = P_{_{D}}P_{_{F}}^{3} + 2P_{_{F}}^{4} + P_{_{F}}^{3}(1-P_{_{D}}) + 2P_{_{F}}^{3}(1-P_{_{F}}) + 3P_{_{D}}P_{_{F}}^{2}(1-P_{_{F}}) + 6P_{_{F}}^{3}(1-P_{_{F}}) \nonumber 
\end{equation}
\begin{equation}
\displaystyle + 3P_{_{F}}^{2}(1-P_{_{D}})(1-P_{_{F}}) + 6P_{_{F}}^{2}(1-P_{_{F}})^{2} + 3P_{_{D}}P_{_{F}}(1-P_{_{D}})^{2} + 6P_{_{F}}^{2}(1-P_{_{F}})^{2} + 3P_{_{D}}(1-P_{_{F}})(1-P_{_{D}})^{2} + \nonumber
\end{equation}
\begin{equation}
\displaystyle 3P_{_{F}}(1-P_{_{D}})(1-P_{_{F}})^{2} + 3P_{_{F}}(1-P_{_{F}})^{3} + P_{_{F}}(1-P_{_{D}})^{3} + 2P_{_{F}}(1-P_{_{F}})^{3} + (1-P_{_{D}})^{3}(1-P_{_{F}}) \nonumber
\end{equation}
\begin{equation}
\label{Pe3100CX}
\displaystyle  + (1-P_{_{D}})(1-P_{_{F}})^{3} + (1-P_{_{F}})^{4}
\end{equation}
\par
\textbf{Case Y:} $\displaystyle \bf P_{_{D}}(1-P_{_{D}}) < P_{_{F}}(1-P_{_{F}}), P_{_{D}}^{2}(1-P_{_{D}})\geq P_{_{F}}^{2}(1-P_{_{F}}), P_{_{D}}(1-P_{_{D}})^{2}< P_{_{F}}(1-P_{_{F}})^{2}$
\begin{equation}
\displaystyle 4P_{e}(3,1,0,0) = P_{_{D}}P_{_{F}}^{3} + 2P_{_{F}}^{4} + P_{_{F}}^{3}(1-P_{_{D}}) + 2P_{_{F}}^{3}(1-P_{_{F}}) + 3P_{_{D}}^{2}P_{_{F}}(1-P_{_{D}}) + 6P_{_{F}}^{3}(1-P_{_{F}}) 
\nonumber
\end{equation}
\begin{equation}
\displaystyle +3P_{_{F}}^{2}(1-P_{_{D}})(1-P_{_{F}}) + 6P_{_{F}}^{2}(1-P_{_{F}})^{2} + 3P_{_{D}}P_{_{F}}(1-P_{_{D}})^{2} + 6P_{_{F}}^{2}(1-P_{_{F}})^{2} + 3P_{_{D}}(1-P_{_{F}})(1-P_{_{D}})^{2} \nonumber
\end{equation}
\begin{equation}
\displaystyle 3P_{_{F}}(1-P_{_{D}})(1-P_{_{F}})^{2} + 3P_{_{F}}(1-P_{_{F}})^{3} + P_{_{F}}(1-P_{_{D}})^{3} + 2P_{_{F}}(1-P_{_{F}})^{3} + (1-P_{_{D}})^{3}(1-P_{_{F}}) \nonumber
\end{equation}
\begin{equation}
\label{Pe3100CY}
\displaystyle +(1-P_{_{D}})(1-P_{_{F}})^{3} + (1-P_{_{F}})^{4}
\end{equation}
\par
\textbf{Case Z:} $\displaystyle \bf P_{_{D}}(1-P_{_{D}}) < P_{_{F}}(1-P_{_{F}}), P_{_{D}}^{2}(1-P_{_{D}})< P_{_{F}}^{2}(1-P_{_{F}}), P_{_{D}}(1-P_{_{D}})^{2}< P_{_{F}}(1-P_{_{F}})^{2}$
\begin{equation}
\displaystyle 4P_{e}(3,1,0,0) = P_{_{D}}P_{_{F}}^{3} + 2P_{_{F}}^{4} + P_{_{F}}^{3}(1-P_{_{D}}) + 2P_{_{F}}^{3}(1-P_{_{F}}) + 3P_{_{D}}^{2}P_{_{F}}(1-P_{_{D}}) + 6P_{_{F}}^{3}(1-P_{_{F}}) \nonumber
\end{equation}
\begin{equation}
\displaystyle 3P_{_{D}}^{2}(1-P_{_{D}})(1-P_{_{F}}) + 3P_{_{F}}^{2}(1-P_{_{D}})(1-P_{_{F}}) + 3P_{_{F}}^{2}(1-P_{_{F}})^{2} + 3P_{_{D}}P_{_{F}}(1-P_{_{D}})^{2} + 6P_{_{F}}^{2}(1-P_{_{F}})^{2} \nonumber
\end{equation}
\begin{equation}
\displaystyle 3P_{_{D}}(1-P_{_{F}})(1-P_{_{D}})^{2} + 3P_{_{F}}(1-P_{_{D}})(1-P_{_{F}})^{2} + 3P_{_{F}}(1-P_{_{F}})^{3} + P_{_{F}}(1-P_{_{D}})^{3} + 2P_{_{F}}(1-P_{_{F}})^{3}  \nonumber
\end{equation}
\begin{equation}
\label{Pe3100CZ}
\displaystyle + (1-P_{_{D}})^{3}(1-P_{_{F}})+(1-P_{_{D}})(1-P_{_{F}})^{3} + (1-P_{_{F}})^{4}
\end{equation}
Now we provide an expression for $P_{e}(4,0,0,0)$ using (\ref{Pewf}), (\ref{I(y)}), and (\ref{I4000}):
\begin{equation}
\displaystyle 4P_{e}(4,0) = \min\{3P_{_{F}}^{4}, P_{_{D}}^{4} + 2P_{_{F}}^{4}\} + 4\min\{3P_{_{F}}^{3}(1-P_{_{F}}), P_{_{D}}^{3}(1-P_{_{D}}) + 2P_{_{F}}^{3}(1-P_{_{F}})\} \nonumber
\end{equation}
\begin{equation}
\displaystyle +6\min\{3P_{_{F}}^{2}(1-P_{_{F}})^{2}, P_{_{D}}^{2}(1-P_{_{D}})^{2} + 2P_{_{F}}^{2}(1-P_{_{F}})^{2}\} + 4\min\{3P_{_{F}}(1-P_{_{F}})^{3}, P_{_{D}}(1-P_{_{D}})^{3} \nonumber
\end{equation}
\begin{equation}
\label{Pe4000}
\displaystyle +2P_{_{F}}(1-P_{_{F}})^{3}\} + \min\{3(1-P_{_{F}})^{4}, (1-P_{_{D}})^{4} + 2(1-P_{_{F}})^{4}\}
\end{equation}
Using the fact $P_{_{D}} \geq P_{_{F}}$, the above expression can be simplified to give:
\begin{equation}
\displaystyle 4P_{e}(4,0) = 3P_{_{F}}^{4} + 4\min\{3P_{_{F}}^{3}(1-P_{_{F}}), P_{_{D}}^{3}(1-P_{_{D}}) + 2P_{_{F}}^{3}(1-P_{_{F}})\} \nonumber
\end{equation}
\begin{equation}
\displaystyle +6\min\{3P_{_{F}}^{2}(1-P_{_{F}})^{2}, P_{_{D}}^{2}(1-P_{_{D}})^{2} + 2P_{_{F}}^{2}(1-P_{_{F}})^{2}\} + 4\min\{3P_{_{F}}(1-P_{_{F}})^{3}, P_{_{D}}(1-P_{_{D}})^{3} \nonumber
\end{equation}
\begin{equation}
\label{Pe4000Sim}
\displaystyle +2P_{_{F}}(1-P_{_{F}})^{3}\} + (1-P_{_{D}})^{4} + 2(1-P_{_{F}})^{4}
\end{equation}
We have four cases here which are given as follows:
\par
\textbf{Case A:} $\displaystyle \bf P_{_{D}}^{3}(1-P_{_{D}}) \geq P_{_{F}}^{3}(1-P_{_{F}}), P_{_{D}}^{2}(1-P_{_{D}})^{2} \geq P_{_{F}}^{2}(1-P_{_{F}})^{2}, P_{_{D}}(1-P_{_{D}})^{3} \geq P_{_{F}}(1-P_{_{F}})^{3}$
\begin{equation}
\label{Pe4000CA}
\displaystyle 4P_{e}(4,0,0,0) = 3P_{_{F}}^{4} + 12P_{_{F}}^{3}(1-P_{_{F}}) + 18P_{_{F}}^{2}(1-P_{_{F}})^{2} + 12P_{_{F}}(1-P_{_{F}})^{3} + (1-P_{_{D}})^{4} + 2(1-P_{_{F}})^{4}
\end{equation}
\par
\textbf{Case B:} $\displaystyle \bf P_{_{D}}^{3}(1-P_{_{D}}) \geq P_{_{F}}^{3}(1-P_{_{F}}), P_{_{D}}^{2}(1-P_{_{D}})^{2} \geq P_{_{F}}^{2}(1-P_{_{F}})^{2}, P_{_{D}}(1-P_{_{D}})^{3} < P_{_{F}}(1-P_{_{F}})^{3}$
\begin{equation}
\displaystyle 4P_{e}(4,0,0,0) = 3P_{_{F}}^{4} + 12P_{_{F}}^{3}(1-P_{_{F}}) + 18P_{_{F}}^{2}(1-P_{_{F}})^{2} + 4P_{_{D}}(1-P_{_{D}})^{3} + 8P_{_{F}}(1-P_{_{F}})^{3} \nonumber
\end{equation}
\begin{equation}
\label{Pe4000CB}
\displaystyle +(1-P_{_{D}})^{4} + 2(1-P_{_{F}})^{4}
\end{equation}
\par
\textbf{Case C:} $\displaystyle \bf P_{_{D}}^{3}(1-P_{_{D}}) \geq P_{_{F}}^{3}(1-P_{_{F}}), P_{_{D}}^{2}(1-P_{_{D}})^{2} < P_{_{F}}^{2}(1-P_{_{F}})^{2}, P_{_{D}}(1-P_{_{D}})^{3} < P_{_{F}}(1-P_{_{F}})^{3}$
\begin{equation}
\displaystyle 4P_{e}(4,0,0,0) = 3P_{_{F}}^{4} + 12P_{_{F}}^{3}(1-P_{_{F}}) + 6P_{_{D}}^{2}(1-P_{_{D}})^{2} + 12P_{_{F}}^{2}(1-P_{_{F}})^{2} + 4P_{_{D}}(1-P_{_{D}})^{3} \nonumber
\end{equation}
\begin{equation}
\label{Pe4000CC}
\displaystyle +8P_{_{F}}(1-P_{_{F}})^{3} + (1-P_{_{D}})^{4} + 2(1-P_{_{F}})^{4}
\end{equation}
\par
\textbf{Case C:} $\displaystyle \bf P_{_{D}}^{3}(1-P_{_{D}}) < P_{_{F}}^{3}(1-P_{_{F}}), P_{_{D}}^{2}(1-P_{_{D}})^{2} < P_{_{F}}^{2}(1-P_{_{F}})^{2}, P_{_{D}}(1-P_{_{D}})^{3} < P_{_{F}}(1-P_{_{F}})^{3}$
\begin{equation}
\displaystyle 4P_{e}(4,0,0,0) = 3P_{_{F}}^{4} + 4P_{_{D}}^{3}(1-P_{_{D}}) + 8P_{_{F}}^{3}(1-P_{_{F}}) + 6P_{_{D}}^{2}(1-P_{_{D}})^{2} + 12P_{_{F}}^{2}(1-P_{_{F}})^{2} + \nonumber
\end{equation}
\begin{equation}
\label{Pe4000CD}
\displaystyle 4P_{_{D}}(1-P_{_{D}})^{3} + 8P_{_{F}}(1-P_{_{F}})^{3} + (1-P_{_{D}})^{4} + 2(1-P_{_{F}})^{4}
\end{equation}
Combining the aforementioned cases for equations (\ref{Pe2110}), (\ref{Pe2200}), (\ref{Pe3100}), and (\ref{Pe4000}) we obtain the following distinct cases:
\vspace{2cm}
\par
Case AW1: $\displaystyle P_{_{D}}^{3}(1-P_{_{D}}) \geq P_{_{F}}^{3}(1-P_{_{F}}), P_{_{D}}^{2}(1-P_{_{D}})^{2} \geq P_{_{F}}^{2}(1-P_{_{F}})^{2}, P_{_{D}}(1-P_{_{D}})^{3} \geq P_{_{F}}(1-P_{_{F}})^{3}, P_{_{D}}(1-P_{_{D}}) \geq P_{_{F}}(1-P_{_{F}}), P_{_{D}}^{2}(1-P_{_{D}})\geq P_{_{F}}^{2}(1-P_{_{F}}), P_{_{D}}(1-P_{_{D}})^{2}\geq P_{_{F}}(1-P_{_{F}})^{2}$ 
\par
Case BW1: $\displaystyle P_{_{D}}^{3}(1-P_{_{D}}) \geq P_{_{F}}^{3}(1-P_{_{F}}), P_{_{D}}^{2}(1-P_{_{D}})^{2} \geq P_{_{F}}^{2}(1-P_{_{F}})^{2}, P_{_{D}}(1-P_{_{D}})^{3} < P_{_{F}}(1-P_{_{F}})^{3}, P_{_{D}}(1-P_{_{D}}) \geq P_{_{F}}(1-P_{_{F}}), P_{_{D}}^{2}(1-P_{_{D}}) \geq P_{_{F}}^{2}(1-P_{_{F}}), P_{_{D}}(1-P_{_{D}})^{2} \geq P_{_{F}}(1-P_{_{F}})^{2}$
\par
Case BX1: $\displaystyle P_{_{D}}^{3}(1-P_{_{D}})\geq P_{_{F}}^{3}(1-P_{_{F}}), P_{_{D}}^{2}(1-P_{_{D}})^{2} \geq P_{_{F}}^{2}(1-P_{_{F}})^{2}, P_{_{D}}(1-P_{_{D}})^{3}< P_{_{F}}(1-P_{_{F}})^{3}, P_{_{D}}(1-P_{_{D}}) \geq P_{_{F}}(1-P_{_{F}}), P_{_{D}}^{2}(1-P_{_{D}}) \geq P_{_{F}}^{2}(1-P_{_{F}}), P_{_{D}}(1-P_{_{D}})^{2} < P_{_{F}}(1-P_{_{F}})^{2}$
\par
Case CX1: $\displaystyle P_{_{D}}^{3}(1-P_{_{D}}) \geq P_{_{F}}^{3}(1-P_{_{F}}), P_{_{D}}^{2}(1-P_{_{D}})^{2} < P_{_{F}}^{2}(1-P_{_{F}})^{2}, P_{_{D}}(1-P_{_{D}})^{3} < P_{_{F}}(1-P_{_{F}})^{3}, P_{_{D}}(1-P_{_{D}}) \geq P_{_{F}}(1-P_{_{F}}), P_{_{D}}^{2}(1-P_{_{D}}) \geq P_{_{F}}^{2}(1-P_{_{F}}), P_{_{D}}(1-P_{_{D}})^{2} < P_{_{F}}(1-P_{_{F}})^{2}$
\par
Case CY2: $\displaystyle P_{_{D}}^{3}(1-P_{_{D}}) \geq P_{_{F}}^{3}(1-P_{_{F}}), P_{_{D}}^{2}(1-P_{_{D}})^{2} < P_{_{F}}^{2}(1-P_{_{F}})^{2}, P_{_{D}}(1-P_{_{D}})^{3} < P_{_{F}}(1-P_{_{F}})^{3}, P_{_{D}}(1-P_{_{D}}) < P_{_{F}}(1-P_{_{F}}), P_{_{D}}^{2}(1-P_{_{D}}) \geq P_{_{F}}^{2}(1-P_{_{F}}), P_{_{D}}(1-P_{_{D}})^{2} < P_{_{F}}(1-P_{_{F}})^{2}$
\par
Case CZ2: $\displaystyle P_{_{D}}^{3}(1-P_{_{D}}) \geq P_{_{F}}^{3}(1-P_{_{F}}), P_{_{D}}^{2}(1-P_{_{D}})^{2} < P_{_{F}}^{2}(1-P_{_{F}})^{2}, P_{_{D}}(1-P_{_{D}})^{3} < P_{_{F}}(1-P_{_{F}})^{3}, P_{_{D}}(1-P_{_{D}}) < P_{_{F}}(1-P_{_{F}}), P_{_{D}}^{2}(1-P_{_{D}}) < P_{_{F}}^{2}(1-P_{_{F}}), P_{_{D}}(1-P_{_{D}})^{2} < P_{_{F}}(1-P_{_{F}})^{2}$
\par
Case DZ2: $\displaystyle P_{_{D}}^{3}(1-P_{_{D}}) < P_{_{F}}^{3}(1-P_{_{F}}), P_{_{D}}^{2}(1-P_{_{D}})^{2} < P_{_{F}}^{2}(1-P_{_{F}})^{2}, P_{_{D}}(1-P_{_{D}})^{3} < P_{_{F}}(1-P_{_{F}})^{3}, P_{_{D}}(1-P_{_{D}}) < P_{_{F}}(1-P_{_{F}}), P_{_{D}}^{2}(1-P_{_{D}}) < P_{_{F}}^{2}(1-P_{_{F}}), P_{_{D}}(1-P_{_{D}})^{2} < P_{_{F}}(1-P_{_{F}})^{2}$
\vspace{2cm}
\par
We will analyze each of these seven cases in detail and determine their respective optimal placements.  

\newpage
\par
\textbf{Case AW1:} $\displaystyle \bf P_{_{D}}^{3}(1-P_{_{D}}) \geq P_{_{F}}^{3}(1-P_{_{F}}), P_{_{D}}^{2}(1-P_{_{D}})^{2} \geq P_{_{F}}^{2}(1-P_{_{F}})^{2}, P_{_{D}}(1-P_{_{D}})^{3} \geq P_{_{F}}(1-P_{_{F}})^{3}, P_{_{D}}(1-P_{_{D}}) \geq P_{_{F}}(1-P_{_{F}}), P_{_{D}}^{2}(1-P_{_{D}})\geq P_{_{F}}^{2}(1-P_{_{F}}), P_{_{D}}(1-P_{_{D}})^{2}\geq P_{_{F}}(1-P_{_{F}})^{2}$ 
\par
For this case the probability of error for the placements are given by equations (\ref{Pe2110C1}), (\ref{Pe2200C1}), (\ref{Pe3100CW}), and (\ref{Pe4000CA}) respectively. For this case we will show that $P_{e}(2,1,1,0) \leq P_{e}(2,2,0,0), P_{e}(3,1,0,0), P_{e}(4,0,0,0)$.
Now using equations (\ref{Pe2110C1}) and (\ref{Pe2200C1}) we can write:
\begin{equation}
\displaystyle 4P_{e}(2,2,0,0) - 4P_{e}(2,1,1,0) = P_{_{D}}^{2}P_{_{F}}^{2} + P_{_{F}}^{4} + 2P_{_{D}}P_{_{F}}^{2}(1-P_{_{D}}) + 4P_{_{F}}^{3}(1-P_{_{F}}) + P_{_{F}}^{2}(1-P_{_{D}})^{2}\nonumber
\end{equation}
\begin{equation}
\displaystyle + 6P_{_{F}}^{2}(1-P_{_{F}})^{2} + 2P_{_{F}}(1-P_{_{F}})(1-P_{_{D}})^{2} + 4P_{_{F}}(1-P_{_{F}})^{3} + (1-P_{_{D}})^{2}(1-P_{_{F}})^{2} + (1-P_{_{F}})^{4} - 2P_{_{D}}P_{_{F}}^{3} \nonumber
\end{equation}
\begin{equation}
\displaystyle -2P_{_{F}}^{3}(1-P_{_{D}}) - 4P_{_{D}}P_{_{F}}^{2}(1-P_{_{F}})  - 6P_{_{F}}^{2}(1-P_{_{D}})(1-P_{_{F}}) - 6P_{_{F}}(1-P_{_{D}})(1-P_{_{F}})^{2}-  \nonumber
\end{equation}
\begin{equation}
\displaystyle P_{_{D}}P_{_{F}}(1-P_{_{F}})^{2} -2(1-P_{_{D}})(1-P_{_{F}})^{3} \nonumber
\end{equation}
Simplifying this equation we get:
\begin{equation}
\displaystyle 4P_{e}(2,2,0,0) - 4P_{e}(2,1,1,0) = P_{_{F}}^{2}(P_{_{D}}-P_{_{F}})^{2} + P_{_{F}}(P_{_{D}}-P_{_{F}})\{1+2(P_{_{D}}-P_{_{F}})\} + (1-P_{_{F}})^{2}(P_{_{D}}-P_{_{F}})^{2} \nonumber
\end{equation}
\begin{equation}
\displaystyle \hspace{3cm}+~ P_{_{F}}^{2}(1-P_{_{D}})(1-P_{_{F}}) + P_{_{F}}^{2}(1-P_{_{D}})^{2} ~\geq~ 0 \nonumber
\end{equation}
\begin{equation}
\displaystyle \Rightarrow~ P_{e}(2,2,0,0) \geq P_{e}(2,1,1,0) \nonumber
\end{equation}
Using equations (\ref{Pe2110C1}) and (\ref{Pe3100CW}) we can write:
\begin{equation}
\displaystyle 4P_{e}(3,1,0,0) - 4P_{e}(2,1,1,0) = P_{_{D}}P_{_{F}}^{3} + 2P_{_{F}}^{4} + P_{_{F}}^{3}(1-P_{_{D}}) + 2P_{_{F}}^{3}(1-P_{_{F}}) + 3P_{_{D}}P_{_{F}}^{2}(1-P_{_{F}}) \nonumber
\end{equation}
\begin{equation}
\displaystyle + 6P_{_{F}}^{3}(1-P_{_{F}}) + 3P_{_{F}}^{2}(1-P_{_{D}})(1-P_{_{F}}) + 12P_{_{F}}^{2}(1-P_{_{F}})^{2} + 3P_{_{D}}P_{_{F}}(1-P_{_{D}})^{2} + 3P_{_{F}}(1-P_{_{D}})(1-P_{_{F}})^{2} \nonumber
\end{equation}
\begin{equation}
\displaystyle 8P_{_{F}}(1-P_{_{F}})^{3} + P_{_{F}}(1-P_{_{D}})^{3} + (1-P_{_{D}})^{3}(1-P_{_{F}}) + (1-P_{_{D}})(1-P_{_{F}})^{3} + (1-P_{_{F}})^{4} - 2P_{_{D}}P_{_{F}}^{3}- P_{_{F}}^{4} \nonumber
\end{equation}
\begin{equation}
\displaystyle -2P_{_{D}}P_{_{F}}^{2}(1-P_{_{F}}) - 2P_{_{F}}^{3}(1-P_{_{D}}) - 2P_{_{F}}^{3}(1-P_{_{F}}) - 2P_{_{F}}^{2}(1-P_{_{D}})(1-P_{_{F}}) - P_{_{F}}^{2}(1-P_{_{F}})^{2} - 2P_{_{D}}P_{_{F}}^{2}(1-P_{_{D}}) \nonumber
\end{equation}
\begin{equation}
\displaystyle -2P_{_{D}}P_{_{F}}^{2}(1-P_{_{F}}) - 2P_{_{F}}^{3}(1-P_{_{F}}) - 4P_{_{D}}P_{_{F}}(1-P_{_{D}})(1-P_{_{F}}) - 4P_{_{F}}^{2}(1-P_{_{D}})(1-P_{_{F}}) - 4P_{_{F}}^{2}(1-P_{_{F}})^{2}
\nonumber
\end{equation}
\begin{equation}
\displaystyle -4P_{_{F}}(1-P_{_{D}})(1-P_{_{F}})^{2} - 2P_{_{F}}(1-P_{_{F}})^{3} - P_{_{F}}^{2}(1-P_{_{D}})^{2} - P_{_{D}}P_{_{F}}(1-P_{_{F}})^{2} - P_{_{F}}^{2}(1-P_{_{F}})^{2} \nonumber
\end{equation}
\begin{equation}
\displaystyle -2P_{_{F}}(1-P_{_{F}})(1-P_{_{D}})^{2} - 2P_{_{F}}(1-P_{_{D}})(1-P_{_{F}})^{2} - 2P_{_{F}}(1-P_{_{F}})^{3} - (1-P_{_{D}})^{2}(1-P_{_{F}})^{2} - 2(1-P_{_{D}})(1-P_{_{F}})^{3} \nonumber
\end{equation}
Simplifying this expression we get:
\begin{equation}
\displaystyle 4P_{e}(3,1,0,0) - 4P_{e}(2,1,1,0) = P_{_{D}}(P_{_{D}}-P_{_{F}})\{ 1 - (P_{_{D}}+P_{_{F}}) + (1-2P_{_{F}}^{2})\} + 2P_{_{F}}^{4} \nonumber
\end{equation}
Using the fact that $P_{a}(1-P_{a}) \geq P_{b}(1-P_{b})$ we get:
\begin{equation}
\displaystyle P_{e}(3,1,0,0) ~\geq~ P_{e}(2,1,1,0) \nonumber
\end{equation}
Using equations (\ref{Pe2110C1}) and (\ref{Pe4000CA}) we can write:
\begin{equation}
\displaystyle 4P_{e}(4,0,0,0) - 4P_{e}(2,1,10) = 3P_{_{F}}^{4} + 12P_{_{F}}^{3}(1-P_{_{F}}) + 18P_{_{F}}^{2}(1-P_{_{F}})^{2} + 12P_{_{F}}(1-P_{_{F}})^{3} + (1-P_{_{D}})^{4} + 2(1-P_{_{F}})^{4} \nonumber
\end{equation}
\begin{equation}
\displaystyle -2P_{_{D}}P_{_{F}}^{3} - P_{_{F}}^{4} - 2P_{_{D}}P_{_{F}}^{2}(1-P_{_{F}}) - 2P_{_{F}}^{3}(1-P_{_{D}}) - 2P_{_{F}}^{3}(1-P_{_{F}}) - 2P_{_{F}}^{2}(1-P_{_{D}})(1-P_{_{F}}) - P_{_{F}}^{2}(1-P_{_{F}})^{2} \nonumber
\end{equation}
\begin{equation}
\displaystyle -2P_{_{D}}P_{_{F}}^{2}(1-P_{_{D}}) - 2P_{_{D}}P_{_{F}}^{2}(1-P_{_{F}}) - 2P_{_{F}}^{3}(1-P_{_{F}}) - 4P_{_{D}}P_{_{F}}(1-P_{_{D}})(1-P_{_{F}}) - 4P_{_{F}}^{2}(1-P_{_{D}})(1-P_{_{F}}) - \nonumber
\end{equation} 
\begin{equation}
\displaystyle 4P_{_{F}}^{2}(1-P_{_{F}})^{2} - 4P_{_{F}}(1-P_{_{D}})(1-P_{_{F}})^{2} - 2P_{_{F}}(1-P_{_{F}})^{3} - P_{_{F}}^{2}(1-P_{_{D}})^{2} - P_{_{D}}P_{_{F}}(1-P_{_{F}})^{2} - P_{_{F}}^{2}(1-P_{_{F}})^{2} \nonumber
\end{equation}
\begin{equation}
\displaystyle -2P_{_{F}}(1-P_{_{F}})(1-P_{_{D}})^{2} - 2P_{_{F}}(1-P_{_{D}})(1-P_{_{F}})^{2} - 2P_{_{F}}(1-P_{_{F}})^{3} - (1-P_{_{D}})^{2}(1-P_{_{F}})^{2} - 2(1-P_{_{D}})(1-P_{_{F}})^{3} \nonumber
\end{equation}  
Simplifying the above expression we get:
\begin{equation}
\displaystyle 4P_{e}(4,0,0,0) - 4P_{e}(2,1,1,0) = P_{_{D}}(P_{_{D}}-P_{_{F}})(5-4P_{_{D}}) + (P_{_{D}}^{4}-P_{_{F}}^{4}) + 2P_{_{D}}^{2}(P_{_{D}}^{2}-P_{_{F}}^{2}) \geq 0 \nonumber
\end{equation}
\begin{equation}
\displaystyle \Rightarrow P_{e}(4,0,0,0) \geq P_{e}(2,1,1,0) \nonumber
\end{equation}
This proves that for this case:
\begin{equation}
\displaystyle P_{e}(2,1,1,0) \leq \min\{P_{e}(2,2,0,0), P_{e}(3,1,0,0), P_{e}(4,0,0,0)\} \nonumber
\end{equation}
\newpage
\par
\textbf{Case BW1:} $\displaystyle \bf P_{_{D}}^{3}(1-P_{_{D}}) \geq P_{_{F}}^{3}(1-P_{_{F}}), P_{_{D}}^{2}(1-P_{_{D}})^{2} \geq P_{_{F}}^{2}(1-P_{_{F}})^{2}, P_{_{D}}(1-P_{_{D}})^{3} < P_{_{F}}(1-P_{_{F}})^{3}, P_{_{D}}^{2}(1-P_{_{D}}) \geq P_{_{F}}^{2}(1-P_{_{F}}), P_{_{D}}(1-P_{_{D}})^{2}\geq P_{_{F}}(1-P_{_{F}})^{2}, P_{_{D}}(1-P_{_{D}}) \geq P_{_{F}}(1-P_{_{F}})$ 
\vspace{1cm}
\par
For this case the probability of error for the placements are given by equations (\ref{Pe2110C1}), (\ref{Pe2200C1}), (\ref{Pe3100CW}), and (\ref{Pe4000CB}) respectively. Using the results of case AW1 we can conclude that $P_{e}(2,1,1,0) \leq P_{e}(2,2,0,0), P_{e}(3,1,0,0)$. We need to show that $P_{e}(2,1,1,0) \leq P_{e}(4,0,0,0)$. Using equations (\ref{Pe2110C1}) and (\ref{Pe4000CB}) we can write:  
\begin{equation}
\displaystyle 4P_{e}(4,0,0,0) - 4P_{e}(2,1,1,0) = 3P_{_{F}}^{4} + 12P_{_{F}}^{3}(1-P_{_{F}}) + 18P_{_{F}}^{2}(1-P_{_{F}}) + 4P_{_{D}}(1-P_{_{D}})^{3} + 8P_{_{F}}(1-P_{_{F}})^{3} +  \nonumber
\end{equation}
\begin{equation}
\displaystyle (1-P_{_{D}})^{4}+ 2(1-P_{_{F}})^{4} - 2P_{_{D}}P_{_{F}}^{3} - P_{_{F}}^{4} - 2P_{_{D}}P_{_{F}}^{2}(1-P_{_{F}})-  2P_{_{F}}^{3}(1-P_{_{D}}) - 2P_{_{F}}^{3}(1-P_{_{F}}) - 2P_{_{F}}^{2}(1-P_{_{D}})(1-P_{_{F}}) \nonumber
\end{equation}
\begin{equation}
\displaystyle -P_{_{F}}^{2}(1-P_{_{F}})^{2} - 2P_{_{D}}P_{_{F}}^{2}(1-P_{_{D}}) - 2P_{_{D}}P_{_{F}}^{2}(1-P_{_{F}}) - 2P_{_{F}}^{3}(1-P_{_{F}}) - 4P_{_{D}}P_{_{F}}(1-P_{_{D}})(1-P_{_{F}}) \nonumber
\end{equation}
\begin{equation}
\displaystyle -4P_{_{F}}^{2}(1-P_{_{D}})(1-P_{_{F}}) - 4P_{_{F}}^{2}(1-P_{_{F}})^{2} - 4P_{_{F}}(1-P_{_{D}})(1-P_{_{F}})^{2} - 2P_{_{F}}(1-P_{_{F}})^{3} - P_{_{F}}^{2}(1-P_{_{D}})^{2} \nonumber
\end{equation}
\begin{equation}
\displaystyle -P_{_{D}}P_{_{F}}(1-P_{_{F}})^{2} - P_{_{F}}^{2}(1-P_{_{F}})^{2}-  2P_{_{F}}(1-P_{_{F}})(1-P_{_{D}})^{2} - 2P_{_{F}}(1-P_{_{D}})(1-P_{_{F}})^{2} - 2P_{_{F}}(1-P_{_{F}})^{3} \nonumber
\end{equation}
\begin{equation}
\displaystyle ~-~(1-P_{_{D}})^{2}(1-P_{_{F}})^{2} - 2(1-P_{_{D}})(1-P_{_{F}})^{3} \nonumber
\end{equation}      
Simplifying this expression we get:
\begin{equation}
\displaystyle 4P_{e}(4,0,0,0) - 4P_{e}(2,1,1,0) = (P_{a}-P_{b})\bigg\{4 - 12P_{b}(1-P_{b}) - 7P_{a}(1-P_{a}) + P_{a}^{2} + 12P_{a}P_{b}  \nonumber
\end{equation}
\begin{equation}
\displaystyle  - 3P_{a}^{3}- 5P_{a}P_{b}^{2}-3P_{a}^{2}P_{b} - 4P_{b}^{3}\bigg\} ~\geq~ 0 \nonumber
\end{equation} 
\begin{equation}
\displaystyle ~\Rightarrow~ P_{e}(4,0,0,0) \geq P_{e}(2,1,1,0) \nonumber
\end{equation}
Therefore, we can conclude for this case:
\begin{equation}
\displaystyle P_{e}(2,1,1,0) \leq \min\{P_{e}(2,2,0,0), P_{e}(3,1,0,0), P_{e}(4,0,0,0)\} \nonumber
\end{equation}
\newpage
\par
\textbf{Case BX1:} $\displaystyle \bf P_{_{D}}^{3}(1-P_{_{D}})\geq P_{_{F}}^{3}(1-P_{_{F}}), P_{_{D}}^{2}(1-P_{_{D}})^{2} \geq P_{_{F}}^{2}(1-P_{_{F}})^{2}, P_{_{D}}(1-P_{_{D}})^{3}< P_{_{F}}(1-P_{_{F}})^{3}, P_{_{D}}^{2}(1-P_{_{D}}) \geq P_{_{F}}^{2}(1-P_{_{F}}), P_{_{D}}(1-P_{_{D}})^{2} < P_{_{F}}(1-P_{_{F}})^{2},P_{_{D}}(1-P_{_{D}}) \geq P_{_{F}}(1-P_{_{F}})$ 
\vspace{1cm}
\par
For this case the probability of error for the placements are given by equations (\ref{Pe2110C1}), (\ref{Pe2200C1}), (\ref{Pe3100CX}), and (\ref{Pe4000CB}) respectively. Using the results of case BW1 we can conclude that $P_{e}(2,1,1,0) \leq P_{e}(2,2,0,0), P_{e}(4,0,0,0)$. We will show that $P_{e}(3,1,0,0) \geq P_{e}(2,1,1,0)$. Using equations (\ref{Pe2110C1}) and (\ref{Pe3100CX}) we can write:
\begin{equation}
\displaystyle 4P_{e}(3,1,0,0) - 4P_{e}(2,1,1,0) = P_{_{D}}P_{_{F}}^{3} + 2P_{_{F}}^{4} + P_{_{F}}^{3}(1-P_{_{D}}) + 2P_{_{F}}^{3}(1-P_{_{F}}) + 3P_{_{D}}P_{_{F}}^{2}(1-P_{_{F}}) + 
\nonumber
\end{equation}
\begin{equation}
\displaystyle + 6P_{_{F}}^{3}(1-P_{_{F}}) + 3P_{_{F}}^{2}(1-P_{_{D}})(1-P_{_{F}}) + 12P_{_{F}}^{2}(1-P_{_{F}})^{2} + 3P_{_{D}}P_{_{F}}(1-P_{_{D}})^{2} + 3P_{_{D}}(1-P_{_{F}})(1-P_{_{D}})^{2} + \nonumber
\end{equation}
\begin{equation}
\displaystyle 3P_{_{F}}(1-P_{_{D}})(1-P_{_{F}})^{2} + 3P_{_{F}}(1-P_{_{F}})^{3} + P_{_{F}}(1-P_{_{D}})^{3} + 2P_{_{F}}(1-P_{_{F}})^{3} + (1-P_{_{D}})^{3}(1-P_{_{F}}) + (1-P_{_{D}})(1-P_{_{F}})^{3} \nonumber
\end{equation}
\begin{equation}
\displaystyle + (1-P_{_{F}})^{4} - 2P_{_{D}}P_{_{F}}^{3} - P_{_{F}}^{4} - 2P_{_{D}}P_{_{F}}^{2}(1-P_{_{F}}) - 2P_{_{F}}^{3}(1-P_{_{D}}) - 2P_{_{F}}^{3}(1-P_{_{F}}) - 2P_{_{F}}^{2}(1-P_{_{D}})(1-P_{_{F}})\nonumber
\end{equation}
\begin{equation}
\displaystyle -P_{_{F}}^{2}(1-P_{_{F}})^{2} - 2P_{_{D}}P_{_{F}}^{2}(1-P_{_{D}}) - 2P_{_{D}}P_{_{F}}^{2}(1-P_{_{F}}) - 2P_{_{F}}^{3}(1-P_{_{F}}) - 4P_{_{D}}P_{_{F}}(1-P_{_{D}})(1-P_{_{F}}) \nonumber
\end{equation}
\begin{equation}
\displaystyle -4P_{_{F}}^{2}(1-P_{_{D}})(1-P_{_{F}}) - 4P_{_{F}}^{2}(1-P_{_{F}})^{2} - 4P_{_{F}}(1-P_{_{D}})(1-P_{_{F}})^{2} - 2P_{_{F}}(1-P_{_{F}})^{3} - P_{_{F}}^{2}(1-P_{_{D}})^{2} \nonumber
\end{equation}
\begin{equation}
\displaystyle -P_{_{D}}P_{_{F}}(1-P_{_{F}})^{2} - P_{_{F}}^{2}(1-P_{_{F}})^{2} - 2P_{_{F}}(1-P_{_{F}})(1-P_{_{D}})^{2} - 2P_{_{F}}(1-P_{_{D}})(1-P_{_{F}})^{2} - 2P_{_{F}}(1-P_{_{F}})^{3} \nonumber
\end{equation}
\begin{equation}
\displaystyle -(1-P_{_{D}})^{2}(1-P_{_{F}})^{2} - 2(1-P_{_{D}})(1-P_{_{F}})^{3} \nonumber
\end{equation}
Simplifying the above expression we get:
\begin{equation}
\displaystyle 4P_{e}(3,1,0,0) - 4P_{e}(2,1,1,0) = (1-P_{_{D}})(3P_{_{D}}-2P_{_{D}}P_{_{F}} - P_{_{D}}^{2}) + P_{_{D}}(P_{_{D}}^{2}-P_{_{F}})^{3} + (1-P_{_{F}})(9P_{_{F}}-3P_{_{D}}P_{_{F}} + 2P_{_{D}}^{2}P_{_{F}}) \nonumber
\end{equation}
\begin{equation}
\displaystyle +3P_{_{F}}^{4} + 3P_{_{D}}P_{_{F}}^{2} ~ \geq ~ 0 \nonumber
\end{equation}
This implies that:
\begin{equation}
P_{e}(2,1,1,0) \leq \min\{P_{e}(2,2,0,0), P_{e}(3,1,0,0), P_{e}(4,0,0,0)\} \nonumber
\end{equation}
\newpage
\par
\textbf{Case CX1:} $\displaystyle \bf P_{_{D}}^{3}(1-P_{_{D}}) \geq P_{_{F}}^{3}(1-P_{_{F}}), P_{_{D}}^{2}(1-P_{_{D}})^{2} < P_{_{F}}^{2}(1-P_{_{F}})^{2}, P_{_{D}}(1-P_{_{D}})^{3} < P_{_{F}}(1-P_{_{F}})^{3}, P_{_{D}}^{2}(1-P_{_{D}})\geq P_{_{F}}^{2}(1-P_{_{F}}), P_{_{D}}(1-P_{_{D}})^{2} < P_{_{F}}(1-P_{_{F}})^{2}, P_{_{D}}(1-P_{_{D}}) \geq P_{_{F}}(1-P_{_{F}})$ 
\vspace{1cm}
\par    
For this case the probability of error for the placements are given by equations (\ref{Pe2110C1}), (\ref{Pe2200C1}), (\ref{Pe3100CX}), and (\ref{Pe4000CC}) respectively. From case BX1 we get $P_{e}(2,1,1,0) \leq P_{e}(2,2,0,0), P_{e}(3,1,0,0)$. We will show that $P_{e}(2,1,1,0) \leq P_{e}(4,0,0,0)$. Using equations (\ref{Pe2110C1}) and (\ref{Pe4000CC}) we can write:
\begin{equation}
\displaystyle 4P_{e}(4,0,0,0) - 4P_{e}(2,1,1,0) = 3P_{_{F}}^{4} + 12P_{_{F}}^{3}(1-P_{_{F}}) + 6P_{_{D}}^{2}(1-P_{_{D}})^{2} + 12P_{_{F}}^{2}(1-P_{_{F}})^{2} + 4P_{_{D}}(1-P_{_{D}})^{3} \nonumber
\end{equation}
\begin{equation}
\displaystyle + 8P_{_{F}}(1-P_{_{F}})^{3} + (1-P_{_{D}})^{4} + 2(1-P_{_{F}})^{4} - 2P_{_{D}}P_{_{F}}^{3} - P_{_{F}}^{4} - 2P_{_{D}}P_{_{F}}^{2}(1-P_{_{F}}) - 2P_{_{F}}^{3}(1-P_{_{D}}) \nonumber
\end{equation}
\begin{equation}
\displaystyle -2P_{_{F}}^{3}(1-P_{_{F}}) - 2P_{_{F}}^{2}(1-P_{_{D}})(1-P_{_{F}}) - P_{_{F}}^{2}(1-P_{_{F}})^{2} - 2P_{_{D}}P_{_{F}}^{2}(1-P_{_{D}}) - 2P_{_{D}}P_{_{F}}^{2}(1-P_{_{F}}) \nonumber
\end{equation}
\begin{equation}
\displaystyle -2P_{_{F}}^{3}(1-P_{_{F}}) - 4P_{_{D}}P_{_{F}}(1-P_{_{D}})(1-P_{_{F}}) - 4P_{_{F}}^{2}(1-P_{_{D}})(1-P_{_{F}}) - 4P_{_{F}}^{2}(1-P_{_{F}})^{2} - 4P_{_{F}}(1-P_{_{D}})(1-P_{_{F}})^{2} \nonumber
\end{equation}
\begin{equation}
\displaystyle -2P_{_{F}}(1-P_{_{F}})^{3} - P_{_{F}}^{2}(1-P_{_{D}})^{2} - P_{_{D}}P_{_{F}}(1-P_{_{F}})^{2} - P_{_{F}}^{2}(1-P_{_{F}})^{2} - 2P_{_{F}}(1-P_{_{F}})(1-P_{_{D}})^{2} -2P_{_{F}}(1-P_{_{D}})(1-P_{_{F}})^{2} \nonumber
\end{equation}
\begin{equation}
\displaystyle -2P_{_{F}}(1-P_{_{F}})^{3} - (1-P_{_{D}})^{2}(1-P_{_{F}})^{2} - 2(1-P_{_{D}})(1-P_{_{F}})^{3} \nonumber
\end{equation} 
Simplifying this expression we get:
\begin{equation}
\displaystyle 4P_{e}(4,0,0,0) - 4P_{e}(2,1,1,0) = (P_{_{D}}-P_{_{F}})\{4-5P_{_{F}}-(P_{_{D}}+P_{_{F}})\} + 2(P_{_{D}}^{2}-P_{_{F}}^{2})(P_{_{D}}^{2}+ P_{_{F}}^{2}) + P_{_{D}}^{4} \nonumber
\end{equation}
\begin{equation}
\displaystyle +P_{_{D}}P_{_{F}}^{3} + 2P_{_{D}}^{2}P_{_{F}} + 2P_{_{D}}^{2}P_{_{F}}(1-P_{_{F}}) \nonumber
\end{equation}
Using the fact $\displaystyle P_{_{D}}(1-P_{_{D}}) \geq P_{_{F}}(1-P_{_{F}})$ we get:
\begin{equation}
\displaystyle 4P_{e}(4,0,0,0) \geq 4P_{e}(2,1,1,0) \nonumber
\end{equation}
\begin{equation}
\displaystyle \Rightarrow ~ P_{e}(2,1,1,0) \leq \min\{P_{e}(2,2,0,0), P_{e}(3,1,0,0), P_{e}(4,0,0,0)\} \nonumber
\end{equation}
\newpage
\par
\textbf{Case CY2:} $\displaystyle \bf P_{_{D}}^{3}(1-P_{_{D}}) \geq P_{_{F}}^{3}(1-P_{_{F}}), P_{_{D}}^{2}(1-P_{_{D}})^{2} < P_{_{F}}^{2}(1-P_{_{F}})^{2}, P_{_{D}}(1-P_{_{D}})^{3} < P_{_{F}}(1-P_{_{F}})^{3}, P_{_{D}}^{2}(1-P_{_{D}})\geq P_{_{F}}^{2}(1-P_{_{F}}), P_{_{D}}(1-P_{_{D}})^{2} < P_{_{F}}(1-P_{_{F}})^{2}, P_{_{D}}(1-P_{_{D}}) < P_{_{F}}(1-P_{_{F}})$ 
\vspace{1cm}
\par     
For this case the probability of error for the placements are given by equations (\ref{Pe2110C2}), (\ref{Pe2200C2}), (\ref{Pe3100CY}), and (\ref{Pe4000CC}) respectively. We will show that $P_{e}(2,1,1,0) \leq P_{e}(2,2,0,0),P_{e}(3,1,0,0), P_{e}(4,0,0,0)$ respectively. Using equations (\ref{Pe2110C2}) and (\ref{Pe4000CC}) we can write:
\begin{equation}
\displaystyle 4P_{e}(4,0,0,0) - 4P_{e}(2,1,1,0) = 3P_{_{F}}^{4} + 12P_{_{F}}^{3}(1-P_{_{F}}) + 6P_{_{D}}^{2}(1-P_{_{D}})^{2} + 12P_{_{F}}^{2}(1-P_{_{F}})^{2} + 4P_{_{D}}(1-P_{_{D}})^{3} \nonumber
\end{equation}
\begin{equation}
\displaystyle + 8P_{_{F}}(1-P_{_{F}})^{3} + (1-P_{_{D}})^{4} + 2(1-P_{_{F}})^{4} - 2P_{_{D}}P_{_{F}}^{3} - P_{_{F}}^{4} - 2P_{_{D}}P_{_{F}}^{2}(1-P_{_{F}}) - 2P_{_{F}}^{3}(1-P_{_{D}}) \nonumber
\end{equation}
\begin{equation}
\displaystyle -2P_{_{F}}^{3}(1-P_{_{F}}) - 2P_{_{F}}^{2}(1-P_{_{D}})(1-P_{_{F}}) - P_{_{F}}^{2}(1-P_{_{F}})^{2} - 2P_{_{D}}P_{_{F}}^{2}(1-P_{_{D}}) - 2P_{_{D}}P_{_{F}}^{2}(1-P_{_{F}}) - 2P_{_{F}}^{3}(1-P_{_{F}}) \nonumber
\end{equation}
\begin{equation}
\displaystyle -4P_{_{D}}P_{_{F}}(1-P_{_{D}})(1-P_{_{F}}) - 4P_{_{F}}^{2}(1-P_{_{D}})(1-P_{_{F}}) - 4P_{_{F}}^{2}(1-P_{_{F}})^{2} - 2P_{_{D}}(1-P_{_{D}})(1-P_{_{F}})^{2}  \nonumber
\end{equation}
\begin{equation}
\displaystyle - 4P_{_{F}}(1-P_{_{D}})(1-P_{_{F}})^{2}-P_{_{F}}^{2}(1-P_{_{D}})^{2} - P_{_{D}}P_{_{F}}(1-P_{_{F}})^{2} - P_{_{F}}^{2}(1-P_{_{F}})^{2} - 2P_{_{F}}(1-P_{_{F}})(1-P_{_{D}})^{2}  \nonumber
\end{equation}
\begin{equation}
\displaystyle  - 2P_{_{F}}(1-P_{_{D}})(1-P_{_{F}})^{2}- 2P_{_{F}}(1-P_{_{F}})^{3} - (1-P_{_{D}})^{2}(1-P_{_{F}})^{2} - 2(1-P_{_{D}})(1-P_{_{F}})^{3} \nonumber
\end{equation}
Simplifying this expression we get:
\begin{equation}
\displaystyle 4P_{e}(4,0,0,0) - 4P_{e}(2,1,1,0) = (P_{_{D}}-P_{_{F}})\{ 2 + P_{_{D}} + 3(P_{_{D}}+P_{_{F}})(P_{_{D}}^{2}+ P_{_{F}}^{2}) + P_{_{F}}^{3}-2P_{_{F}}^{2}- \nonumber
\end{equation}
\begin{equation}
\displaystyle 4P_{_{D}}^{2}-4P_{_{F}}^{2} - 4P_{_{D}}P_{_{F}}\} ~ \geq ~ 0 \nonumber
\end{equation}
\begin{equation}
\displaystyle \Rightarrow ~ P_{e}(4,0,0,0) \geq P_{e}(2,1,1,0) \nonumber
\end{equation}
Using equations (\ref{Pe2110C2}) and (\ref{Pe3100CY}) we can write:
\begin{equation}
\displaystyle 4P_{e}(3,1,0,0) - 4P_{e}(2,1,1,0) = P_{_{F}}^{4} + 3P_{_{D}}^{2}P_{_{F}}(1-P_{_{D}}) + 4P_{_{F}}^{3}(1-P_{_{F}}) + 6P_{_{F}}^{2}(1-P_{_{F}})^{2} + 3P_{_{D}}P_{_{F}}(1-P_{_{D}})^{2} \nonumber
\end{equation}
\begin{equation}
\displaystyle + 3P_{_{D}}(1-P_{_{F}})(1-P_{_{D}})^{2} + 3P_{_{F}}(1-P_{_{F}})^{3} + P_{_{F}}(1-P_{_{D}})^{3} + (1-P_{_{D}})^{3}(1-P_{_{F}}) + (1-P_{_{F}})^{4} - P_{_{D}}P_{_{F}}^{3} \nonumber
\end{equation}
\begin{equation}
\displaystyle -2P_{_{D}}P_{_{F}}^{2}(1-P_{_{F}}) - P_{_{F}}^{3}(1-P_{_{D}}) - 2P_{_{D}}P_{_{F}}^{2}(1-P_{_{D}}) - 2P_{_{D}}P_{_{F}}^{2}(1-P_{_{F}}) - 4P_{_{D}}P_{_{F}}(1-P_{_{D}})(1-P_{_{F}}) \nonumber
\end{equation}
\begin{equation}
\displaystyle -3P_{_{F}}^{2}(1-P_{_{D}})(1-P_{_{F}}) - 2P_{_{D}}(1-P_{_{D}})(1-P_{_{F}})^{2} - 3P_{_{F}}(1-P_{_{D}})(1-P_{_{F}})^{2} - P_{_{F}}^{2}(1-P_{_{D}})^{2} - P_{_{D}}P_{_{F}}(1-P_{_{F}})^{2} \nonumber
\end{equation}
\begin{equation}
\displaystyle -2P_{_{F}}(1-P_{_{F}})(1-P_{_{D}})^{2} - (1-P_{_{D}})^{2}(1-P_{_{F}})^{2} - (1-P_{_{D}})(1-P_{_{F}})^{3} \nonumber
\end{equation}
This expression can be simplified to give:
\begin{equation}
\displaystyle 4P_{e}(3,1,0,0) - 4P_{e}(2,1,1,0) = (P_{_{D}}-P_{_{F}})\{1 - 2(P_{_{D}}+P_{_{F}})(1-P_{_{D}})(1-P_{_{F}})\} + P_{_{D}}^{2}P_{_{F}}(1-P_{_{D}}) + \nonumber
\end{equation}
\begin{equation}
\displaystyle P_{_{F}}^{2}(1-P_{_{F}}) + P_{_{F}}^{4} - P_{_{D}}P_{_{F}} ~ \geq ~ 0 \nonumber
\end{equation}
\begin{equation}
\displaystyle \Rightarrow ~ P_{e}(3,1,0,0) \geq P_{e}(2,1,1,0) \nonumber
\end{equation}
Using equations (\ref{Pe2110C2}) and (\ref{Pe2200C2}) we can write:
\begin{equation}
\displaystyle P_{e}(2,2,0,0) - P_{e}(2,1,1,0) = P_{_{D}}^{2}P_{_{F}}^{2} + 2P_{_{F}}^{4} + 4P_{_{D}}P_{_{F}}^{2}(1-P_{_{D}}) + 8P_{_{F}}^{3}(1-P_{_{F}}) + 2P_{_{F}}^{2}(1-P_{_{D}})^{2} \nonumber
\end{equation}
\begin{equation}
\displaystyle  + 4P_{_{F}}^{2}(1-P_{_{F}})^{2} +8P_{_{D}}P_{_{F}}(1-P_{_{D}})(1-P_{_{F}}) + 4P_{_{F}}^{2}(1-P_{_{F}})^{2} + 4P_{_{F}}(1-P_{_{F}})(1-P_{_{D}})^{2} + 4P_{_{D}}(1-P_{_{D}})(1-P_{_{F}})^{2} \nonumber
\end{equation}
\begin{equation}
\displaystyle + 4P_{_{F}}(1-P_{_{_F}})^{3} + 2(1-P_{_{D}})^{2}(1-P_{_{F}})^{2} + (1-P_{_{F}})^{4} - 2P_{_{D}}P_{_{F}}^{3} - P_{_{F}}^{4} - 2P_{_{D}}P_{_{F}}^{2}(1-P_{_{F}}) \nonumber
\end{equation}
\begin{equation}
\displaystyle -2P_{_{F}}^{3}(1-P_{_{D}}) - 2P_{_{F}}^{3}(1-P_{_{F}}) - 2P_{_{F}}^{2}(1-P_{_{D}})(1-P_{_{F}}) - P_{_{F}}^{2}(1-P_{_{F}})^{2} - 2P_{_{D}}P_{_{F}}^{2}(1-P_{_{D}})  \nonumber
\end{equation}
\begin{equation}
\displaystyle - 2P_{_{D}}P_{_{F}}^{2}(1-P_{_{F}}) - 2P_{_{F}}^{3}(1-P_{_{F}}) - 4P_{_{D}}P_{_{F}}(1-P_{_{D}})(1-P_{_{F}}) - 4P_{_{F}}^{2}(1-P_{_{D}})(1-P_{_{F}}) - 4P_{_{F}}^{2}(1-P_{_{F}})^{2} \nonumber
\end{equation}
\begin{equation}
\displaystyle -2P_{_{D}}(1-P_{_{D}})(1-P_{_{F}})^{2} - 4P_{_{F}}(1-P_{_{D}})(1-P_{_{F}})^{2} - P_{_{F}}^{2}(1-P_{_{D}})^{2} - P_{_{D}}P_{_{F}}(1-P_{_{F}})^{2} - P_{_{F}}^{2}(1-P_{_{F}})^{2} \nonumber
\end{equation}
\begin{equation}
\displaystyle -2P_{_{F}}(1-P_{_{F}})(1-P_{_{D}})^{2} - 2P_{_{F}}(1-P_{_{D}})(1-P_{_{F}})^{2} - 2P_{_{F}}(1-P_{_{F}})^{3} - (1-P_{_{D}})^{2}(1-P_{_{F}})^{2} - 2(1-P_{_{D}})(1-P_{_{F}})^{3} \nonumber
\end{equation}
This expression can be simplified to get:
\begin{equation}
\displaystyle 4P_{e}(2,2,0,0) - 4P_{e}(2,1,1,0) = (P_{_{D}}-P_{_{F}})(1-P_{_{F}}^{2})\{2(1-P_{_{F}}) - P_{_{D}}\} ~\geq ~0\nonumber
\end{equation}
\begin{equation}
\displaystyle \Rightarrow P_{e}(2,2,0,0) \geq P_{e}(2,1,1,0) \nonumber
\end{equation}
From these results we can conclude that:
\begin{equation}
\displaystyle P_{e}(2,1,1,0) \leq \min\{P_{e}(2,2,0,0), P_{e}(3,1,0,0), P_{e}(4,0,0,0)\} \nonumber
\end{equation}
\newpage
\newpage
\par
\textbf{Case CZ2:} $\displaystyle \bf P_{_{D}}^{3}(1-P_{_{D}}) \geq P_{_{F}}^{3}(1-P_{_{F}}), P_{_{D}}^{2}(1-P_{_{D}})^{2} < P_{_{F}}^{2}(1-P_{_{F}})^{2}, P_{_{D}}(1-P_{_{D}})^{3} < P_{_{F}}(1-P_{_{F}})^{3}, P_{_{D}}^{2}(1-P_{_{D}}) < P_{_{F}}^{2}(1-P_{_{F}}), P_{_{D}}(1-P_{_{D}})^{2} < P_{_{F}}(1-P_{_{F}})^{2}, P_{_{D}}(1-P_{_{D}}) < P_{_{F}}(1-P_{_{F}})$ 
\vspace{1cm}
\par     
For this case the probability of error for the placements are given by equations (\ref{Pe2110C2}), (\ref{Pe2200C2}), (\ref{Pe3100CZ}), and (\ref{Pe4000CC}) respectively. From CY2 we already know that $P_{e}(2,1,1,0) \leq 4P_{e}(4,0,0,0)$. Also using CY2 we can write: 
\begin{equation}
\displaystyle 4P_{e}(2,2,0,0) - 4P_{e}(2,1,1,0) = (P_{_{D}}-P_{_{F}})(1-P_{_{F}}^{2})\{2(1-P_{_{F}}) - P_{_{D}}\} \nonumber
\end{equation}
For $2(1-P_{_{F}}) \geq P_{_{D}}$:
\begin{equation}
\displaystyle P_{e}(2,1,1,0) \leq P_{e}(2,2,0,0) \nonumber
\end{equation}
For $2(1-P_{_{F}}) < P_{_{D}}$:
\begin{equation}
\displaystyle P_{e}(2,1,1,0) \geq P_{e}(2,2,0,0) \nonumber
\end{equation}
Using equations (\ref{Pe3100CZ}) and (\ref{Pe4000CC}) we can write:
\begin{equation}
\displaystyle 4P_{e}(3,1,0,0) - 4P_{e}(2,2,0,0) = P_{_{D}}P_{_{F}}^{3} + 2P_{_{F}}^{4} + P_{_{F}}^{3}(1-P_{_{D}}) + 2P_{_{F}}^{3}(1-P_{_{F}}) + 3P_{_{D}}^{2}P_{_{F}}(1-P_{_{D}}) \nonumber
\end{equation}
\begin{equation}
\displaystyle  + 6P_{_{F}}^{3}(1-P_{_{F}}) + 3P_{_{D}}^{2}(1-P_{_{D}})(1-P_{_{F}}) + 3P_{_{F}}^{2}(1-P_{_{D}})(1-P_{_{F}}) + 3P_{_{F}}^{2}(1-P_{_{F}})^{2} + 3P_{_{D}}P_{_{F}}(1-P_{_{D}})^{2} \nonumber
\end{equation}
\begin{equation}
\displaystyle + 6P_{_{F}}^{2}(1-P_{_{F}})^{2} + 3P_{_{D}}(1-P_{_{F}})(1-P_{_{D}})^{2} + 3P_{_{F}}(1-P_{_{D}})(1-P_{_{F}})^{2} + 3P_{_{F}}(1-P_{_{F}})^{3}+ P_{_{F}}(1-P_{_{D}})^{3} \nonumber
\end{equation}
\begin{equation}
\displaystyle  + 2P_{_{F}}(1-P_{_{F}})^{3} + (1-P_{_{D}})^{3}(1-P_{_{F}}) + (1-P_{_{D}})(1-P_{_{F}})^{3} + (1-P_{_{F}})^{4}  - P_{_{D}}^{2}P_{_{F}}^{2}-2P_{_{F}}^{4}\nonumber
\end{equation}
\begin{equation}
\displaystyle  - 4P_{_{D}}P_{_{F}}^{2}(1-P_{_{D}}) - 8P_{_{F}}^{3}(1-P_{_{F}}) - 2P_{_{F}}^{2}(1-P_{_{D}})^{2} - 4P_{_{F}}^{2}(1-P_{_{F}})^{2}-8P_{_{D}}P_{_{F}}(1-P_{_{D}})(1-P_{_{F}}) \nonumber
\end{equation}
\begin{equation}
\displaystyle  - 4P_{_{F}}^{2}(1-P_{_{F}})^{2} - 4P_{_{F}}(1-P_{_{F}})(1-P_{_{D}})^{2} - 4P_{_{D}}(1-P_{_{D}})(1-P_{_{F}})^{2} -4P_{_{F}}(1-P_{_{F}})^{3}\nonumber
\end{equation}
\begin{equation}
\displaystyle  - 2(1-P_{_{D}})^{2}(1-P_{_{F}})^{2} - (1-P_{_{F}})^{4} \nonumber
\end{equation}
This expression can be simplified to get:
\begin{equation}
\displaystyle 4P_{e}(3,1,0,0) - 4P_{e}(2,2,0,0) = 2(P_{_{D}}^{2}-P_{_{F}}^{2}) - (P_{_{D}}-P_{_{F}}) - (P_{_{D}}^{3}-P_{_{F}}^{3}) - P_{_{D}}P_{_{F}}^{2}(P_{_{D}}-P_{_{F}}) \nonumber
\end{equation}
It should be noted that:
\begin{equation}
\displaystyle 2(P_{_{D}}^{2} - P_{_{F}}^{2})-(P_{_{D}}-P_{_{F}}) - (P_{_{D}}^{3}-P_{_{F}}^{3}) - P_{_{D}}P_{_{F}}^{2}(P_{_{D}}-P_{_{F}}) \geq 0 \nonumber
\end{equation}
\begin{equation}
\displaystyle \Leftrightarrow (P_{_{D}}+P_{_{F}}-1)^{2} \leq P_{_{D}}P_{_{F}}(1-P_{_{F}}) \nonumber
\end{equation}
For $(P_{_{D}}+P_{_{F}}-1)^{2} \leq P_{_{D}}P_{_{F}}(1-P_{_{F}})$:
\begin{equation}
\displaystyle P_{e}(3,1,0,0) \geq P_{e}(2,2,0,0) \nonumber
\end{equation}
For $(P_{_{D}}+P_{_{F}}-1)^{2} > P_{_{D}}P_{_{F}}(1-P_{_{F}})$:
\begin{equation}
\displaystyle P_{e}(3,1,0,0) < P_{e}(2,2,0,0) \nonumber
\end{equation}
Using equations (\ref{Pe3100CZ}) and (\ref{Pe2110C2}) we can write:
\begin{equation}
\displaystyle 4P_{e}(3,1,0,0) - 4P_{e}(2,1,1,0) = P_{_{D}}P_{_{F}}^{3} + 2P_{_{F}}^{4} + P_{_{F}}^{3}(1-P_{_{D}}) + 3P_{_{D}}^{2}P_{_{F}}(1-P_{_{D}}) \nonumber 
\end{equation}
\begin{equation}
\displaystyle  + 6P_{_{F}}^{3}(1-P_{_{F}})  + 3P_{_{D}}^{2}(1-P_{_{D}})(1-P_{_{F}}) + 3P_{_{F}}^{2}(1-P_{_{D}})(1-P_{_{F}}) + 3P_{_{F}}^{2}(1-P_{_{F}})^{2} + 3P_{_{D}}P_{_{F}}(1-P_{_{D}})^{2} \nonumber
\end{equation}
\begin{equation}
\displaystyle  + 6P_{_{F}}^{2}(1-P_{_{F}})^{2} + 3P_{_{D}}(1-P_{_{F}})(1-P_{_{D}})^{2} + 3P_{_{F}}(1-P_{_{D}})(1-P_{_{F}})^{2} + 3P_{_{F}}(1-P_{_{F}})^{3}+ P_{_{F}}(1-P_{_{D}})^{3} \nonumber
\end{equation}
\begin{equation}
\displaystyle   + 2P_{_{F}}(1-P_{_{F}})^{3} + (1-P_{_{D}})^{3}(1-P_{_{F}}) + (1-P_{_{D}})(1-P_{_{F}})^{3} + (1-P_{_{F}})^{4} -2P_{_{D}}P_{_{F}}^{3} - P_{_{F}}^{4}\nonumber
\end{equation}
\begin{equation}
\displaystyle -2P_{_{D}}P_{_{F}}^{2}(1-P_{_{D}}) - 2P_{_{D}}P_{_{F}}^{2}(1-P_{_{F}}) - 2P_{_{F}}^{3}(1-P_{_{F}}) - 2P_{_{F}}^{2}(1-P_{_{D}})(1-P_{_{F}}) - P_{_{F}}^{2}(1-P_{_{F}})^{2} \nonumber
\end{equation}
\begin{equation}
\displaystyle -2P_{_{D}}P_{_{F}}^{2}(1-P_{_{D}}) - 2P_{_{D}}P_{_{F}}^{2}(1-P_{_{F}}) - 2P_{_{F}}^{3}(1-P_{_{F}}) - 4P_{_{D}}P_{_{F}}(1-P_{_{D}})(1-P_{_{F}}) - 4P_{_{F}}^{2}(1-P_{_{D}})(1-P_{_{F}}) \nonumber
\end{equation}
\begin{equation}
\displaystyle -4P_{_{F}}^{2}(1-P_{_{F}})^{2} - 2P_{_{D}}(1-P_{_{D}})(1-P_{_{F}})^{2} - 4P_{_{F}}(1-P_{_{D}})(1-P_{_{F}})^{2} - P_{_{F}}^{2}(1-P_{_{D}})^{2} - P_{_{D}}P_{_{F}}(1-P_{_{F}})^{2} \nonumber
\end{equation}
\begin{equation}
\displaystyle -P_{_{F}}^{2}(1-P_{_{F}})^{2} - 2P_{_{F}}(1-P_{_{F}})(1-P_{_{D}})^{2} - 2P_{_{F}}(1-P_{_{D}})(1-P_{_{F}})^{2} - 2P_{_{F}}(1-P_{_{F}})^{3}\nonumber
\end{equation}
\begin{equation}
\displaystyle -(1-P_{_{D}})^{2}(1-P_{_{F}})^{2} - 2(1-P_{_{D}})(1-P_{_{F}})^{3} \nonumber
\end{equation}
Simplifying this expression we get:
\begin{equation}
\displaystyle 4P_{e}(3,1,0,0) - 4P_{e}(2,1,1,0) = (P_{_{D}}-P_{_{F}})\bigg\{1 + P_{_{D}}(1-P_{_{D}}) - 2P_{_{F}}^{2}(1-P_{_{F}}) - P_{_{D}}P_{_{F}} - P_{_{F}}^{2} \bigg\} \nonumber
\end{equation}
It should be noted that for this case $P_{_{D}} > P_{_{F}}$. Using the above expression we get:
\\
For $\displaystyle \{1 + P_{_{D}}(1-P_{_{D}}) - 2P_{_{F}}^{2}(1-P_{_{F}}) - P_{_{D}}P_{_{F}} - P_{_{F}}^{2}\} ~\geq ~0$
\begin{equation}
\displaystyle P_{e}(3,1,0,0) ~\geq~ P_{e}(2,1,1,0) \nonumber
\end{equation}
For $\displaystyle \{1 + P_{_{D}}(1-P_{_{D}}) - 2P_{_{F}}^{2}(1-P_{_{F}}) - P_{_{D}}P_{_{F}} - P_{_{F}}^{2}\} ~< ~0$
\begin{equation}
\displaystyle P_{e}(3,1,0,0) ~<~ P_{e}(2,1,1,0) \nonumber
\end{equation} 
\vspace{1cm}   
From these comparison equations we obtain 3 distinct subcases which are provided below:
\vspace{1cm}
\par
\textbf{Case CZ2.1:} $~\displaystyle 2(1-P_{_{F}})\geq P_{_{D}}, (P_{_{D}}+P_{_{F}}-1)^{2} \leq P_{_{D}}P_{_{F}}(1-P_{_{F}})$
\begin{equation}
\displaystyle P_{e}(2,1,1,0) \leq P_{e}(2,2,0,0), P_{e}(3,1,0,0) \nonumber
\end{equation}
\par
\textbf{Case CZ2.2:} $~\displaystyle 2(1-P_{_{F}})< P_{_{D}}, (P_{_{D}}+P_{_{F}}-1)^{2} \leq P_{_{D}}P_{_{F}}(1-P_{_{F}})$
\begin{equation}
\displaystyle P_{e}(2,2,0,0) \leq P_{e}(2,1,1,0), P_{e}(3,1,0,0) \nonumber
\end{equation}
\par
\textbf{Case CZ2.3:} $~\displaystyle 2(1-P_{_{F}})< P_{_{D}}, (P_{_{D}}+P_{_{F}}-1)^{2} > P_{_{D}}P_{_{F}}(1-P_{_{F}})$
\begin{equation}
\displaystyle P_{e}(3,1,0,0) \leq P_{e}(2,1,1,0), P_{e}(2,2,0,0) \nonumber
\end{equation}
\newpage
\textbf{Case DZ2:} $\displaystyle \bf P_{_{D}}^{3}(1-P_{_{D}}) < P_{_{F}}^{3}(1-P_{_{F}}), P_{_{D}}^{2}(1-P_{_{D}})^{2} < P_{_{F}}^{2}(1-P_{_{F}})^{2}, P_{_{D}}(1-P_{_{D}})^{3} < P_{_{F}}(1-P_{_{F}})^{3}, P_{_{D}}^{2}(1-P_{_{D}}) < P_{_{F}}^{2}(1-P_{_{F}}), P_{_{D}}(1-P_{_{D}})^{2} < P_{_{F}}(1-P_{_{F}})^{2}, P_{_{D}}(1-P_{_{D}}) < P_{_{F}}(1-P_{_{F}})$  
\vspace{1cm}
\par     
For this case the probability of error for the placements are given by equations (\ref{Pe2110C2}), (\ref{Pe2200C2}), (\ref{Pe3100CZ}), and (\ref{Pe4000CD}) respectively. Using equations (\ref{Pe4000CD}) and (\ref{Pe3100CZ}) we can write:
\begin{equation}
\displaystyle 4P_{e}(4,0,0,0) - 4P_{e}(3,1,0,0) = 3P_{_{F}}^{4} + 4P_{_{D}}^{3}(1-P_{_{D}}) + 8P_{_{F}}^{3}(1-P_{_{F}}) + 6P_{_{D}}^{2}(1-P_{_{D}})^{2} + 12P_{_{F}}^{2}(1-P_{_{F}})^{2} \nonumber
\end{equation}
\begin{equation}
\displaystyle  + 4P_{_{D}}(1-P_{_{D}})^{3} + 8P_{_{F}}(1-P_{_{F}})^{3} + (1-P_{_{D}})^{4} + 2(1-P_{_{F}})^{4} - P_{_{D}}P_{_{F}}^{3} - 2P_{_{F}}^{4} - P_{_{F}}^{3}(1-P_{_{D}}) \nonumber
\end{equation}
\begin{equation}
\displaystyle -2P_{_{F}}^{3}(1-P_{_{F}}) - 3P_{_{D}}^{2}P_{_{F}}(1-P_{_{D}}) - 6P_{_{F}}^{3}(1-P_{_{F}}) - 3P_{_{D}}^{2}(1-P_{_{D}})(1-P_{_{F}}) - 3P_{_{F}}^{2}(1-P_{_{D}})(1-P_{_{F}}) \nonumber
\end{equation}
\begin{equation}
\displaystyle -3P_{_{F}}^{2}(1-P_{_{F}})^{2} - 3P_{_{D}}P_{_{F}}(1-P_{_{D}})^{2} - 6P_{_{F}}^{2}(1-P_{_{F}})^{2} - 3P_{_{D}}(1-P_{_{F}})(1-P_{_{D}})^{2}-3P_{_{F}}(1-P_{_{D}})(1-P_{_{F}})^{2} \nonumber
\end{equation}
\begin{equation}
\displaystyle  - 3P_{_{F}}(1-P_{_{F}})^{3} - P_{_{F}}(1-P_{_{D}})^{3} - 2P_{_{F}}(1-P_{_{F}})^{3} - (1-P_{_{D}})^{3}(1-P_{_{F}}) -(1-P_{_{D}})(1-P_{_{F}})^{3} - (1-P_{_{F}})^{4}\nonumber
\end{equation}
Simplifying this expression we get:
\begin{equation}
\displaystyle 4P_{e}(4,0,0,0) - 4P_{e}(3,1,0,0) = -P_{_{D}}^{4} - P_{_{F}}^{3} + 2P_{_{F}}^{4} + P_{_{D}}-P_{_{F}} + P_{_{D}}^{3}-P_{_{D}}P_{_{F}}^{3} \nonumber
\end{equation}
For $\displaystyle ~\{-P_{_{D}}^{4}-P_{_{F}}^{3}+2P_{_{F}}^{4}+P_{_{D}}-P_{_{F}}+P_{_{D}}^{3}-P_{_{D}}P_{_{F}}^{3}\}~\geq~ 0$
\begin{equation}
\displaystyle P_{e}(4,0,0,0) \geq P_{e}(3,1,0,0) \nonumber
\end{equation}
For $\displaystyle ~\{-P_{_{D}}^{4}-P_{_{F}}^{3}+2P_{_{F}}^{4}+P_{_{D}}-P_{_{F}}+P_{_{D}}^{3}-P_{_{D}}P_{_{F}}^{3}\}~<~ 0$
\begin{equation}
\displaystyle P_{e}(4,0,0,0) < P_{e}(3,1,0,0) \nonumber
\end{equation} 
Using equations (\ref{Pe4000CD}) and (\ref{Pe2200C2}) we can write:
\begin{equation}
\displaystyle 4P_{e}(4,0,0,0) - 4P_{e}(2,2,0,0) = 3P_{_{F}}^{4} + 4P_{_{D}}^{3}(1-P_{_{D}}) + 8P_{_{F}}^{3}(1-P_{_{F}}) + 6P_{_{D}}^{2}(1-P_{_{D}})^{2} + 12P_{_{F}}^{2}(1-P_{_{F}})^{2} \nonumber
\end{equation}
\begin{equation}
\displaystyle  + 4P_{_{D}}(1-P_{_{D}})^{3} + 8P_{_{F}}(1-P_{_{F}})^{3} + (1-P_{_{D}})^{4} + 2(1-P_{_{F}})^{4} - P_{_{D}}^{2}P_{_{F}}^{2} - 2P_{_{F}}^{4}-4P_{_{D}}P_{_{F}}^{2}(1-P_{_{D}}) \nonumber
\end{equation}
\begin{equation}
\displaystyle  - 8P_{_{F}}^{3}(1-P_{_{F}}) - 2P_{_{F}}^{2}(1-P_{_{D}})^{2} - 4P_{_{F}}^{2}(1-P_{_{F}})^{2} - 8P_{_{D}}P_{_{F}}(1-P_{_{D}})(1-P_{_{F}})-4P_{_{F}}^{2}(1-P_{_{F}})^{2} \nonumber
\end{equation}
\begin{equation}
\displaystyle  - 4P_{_{F}}(1-P_{_{F}})(1-P_{_{D}})^{2} - 4P_{_{D}}(1-P_{_{D}})(1-P_{_{F}})^{2} - 4P_{_{F}}(1-P_{_{F}})^{3} -2(1-P_{_{D}})^{2}(1-P_{_{F}})^{2} - (1-P_{_{F}})^{4}\nonumber
\end{equation}
Simplifying this expression we get:
\begin{equation}
\displaystyle 4P_{e}(4,0,0,0) - 4P_{e}(2,2,0,0) = (P_{_{D}}^{2}-P_{_{F}}^{2})(2 - P_{_{D}}^{2} - 2P_{_{F}}^{2}) \nonumber
\end{equation}
For $\displaystyle (P_{_{D}}^{2}-P_{_{F}}^{2})(2 - P_{_{D}}^{2} - 2P_{_{F}}^{2}) \geq 0$
\begin{equation}
\displaystyle P_{e}(4,0,0,0) \geq P_{e}(2,2,0,0) \nonumber
\end{equation}
For $\displaystyle (P_{_{D}}^{2}-P_{_{F}}^{2})(2 - P_{_{D}}^{2} - 2P_{_{F}}^{2}) < 0$
\begin{equation}
\displaystyle P_{e}(4,0,0,0) < P_{e}(2,2,0,0) \nonumber
\end{equation}
Using equations (\ref{Pe2110C2}) and (\ref{Pe4000CD}) we can write:
\begin{equation}
\displaystyle 4P_{e}(4,0,0,0) - 4P_{e}(2,1,1,0) = 3P_{_{F}}^{4} + 4P_{_{D}}^{3}(1-P_{_{D}}) + 8P_{_{F}}^{3}(1-P_{_{F}}) + 6P_{_{D}}^{2}(1-P_{_{D}})^{2} \nonumber
\end{equation}
\begin{equation}
\displaystyle + 12P_{_{F}}^{2}(1-P_{_{F}})^{2} + 4P_{_{D}}(1-P_{_{D}})^{3} + 8P_{_{F}}(1-P_{_{F}})^{3} + (1-P_{_{D}})^{4} + 2(1-P_{_{F}})^{4} - 2P_{_{D}}P_{_{F}}^{3} -P_{_{F}}^{4} \nonumber
\end{equation}
\begin{equation}
\displaystyle  -2P_{_{D}}P_{_{F}}^{2}(1-P_{_{F}}) - 2P_{_{F}}^{3}(1-P_{_{D}}) - 2P_{_{F}}^{3}(1-P_{_{F}}) - 2P_{_{F}}^{2}(1-P_{_{D}})(1-P_{_{F}})-P_{_{F}}^{2}(1-P_{_{F}})^{2} - 2P_{_{D}}P_{_{F}}^{2}(1-P_{_{D}}) \nonumber
\end{equation}
\begin{equation}
\displaystyle  - 2P_{_{D}}P_{_{F}}^{2}(1-P_{_{F}}) - 2P_{_{F}}^{3}(1-P_{_{F}}) - 4P_{_{D}}P_{_{F}}(1-P_{_{D}})(1-P_{_{F}})-4P_{_{F}}^{2}(1-P_{_{D}})(1-P_{_{F}})- 4P_{_{F}}^{2}(1-P_{_{F}})^{2} \nonumber
\end{equation}
\begin{equation}
\displaystyle   - 2P_{_{D}}(1-P_{_{D}})(1-P_{_{F}})^{2} - 4P_{_{F}}(1-P_{_{D}})(1-P_{_{F}})^{2} -P_{_{F}}^{2}(1-P_{_{D}})^{2}- P_{_{D}}P_{_{F}}(1-P_{_{F}})^{2}- P_{_{F}}^{2}(1-P_{_{F}})^{2} \nonumber
\end{equation}
\begin{equation}
\displaystyle   - 2P_{_{F}}(1-P_{_{F}})(1-P_{_{D}})^{2} - 2P_{_{F}}(1-P_{_{D}})(1-P_{_{F}})^{2} -2P_{_{F}}(1-P_{_{F}})^{3} -(1-P_{_{D}})^{2}(1-P_{_{F}})^{2}- 2(1-P_{_{D}})(1-P_{_{F}})^{3}\nonumber
\end{equation}
This expression can be simplified to give:
\begin{equation}
\displaystyle  4P_{e}(4,0,0,0) - 4P_{e}(2,1,1,0) = 2(P_{_{D}}-P_{_{F}}) + P_{_{D}}(P_{_{D}}-P_{_{F}}) - P_{_{D}}(P_{_{D}}^{3}-P_{_{F}}^{3}) - 2P_{_{F}}^{2}(P_{_{D}}-P_{_{F}})  \nonumber
\end{equation}
For: $\displaystyle 2(P_{_{D}}-P_{_{F}}) + P_{_{D}}(P_{_{D}}-P_{_{F}}) - P_{_{D}}(P_{_{D}}^{3}-P_{_{F}}^{3}) - 2P_{_{F}}^{2}(P_{_{D}}-P_{_{F}}) \geq 0$
\begin{equation}
\displaystyle P_{e}(4,0,0,0) \geq P_{e}(2,1,1,0) \nonumber
\end{equation}
For: $\displaystyle 2(P_{_{D}}-P_{_{F}}) + P_{_{D}}(P_{_{D}}-P_{_{F}}) - P_{_{D}}(P_{_{D}}^{3}-P_{_{F}}^{3}) - 2P_{_{F}}^{2}(P_{_{D}}-P_{_{F}}) < 0$
\begin{equation}
\displaystyle P_{e}(4,0,0,0) < P_{e}(2,1,1,0) \nonumber
\end{equation} 
Using equations (\ref{Pe2200C2}) and (\ref{Pe3100CZ}) we can write:
\begin{equation}
\displaystyle 4P_{e}(3,1,0,0) - 4P_{e}(2,2,0,0) = P_{_{D}}P_{_{F}}^{3} + 2P_{_{F}}^{4} + P_{_{F}}^{3}(1-P_{_{D}}) + 2P_{_{F}}^{3}(1-P_{_{F}}) + 3P_{_{D}}^{2}P_{_{F}}(1-P_{_{D}}) \nonumber
\end{equation}
\begin{equation}
\displaystyle + 6P_{_{F}}^{3}(1-P_{_{F}}) + 3P_{_{D}}^{2}(1-P_{_{D}})(1-P_{_{F}}) + 3P_{_{F}}^{2}(1-P_{_{D}})(1-P_{_{F}}) + 3P_{_{F}}^{2}(1-P_{_{F}})^{2} + 3P_{_{D}}P_{_{F}}(1-P_{_{D}})^{2} \nonumber
\end{equation}
\begin{equation}
\displaystyle +6P_{_{F}}^{2}(1-P_{_{F}})^{2} + 3P_{_{D}}(1-P_{_{F}})(1-P_{_{D}})^{2} + 3P_{_{F}}(1-P_{_{D}})(1-P_{_{F}})^{2} + 3P_{_{F}}(1-P_{_{F}})^{3} + P_{_{F}}(1-P_{_{D}})^{3} \nonumber
\end{equation}
\begin{equation}
\displaystyle +2P_{_{F}}(1-P_{_{F}})^{3} + (1-P_{_{D}})^{3}(1-P_{_{F}}) + (1-P_{_{D}})(1-P_{_{F}})^{3} + (1-P_{_{F}})^{4} - P_{_{D}}^{2}P_{_{F}}^{2} - 2P_{_{F}}^{4}-4P_{_{D}}P_{_{F}}^{2}(1-P_{_{D}}) \nonumber
\end{equation}
\begin{equation}
\displaystyle  - 8P_{_{F}}^{3}(1-P_{_{F}}) - 2P_{_{F}}^{2}(1-P_{_{D}})^{2} - 4P_{_{F}}^{2}(1-P_{_{F}})^{2} - 4P_{_{F}}(1-P_{_{F}})^{3}- 8P_{_{D}}P_{_{F}}(1-P_{_{D}})(1-P_{_{F}}) -4P_{_{F}}^{2}(1-P_{_{F}})^{2}\nonumber
\end{equation}
\begin{equation}
\displaystyle -4P_{_{F}}(1-P_{_{F}})(1-P_{_{D}})^{2} - 4P_{_{D}}(1-P_{_{D}})(1-P_{_{F}})^{2}  - 2(1-P_{_{D}})^{2}(1-P_{_{F}})^{2} - (1-P_{_{F}})^{4} \nonumber
\end{equation}
Simplifying this epxression we get:
\begin{equation}
\displaystyle 4P_{e}(3,1,0,0) - 4P_{e}(2,2,0,0) = 2(P_{_{D}}^{2}-P_{_{F}}^{2}) - (P_{_{D}}-P_{_{F}}) - (P_{_{D}}^{3}-P_{_{F}}^{3}) - P_{_{D}}P_{_{F}}^{2}(P_{_{D}}-P_{_{F}}) \nonumber
\end{equation}
For: $\displaystyle 2(P_{_{D}}^{2}-P_{_{F}}^{2}) - (P_{_{D}}-P_{_{F}}) - (P_{_{D}}^{3}-P_{_{F}}^{3}) - P_{_{D}}P_{_{F}}^{2}(P_{_{D}}-P_{_{F}}) \geq 0$
\begin{equation}
\displaystyle P_{e}(3,1,0,0) \geq P_{e}(2,2,0,0) \nonumber
\end{equation}
For: $\displaystyle 2(P_{_{D}}^{2}-P_{_{F}}^{2}) - (P_{_{D}}-P_{_{F}}) - (P_{_{D}}^{3}-P_{_{F}}^{3}) - P_{_{D}}P_{_{F}}^{2}(P_{_{D}}-P_{_{F}}) < 0$
\begin{equation}
\displaystyle P_{e}(3,1,0,0) < P_{e}(2,2,0,0) \nonumber
\end{equation}
Using equations (\ref{Pe2110C2}) and (\ref{Pe3100CZ}) we can write:
\begin{equation}
\displaystyle 4P_{e}(3,1,0,0) - 4P_{e}(2,1,1,0) = P_{_{D}}P_{_{F}}^{3} + 2P_{_{F}}^{4} + P_{_{F}}^{3}(1-P_{_{D}}) + 2P_{_{F}}^{3}(1-P_{_{F}}) + 3P_{_{D}}^{2}P_{_{F}}(1-P_{_{D}}) +  \nonumber
\end{equation}
\begin{equation}
\displaystyle 6P_{_{F}}^{3}(1-P_{_{F}}) + 3P_{_{D}}^{2}(1-P_{_{D}})(1-P_{_{F}}) + 3P_{_{F}}^{2}(1-P_{_{D}})(1-P_{_{F}}) + 3P_{_{F}}^{2}(1-P_{_{F}})^{2} + 3P_{_{D}}P_{_{F}}(1-P_{_{D}})^{2} \nonumber
\end{equation}
\begin{equation}
\displaystyle +6P_{_{F}}^{2}(1-P_{_{F}})^{2} + 3P_{_{D}}(1-P_{_{F}})(1-P_{_{D}})^{2} + 3P_{_{F}}(1-P_{_{D}})(1-P_{_{F}})^{2} + 3P_{_{F}}(1-P_{_{F}})^{3} + P_{_{F}}(1-P_{_{D}})^{3} \nonumber
\end{equation}
\begin{equation}
\displaystyle + 2P_{_{F}}(1-P_{_{F}})^{3} + (1-P_{_{D}})^{3}(1-P_{_{F}}) + (1-P_{_{D}})(1-P_{_{F}})^{3} + (1-P_{_{F}})^{4} -2P_{_{D}}P_{_{F}}^{3} - P_{_{F}}^{4} - 2P_{_{D}}P_{_{F}}^{2}(1-P_{_{F}}) \nonumber
\end{equation}
\begin{equation}
\displaystyle -2P_{_{F}}^{3}(1-P_{_{D}}) - 2P_{_{F}}^{3}(1-P_{_{F}}) - 2P_{_{F}}^{2}(1-P_{_{D}})(1-P_{_{F}}) - P_{_{F}}^{2}(1-P_{_{F}})^{2} - 2P_{_{D}}P_{_{F}}^{2}(1-P_{_{D}}) - 2P_{_{D}}P_{_{F}}^{2}(1-P_{_{F}}) \nonumber
\end{equation}
\begin{equation}
\displaystyle -2P_{_{F}}^{3}(1-P_{_{F}}) - 4P_{_{D}}P_{_{F}}(1-P_{_{D}})(1-P_{_{F}}) - 4P_{_{F}}^{2}(1-P_{_{D}})(1-P_{_{F}}) - 4P_{_{F}}^{2}(1-P_{_{F}})^{2} - 2P_{_{D}}(1-P_{_{D}})(1-P_{_{F}})^{2} \nonumber
\end{equation}
\begin{equation}
\displaystyle -4P_{_{F}}(1-P_{_{D}})(1-P_{_{F}})^{2} - P_{_{F}}^{2}(1-P_{_{D}})^{2} - P_{_{D}}P_{_{F}}(1-P_{_{F}})^{2} - P_{_{F}}^{2}(1-P_{_{F}})^{2} - 2P_{_{F}}(1-P_{_{D}})(1-P_{_{F}})^{2} \nonumber
\end{equation}
\begin{equation}
\displaystyle -2P_{_{F}}(1-P_{_{F}})^{3} - (1-P_{_{D}})^{2}(1-P_{_{F}})^{2} - 2(1-P_{_{D}})(1-P_{_{F}})^{3} \nonumber
\end{equation}
Simplifying this expression we get:
\begin{equation}
\displaystyle 4P_{e}(3,1,0,0) - 4P_{e}(2,1,1,0) = 2P_{_{F}}^{3}(P_{_{D}}-P_{_{F}}) - (P_{_{D}}^{3}-P_{_{F}}^{3}) - 2P_{_{F}}^{2}(P_{_{D}}-P_{_{F}}) + (P_{_{D}}-P_{_{F}}) + P_{_{D}}(P_{_{D}}-P_{_{F}}) \nonumber
\end{equation}
For: $\displaystyle 2P_{_{F}}^{3}(P_{_{D}}-P_{_{F}}) - (P_{_{D}}^{3}-P_{_{F}}^{3}) - 2P_{_{F}}^{2}(P_{_{D}}-P_{_{F}}) + (P_{_{D}}-P_{_{F}}) + P_{_{D}}(P_{_{D}}-P_{_{F}}) \geq 0$
\begin{equation}
\displaystyle P_{e}(3,1,0,0) \geq P_{e}(2,1,1,0) \nonumber
\end{equation}
For: $\displaystyle 2P_{_{F}}^{3}(P_{_{D}}-P_{_{F}}) - (P_{_{D}}^{3}-P_{_{F}}^{3}) - 2P_{_{F}}^{2}(P_{_{D}}-P_{_{F}}) + (P_{_{D}}-P_{_{F}}) + P_{_{D}}(P_{_{D}}-P_{_{F}}) < 0$
\begin{equation}
\displaystyle P_{e}(3,1,0,0) < P_{e}(2,1,1,0) \nonumber
\end{equation} 
Using equations (\ref{Pe2110C2}) and (\ref{Pe2200C2}) we can write:
\begin{equation}
\displaystyle 4P_{e}(2,2,0,0) - 4P_{e}(2,1,1,0) = P_{_{D}}^{2}P_{_{F}}^{2} + 2P_{_{F}}^{4} + 4P_{_{D}}P_{_{F}}^{2}(1-P_{_{D}}) + 8P_{_{F}}^{3}(1-P_{_{F}}) + 2P_{_{F}}^{2}(1-P_{_{D}})^{2} \nonumber
\end{equation}
\begin{equation}
\displaystyle +4P_{_{F}}^{2}(1-P_{_{F}})^{2} + 8P_{_{D}}P_{_{F}}(1-P_{_{D}})(1-P_{_{F}}) + 4P_{_{F}}^{2}(1-P_{_{F}})^{2} + 4P_{_{F}}(1-P_{_{F}})(1-P_{_{D}})^{2} + 4P_{_{D}}(1-P_{_{D}})(1-P_{_{F}})^{2} \nonumber
\end{equation}
\begin{equation}
\displaystyle  + 4P_{_{F}}(1-P_{_{F}})^{3} + 2(1-P_{_{D}})^{2}(1-P_{_{F}})^{2} + (1-P_{_{F}})^{4} - 2P_{_{D}}P_{_{F}}^{3} - P_{_{F}}^{4} - 2P_{_{D}}P_{_{F}}^{2}(1-P_{_{F}}) - 2P_{_{F}}^{3}(1-P_{_{D}}) \nonumber
\end{equation}
\begin{equation}
\displaystyle -2P_{_{F}}^{3}(1-P_{_{F}}) - 2P_{_{F}}^{2}(1-P_{_{D}})(1-P_{_{F}}) - P_{_{F}}^{2}(1-P_{_{F}})^{2} - 2P_{_{D}}P_{_{F}}^{2}(1-P_{_{D}}) - 2P_{_{D}}P_{_{F}}^{2}(1-P_{_{F}}) - 2P_{_{F}}^{3}(1-P_{_{F}}) \nonumber
\end{equation}
\begin{equation}
\displaystyle -4P_{_{D}}P_{_{F}}(1-P_{_{D}})(1-P_{_{F}}) - 4P_{_{F}}^{2}(1-P_{_{D}})(1-P_{_{F}}) - 4P_{_{F}}^{2}(1-P_{_{F}})^{2} - 2P_{_{D}}(1-P_{_{D}})(1-P_{_{F}})^{2} - 4P_{_{F}}(1-P_{_{D}})(1-P_{_{F}})^{2} \nonumber
\end{equation}
\begin{equation}
\displaystyle -P_{_{F}}^{2}(1-P_{_{D}})^{2} - P_{_{D}}P_{_{F}}(1-P_{_{F}})^{2} - P_{_{F}}^{2}(1-P_{_{F}})^{2} - 2P_{_{F}}(1-P_{_{F}})(1-P_{_{D}})^{2} - 2P_{_{F}}(1-P_{_{D}})(1-P_{_{F}})^{2} - 2P_{_{F}}(1-P_{_{F}})^{3} \nonumber
\end{equation}
\begin{equation}
\displaystyle -(1-P_{_{D}})^{2}(1-P_{_{F}})^{2} - 2(1-P_{_{D}})(1-P_{_{F}})^{3} \nonumber
\end{equation}
Simplifying this expression we get:
\begin{equation}
\displaystyle 4P_{e}(2,2,0,0) - 4P_{e}(2,1,1,0) = -2P_{_{F}}^{4} - P_{_{D}}^{2} + 2P_{_{F}}^{2} + P_{_{D}}^{2}P_{_{F}}^{2} - 2P_{_{F}} + P_{_{D}}P_{_{F}}^{3} - 2P_{_{D}}P_{_{F}}^{2} \nonumber
\end{equation}
\begin{equation}
\displaystyle -P_{_{D}}P_{_{F}} + 2P_{_{F}}^{3} + 2P_{_{D}}= 2(P_{_{D}}-P_{_{F}}) + 2P_{_{F}}^{3}(1-P_{_{F}}) - P_{_{D}}P_{_{F}}^{2}(1-P_{_{F}})   \nonumber
\end{equation}
\begin{equation}
\displaystyle  -P_{_{D}}P_{_{F}}^{2}(1-P_{_{D}}) - (P_{_{D}}^{2}-P_{_{F}}^{2}) - P_{_{F}}(P_{_{D}}-P_{_{F}}) \nonumber
\end{equation}
For: $\displaystyle  2(P_{_{D}}-P_{_{F}}) + 2P_{_{F}}^{3}(1-P_{_{F}}) - P_{_{D}}P_{_{F}}^{2}(1-P_{_{F}})-P_{_{D}}P_{_{F}}^{2}(1-P_{_{D}}) - (P_{_{D}}^{2}-P_{_{F}}^{2}) - P_{_{F}}(P_{_{D}}-P_{_{F}}) ~\geq~ 0$
\begin{equation}
\displaystyle P_{e}(2,2,0,0) \geq P_{e}(2,1,1,0) \nonumber
\end{equation}
For: $\displaystyle 2(P_{_{D}}-P_{_{F}}) + 2P_{_{F}}^{3}(1-P_{_{F}}) - P_{_{D}}P_{_{F}}^{2}(1-P_{_{F}})-P_{_{D}}P_{_{F}}^{2}(1-P_{_{D}}) - (P_{_{D}}^{2}-P_{_{F}}^{2}) - P_{_{F}}(P_{_{D}}-P_{_{F}}) ~\geq~ 0$
\begin{equation}
\displaystyle P_{e}(2,2,0,0) < P_{e}(2,1,1,0) \nonumber
\end{equation}

Using the results from Cases AW1, BW1, BX1, CX1, CY2, CZ2, and DZ2 we obtain the following optimality regions on the $(P_{_{F}},P_{_{D}})$ plane for $P_{D}>P_{F}$:
\par
$(4,0,0,0)$ is the Optimal Placement if:
\begin{equation}
\displaystyle \bigg[\bigg\{-P_{_{D}}^{4}-P_{_{F}}^{3} + 2P_{_{F}}^{4} + P_{_{D}}-P_{_{F}} + P_{_{D}}^{3} - P_{_{D}}P_{_{F}}^{3} < 0\bigg\}\cap \bigg\{(P_{_{D}}^{2}-P_{_{F}}^{2})(2-P_{_{D}}^{2}-2P_{_{F}}^{2}) < 0\bigg\}\cap \bigg\{ 2(P_{_{D}}-P_{_{F}}) +  \nonumber
\end{equation}
\begin{equation}
\displaystyle  + P_{_{D}}(P_{_{D}}-P_{_{F}}) - P_{_{D}}(P_{_{D}}^{3}-P_{_{F}}^{3}) - 2P_{_{F}}^{2}(P_{_{D}}-P_{_{F}}) < 0\bigg\}\cap\bigg\{ P_{_{D}}^{3}(1-P_{_{D}}) < P_{_{F}}^{3}(1-P_{_{F}})\bigg\}\bigg] \nonumber
\end{equation}
\par
$(3,1,0,0)$ is the Optimal Placement if: 
\begin{equation}
\displaystyle \bigg[\bigg\{-P_{_{D}}^{4} - P_{_{F}}^{3} + 2P_{_{F}}^{4} + P_{_{D}}-P_{_{F}} + P_{_{D}}^{3} - P_{_{D}}P_{_{F}}^{3} \geq 0\bigg\}\cap \bigg\{ 2(P_{_{D}}^{2}-P_{_{F}}^{2}) - (P_{_{D}}-P_{_{F}}) - (P_{_{D}}^{3}-P_{_{F}}^{3}) \nonumber
\end{equation}
\begin{equation}
\displaystyle - P_{_{D}}P_{_{F}}^{2}(P_{_{D}}-P_{_{F}}) < 0\bigg\} \cap \bigg\{ 2P_{_{F}}^{3}(P_{_{D}}-P_{_{F}}) - (P_{_{D}}^{3}-P_{_{F}}^{3}) - 2P_{_{F}}^{2}(P_{_{D}}-P_{_{F}}) + (P_{_{D}}-P_{_{F}}) + P_{_{D}}(P_{_{D}}-P_{_{F}}) < 0\bigg\} \nonumber
\end{equation}
\begin{equation}
\displaystyle \cap\bigg\{ P_{_{D}}^{3}(1-P_{_{D}}) < P_{_{F}}^{3}(1-P_{_{F}})\bigg\}\bigg]\bigcup \bigg[\bigg\{ 2(1-P_{_{F}}) < P_{_{D}}\bigg\}\cap \bigg\{(P_{_{D}}+P_{_{F}}-1)^{2} > P_{_{D}}P_{_{F}}(1-P_{_{F}}) \bigg\} \nonumber
\end{equation}
\begin{equation}
\displaystyle \cap \bigg\{P_{_{D}}^{3}(1-P_{_{D}}) \geq P_{_{F}}^{3}(1-P_{_{F}})\bigg\}\cap \bigg\{P_{_{D}}^{2}(1-P_{_{D}}) < P_{_{F}}^{2}(1-P_{_{F}}) \bigg\} \bigg] \hspace{5cm}\nonumber
\end{equation}
\par
$(2,2,0,0)$ is the Optimal Placement if:
\begin{equation}
\displaystyle \bigg[\bigg\{(P_{_{D}}^{2}-P_{_{F}}^{2})(2-P_{_{D}}^{2}-2P_{_{F}}^{2})\bigg\}\cap\bigg\{2(P_{_{D}}^{2}-P_{_{F}}^{2}) - (P_{_{D}}-P_{_{F}}) - (P_{_{D}}^{3}-P_{_{F}}^{3}) - P_{_{D}}P_{_{F}}^{2}(P_{_{D}}-P_{_{F}}) \geq 0\bigg\} \nonumber
\end{equation}
\begin{equation}
\displaystyle \bigg\{2(P_{_{D}}-P_{_{F}}) + 2P_{_{F}}^{3}(1-P_{_{F}}) - P_{_{D}}P_{_{F}}^{2}(1-P_{_{F}}) - P_{_{D}}P_{_{F}}^{2}(1-P_{_{D}}) - (P_{_{D}}^{2}-P_{_{F}}^{2}) - P_{_{F}}(P_{_{D}}-P_{_{F}}) < 0\bigg\} \nonumber \end{equation}
\begin{equation}
\displaystyle \cap \bigg\{P_{_{D}}^{3}(1-P_{_{D}}) < P_{_{F}}^{3}(1-P_{_{F}})\bigg\}\bigg]\bigcup \bigg[\bigg\{2(1-P_{_{F}}) < P_{_{D}}\bigg\}\cap\bigg\{ (P_{_{D}}+P_{_{F}}-1)^{2} \leq P_{_{D}}P_{_{F}}(1-P_{_{F}})\bigg\}\nonumber 
\end{equation}
\begin{equation}
\displaystyle \cap\bigg\{P_{_{D}}^{3}(1-P_{_{D}}) \geq P_{_{F}}^{3}(1-P_{_{F}})\bigg\}\cap\bigg\{P_{_{D}}^{2}(1-P_{_{D}}) < P_{_{F}}^{2}(1-P_{_{F}}) \bigg\}\bigg] \nonumber
\end{equation}
\par
$(2,1,1,0)$ is the Optimal Placement if:
\begin{equation}
\displaystyle \bigg[P_{_{D}}(1-P_{_{D}}) \geq P_{_{F}}(1-P_{_{F}})\bigg]\bigcup\bigg[\bigg\{P_{_{D}}(1-P_{_{D}}) < P_{_{F}}(1-P_{_{F}})\bigg\}\cap\bigg\{P_{_{D}}^{2}(1-P_{_{D}}) \geq P_{_{F}}^{2}(1-P_{_{F}})\bigg\}\bigg] \nonumber
\end{equation}
\begin{equation}
\displaystyle \bigcup \bigg[\bigg\{ 2(1-P_{_{F}}) \geq P_{_{D}}\bigg\} \cap \bigg\{ (P_{_{D}}+P_{_{F}}-1)^{2} \leq P_{_{D}}P_{_{F}}(1-P_{_{F}}) \bigg\} \cap \bigg\{P_{_{D}}^{3}(1-P_{_{D}}) \geq P_{_{F}}^{3}(1-P_{_{F}})\bigg\} \nonumber
\end{equation}
\begin{equation}
\displaystyle \bigg\{P_{_{D}}^{2}(1-P_{_{D}}) < P_{_{F}}^{2}(1-P_{_{F}})\bigg\}\bigg]\bigcup\bigg[\bigg\{2(P_{_{D}}-P_{_{F}}) + P_{_{D}}(P_{_{D}}-P_{_{F}}) - P_{_{D}}(P_{_{D}}^{3}-P_{_{F}}^{3}) - 2P_{_{F}}^{2}(P_{_{D}}-P_{_{F}}) \geq 0\bigg\} \nonumber
\end{equation}
\begin{equation}
\displaystyle \bigg\{2P_{_{F}}^{3}(P_{_{D}}-P_{_{F}}) - (P_{_{D}}^{3}-P_{_{F}}^{3}) - 2P_{_{F}}^{2}(P_{_{D}}-P_{_{F}}) + (P_{_{D}}-P_{_{F}}) + P_{_{D}}(P_{_{D}}-P_{_{F}}) \geq 0\bigg\}\cap\bigg\{ 2(P_{_{D}}-P_{_{F}}) + 2P_{_{F}}^{3}(1-P_{_{F}}) \nonumber 
\end{equation}
\begin{equation}
\displaystyle -P_{_{D}}P_{_{F}}^{2}(1-P_{_{F}}) - P_{_{D}}P_{_{F}}^{2}(1-P_{_{D}}) - (P_{_{D}}^{2}-P_{_{F}}^{2}) - P_{_{F}}(P_{_{D}}-P_{_{F}}) \geq 0\bigg\}\cap\bigg\{P_{_{D}}^{3}(1-P_{_{D}}) < P_{_{F}}^{3}(1-P_{_{F}}) \bigg\}\bigg]\nonumber
\end{equation}
The aforementioned regions can be simplified to get the following more concise description of these regions:
\par
$(4,0,0,0)$ is the optimal placement if:
\begin{equation}
\label{4Opt}
\displaystyle \bigg[(P_{_{D}}-P_{_{F}})\big[-(P_{_{D}}+P_{_{F}})(P_{_{D}}^{2} + P_{_{F}}^{2}) + (P_{_{D}}^{2} + P_{_{D}}P_{_{F}} + P_{_{F}}^{2})  + (1-P_{_{F}}^{3})\big] < 0 \bigg]
\end{equation}
\par
$(3,1,0,0)$ is the optimal placement if:
\begin{equation}
\displaystyle \bigg[\bigg\{(P_{_{D}}-P_{_{F}})\big[-(P_{_{D}}+P_{_{F}})(P_{_{D}}^{2} + P_{_{F}}^{2}) + (P_{_{D}}^{2} + P_{_{D}}P_{_{F}} +    P_{_{F}}^{2})+ (1-P_{_{F}}^{3})\big] \geq 0\bigg\}\cap\bigg\{2(P_{_{D}}^{2}-P_{_{F}}^{2}) - (P_{_{D}}-P_{_{F}}) -  \nonumber 
\end{equation}
\begin{equation}
\displaystyle (P_{_{D}}^{3}-P_{_{F}}^{3})- P_{_{D}}P_{_{F}}^{2}(P_{_{D}}-P_{_{F}}) < 0 \bigg\}\cap\bigg\{ P_{_{D}}^{3}(1-P_{_{D}}) <  P_{_{F}}^{3}(1-P_{_{F}})\bigg\}\bigg] \bigcup\bigg[\bigg\{(P_{_{D}}+P_{_{F}}-1)^{2} > P_{_{D}}P_{_{F}}(1-P_{_{F}})\bigg\} \nonumber 
\end{equation}
\begin{equation}
\label{3Opt}
\displaystyle  \cap\bigg\{P_{_{D}}^{3}(1-P_{_{D}}) \geq P_{_{F}}^{3}(1-P_{_{F}})\bigg\}\cap\bigg\{P_{_{D}}^{2}(1-P_{_{D}}) <  P_{_{F}}^{2}(1-P_{_{F}})\bigg\}\bigg] \hspace{4cm}
\end{equation}
\par
$(2,2,0,0)$ is the optimal placement if:
\begin{equation}
\displaystyle \bigg[\bigg\{(P_{_{D}}^{2}-P_{_{F}}^{2})(2-P_{_{D}}^{2}-2P_{_{F}}^{2}) \geq 0)\bigg\}\cap \bigg\{2(P_{_{D}}^{2}-P_{_{F}}^{2}) - (P_{_{D}}-P_{_{F}}) -(P_{_{D}}^{3}-P_{_{F}}^{3})-P_{_{D}}P_{_{F}}^{2}(P_{_{D}}-P_{_{F}}) \geq 0 \bigg\} \nonumber \end{equation}
\begin{equation}
\displaystyle \cap\bigg\{2(P_{_{D}}-P_{_{F}}) +2P_{_{F}}^{3}(1-P_{_{F}})-P_{_{D}}P_{_{F}}^{2}(1-P_{_{F}}) - P_{_{D}}P_{_{F}}^{2}(1-P_{_{D}}) - (P_{_{D}}^{2}-P_{_{F}}^{2})-P_{_{F}}(P_{_{D}}-P_{_{F}}) \leq 0\bigg\} \nonumber
\end{equation}
\begin{equation}
\displaystyle \cap\bigg\{P_{_{D}}^{3}(1-P_{_{D}}) <P_{_{F}}^{3}(1-P_{_{F}})\bigg\}\bigg]\bigcup\bigg[\bigg\{2(1-P_{_{F}})<P_{_{D}}\bigg\} \cap\bigg\{(P_{_{D}}+P_{_{F}}-1)^{2} \leq P_{_{D}}P_{_{F}}(1-P_{_{F}})\bigg\}\nonumber
\end{equation}
\begin{equation}
\label{22Opt}
\displaystyle \cap\bigg\{P_{_{D}}^{3}(1-P_{_{D}}) \geq P_{_{F}}^{3}(1-P_{_{F}})\bigg\}\bigg] \hspace{8cm}
\end{equation}
\par
$(2,1,1,0)$ is the optimal placement if:
\begin{equation}
\label{21Opt}
\displaystyle \bigg[2(1-P_{_{F}}) \geq P_{_{D}}\bigg]
\end{equation}
\par 
For $P_{_{D}}=P_{_{F}}$, all the placements have the same probability of error and therefore we choose either the highest placement in the majorization ordering to be the optimal placement on this boundary or select other higher placements which preserve the statement of this proposition. Using such a selection policy at the boundary $P_{_{D}}=P_{_{F}}$ and equations (\ref{4Opt}), (\ref{3Opt}), (\ref{22Opt}), and (\ref{21Opt}) we observe that by fixing $P_{_{D}}$ in the interval $[0,\frac{2}{3}]$ and by increasing $P_{_{F}}$, from 0 to $P_{_{D}}$, $(2,1,1,0)$ will be the optimal placement. For $P_{_{D}}$ fixed in the interval $(\frac{2}{3},\frac{373}{539}]$ and by increasing $P_{_{F}}$ from $0$ to $P_{_{D}}$, the optimal placement will start as the $(2,1,1,0)$ placement and will then switch over to the $(2,2,0,0)$ placement. For $P_{_{D}}$ fixed in the interval $(\frac{373}{539},\frac{947}{1093}]$ and by increasing $P_{_{F}}$ from $0$ to $P_{_{D}}$, the optimal placement will start as the $(2,1,1,0)$ placement, switch over to the $(2,2,0,0)$ placement, and will end up as the $(3,1,0,0)$ placement. Finally for $P_{_{D}}$ fixed in the interval $(\frac{947}{1093},1]$ and for $P_{_{F}}$ increased from 0 to $P_{_{D}}$, the optimal placement will start as the $(2,1,1,0)$ placement, then switch over to the $(2,2,0,0)$ placement first and then to the $(3,1,0,0)$ placement, and finally end up as the $(4,0,0,0)$ placement. Similarly it can be shown, from equations (\ref{4Opt}), (\ref{3Opt}), (\ref{22Opt}), and (\ref{21Opt}), that for fixed $P_{_{F}}$ and increasing $P_{_{D}}$ the optimal placement occurs nondecreasingly on a majorization based placement scale. 
We present a plot of (\ref{4Opt}), (\ref{3Opt}), (\ref{22Opt}), and (\ref{21Opt}) on the $(P_{_{F}},P_{_{D}})$ plane in Section VI which further validates the statement of this proposition for the case $M=N=4$. Other cases where $M \leq 5$ can be proved similarly by using equation (\ref{Pewf}).   
\end{proof}

\par
\textit{For $M>5$ the aforementioned result does not necessarily hold}. Consider the case where $(M,N)=(7,8)$. Clearly $(3,2,1,1,0,0,0) \succ (2,2,2,1,0,0,0)$, so $(3,2,1,1,0,0,0)$ is at a higher level than $(2,2,2,1,0,0,0)$ on the majorization based placement scale for this problem. Using equation (\ref{Pewf}) we get $(3,2,1,1,0,0,0)$ as the strictly optimal placement for $(P_{_{F}}, P_{_{D}}) = (0.46,0.6)$ whereas $(2,2,2,1,0,0,0)$ is the strictly optimal placement for $(P_{_{F}}, P_{_{D}}) = (0.48,0.6)$. Therefore, for fixed $P_{_{D}}$ and increasing $P_{_{F}}$ the optimal placement does not necessarily occur nondecreasingly on a majorization based placement scale for $M > 5$. Also for $(P_{_{F}},P_{_{D}}) = (0.48,0.5)$ the placement $(3,2,1,1,0,0,0)$ is strictly optimal whereas $(2,2,2,1,0,0,0)$ is the strictly optimal placement for $(P_{_{F}}, P_{_{D}}) = (0.48,0.6)$. Therefore, we can conclude that for fixed $P_{_{F}}$ and increasing $P_{_{D}}$ the optimal placement does not necessarily occur nondecreasingly on a majorization based placement scale for $M > 5$. This counterexample shows that the statement of the proposition is not necessarily valid for $M>5$.

\newpage
\section{Simulations}
\par
\textbf{\textit{Example 6.1:}} $\displaystyle \bf M=3$
\par
Let $(M,N)=(3,3)$, $P_{_{D}}>P_{_{F}}$. A $(P_{_{F}},P_{_{D}})$ plane using a set of optimal placements is given in Fig.1.\\
\begin{figure}[h]
\centering
\includegraphics[trim=0cm 1.25cm 0cm 1.5cm, clip=true, width=1\textwidth]{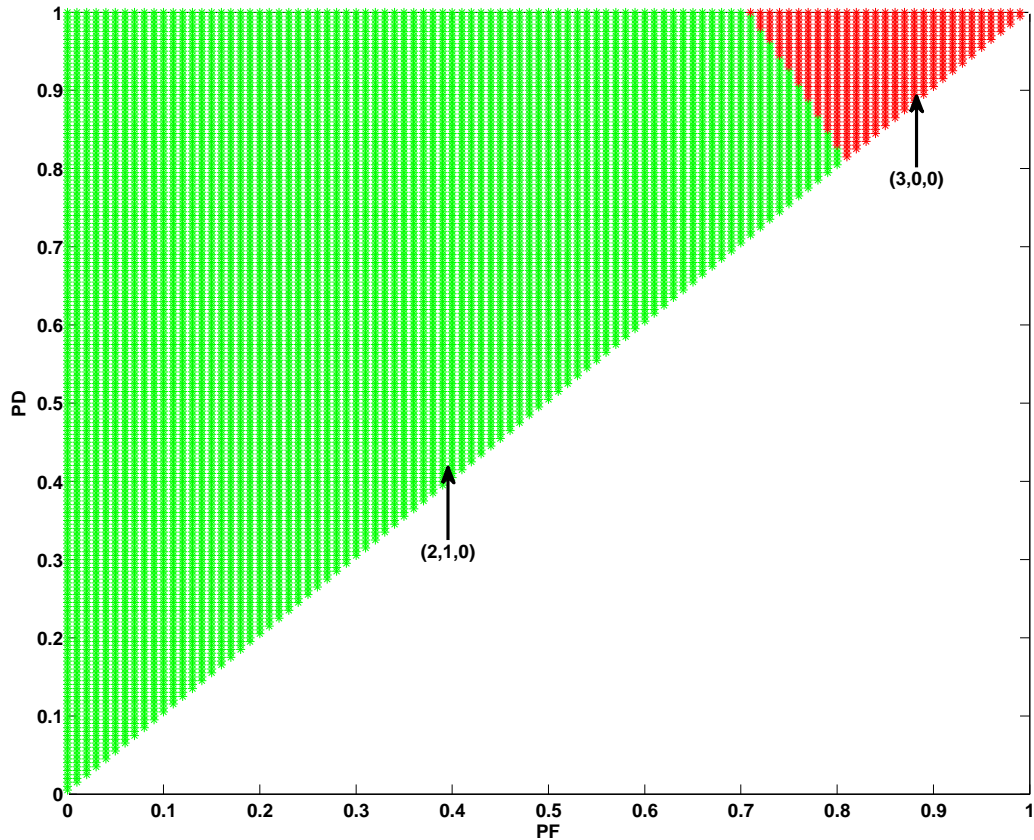}
\caption{Optimal placement structure over $(P_{_{F}},P_{_{D}})$ for $(M,N)=(3,3)$}
\end{figure}
The arrows are used to represent the regions in which a specific placement is optimal. In Fig. 1 we observe that uniform placement is not strictly optimal as specified by Theorem 3.1. By fixing $P_{_{D}}$ and increasing $P_{_{F}}$, we observe that the optimal placement occurs nondecreasingly on the majorization defined scale :$(3,0,0) \succ (2,1,0) \succ (1,1,1)$. Also by fixing $P_{F}$ and increasing $P_{D}$, we observe that the optimal placement occurs nondecreasingly on the majorization defined scale given by (\ref{mscale}).  
\par
Now consider $(M,N)=(3,4)$, $P_{_{D}}>P_{_{F}}$. A $(P_{_{F}},P_{_{D}})$ plane using a set of optimal placements is given in Fig. 2. 
\begin{figure}[h]
\centering
\includegraphics[trim=0cm 1.25cm 0cm 1.5cm, clip=true, width=1\textwidth]{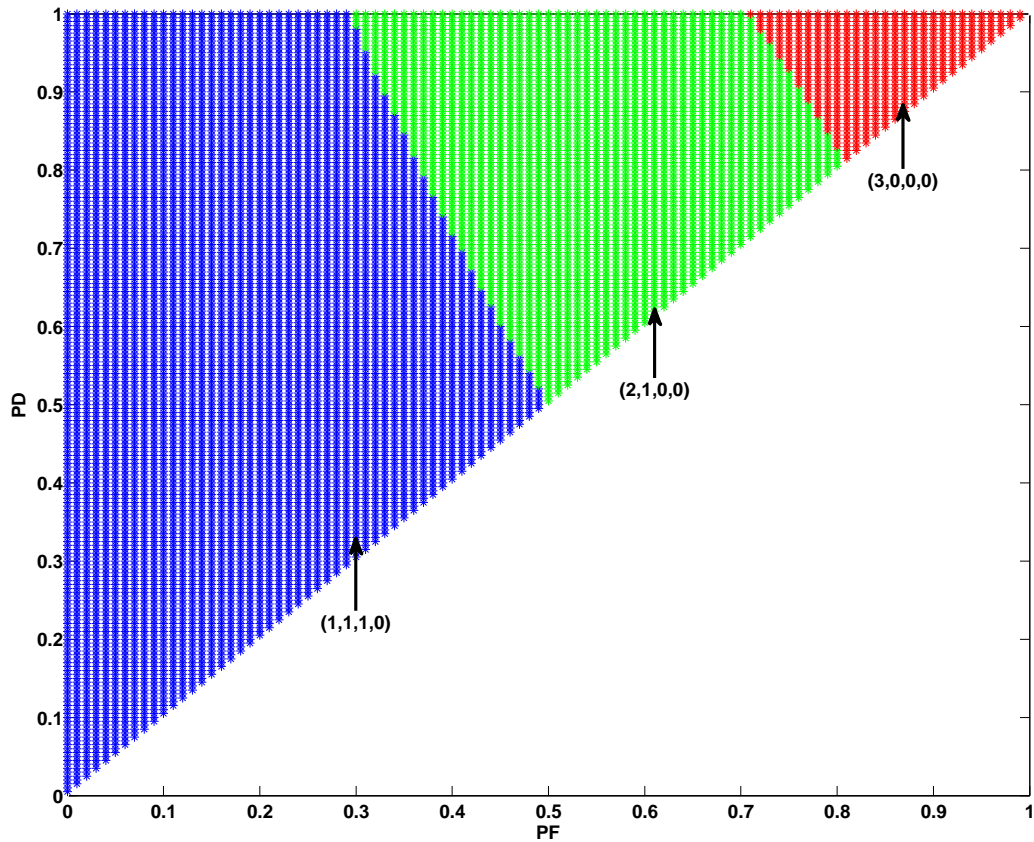}
\caption{Optimal placement structure over $(P_{_{F}},P_{_{D}})$ for $(M,N)=(3,4)$}
\end{figure}
In Fig. 2, we observe that uniform placement is strictly optimal along with the set of optimal placements for the case $(M,N)=(3,3)$. This was previously prescribed by Corollary 4.1 and these simulations provide a practical validation. As in the case $(M,N) = (3,3)$ the result of Proposition 5.1 holds for this case and can easily be verified using Fig. 2. 
\newpage
\par
\textbf{\textit{Example 6.2:}} $\displaystyle \bf M=4$
\par
Let $(M,N)=(4,4)$, $P_{_{D}}>P_{_{F}}$. A $(P_{_{F}},P_{_{D}})$ plane using a set of optimal placements is given in Fig.3. 
\begin{figure}[h]
\centering
\includegraphics[trim=0cm 1.25cm 0cm 1.5cm, clip=true, width=1\textwidth]{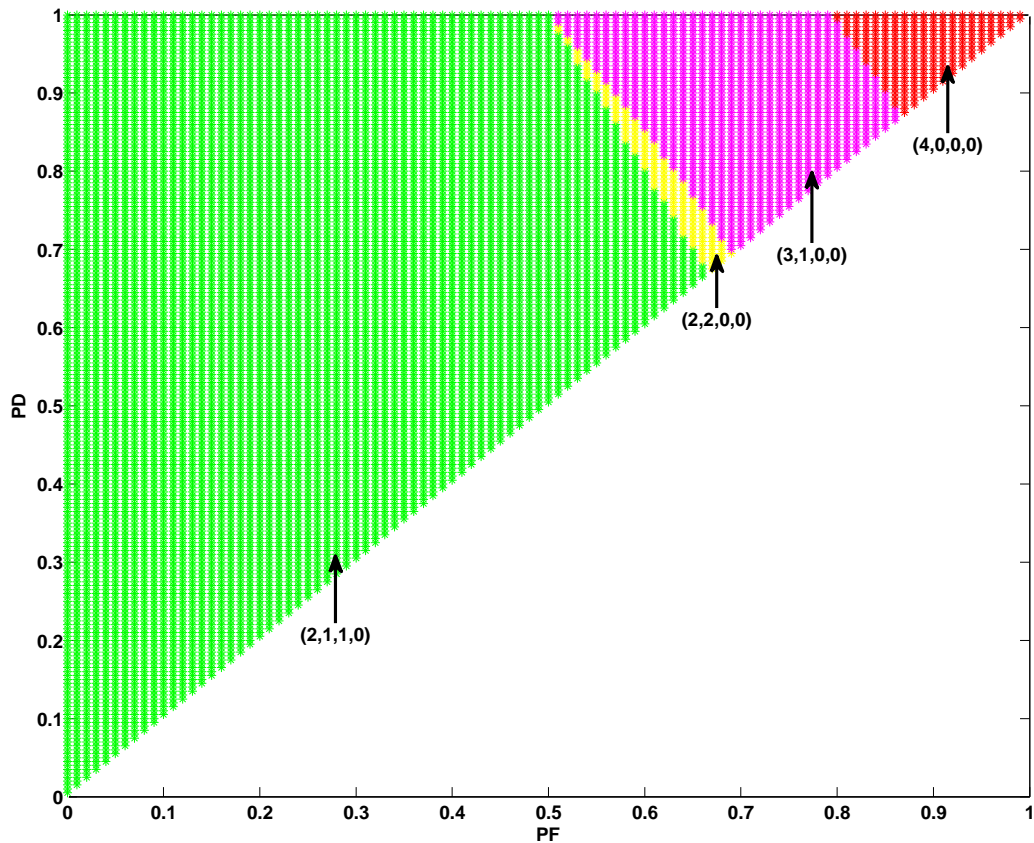}
\caption{Optimal placement structure over $(P_{_{F}},P_{_{D}})$ for $(M,N)=(4,4)$}
\end{figure}
The arrows are used to represent the regions in which a specific placement is optimal. In Fig. 3 we observe that uniform placement is not strictly optimal. By fixing $P_{_{D}}$ and increasing $P_{_{F}}$, we observe that the optimal placement occurs nondecreasingly on the majorization defined scale given by (\ref{mscale}). Also by fixing $P_{F}$ and increasing $P_{D}$, we observe that the optimal placement occurs nondecreasingly on the majorization defined scale given by (\ref{mscale}).  
\par
Now consider $(M,N)=(4,5)$, $P_{_{D}}>P_{_{F}}$. A $(P_{_{F}},P_{_{D}})$ plane using a set of optimal placements is given in Fig. 4.  
\begin{figure}[h]
\centering
\includegraphics[trim=0cm 1.25cm 0cm 1.5cm, clip=true, width=1\textwidth]{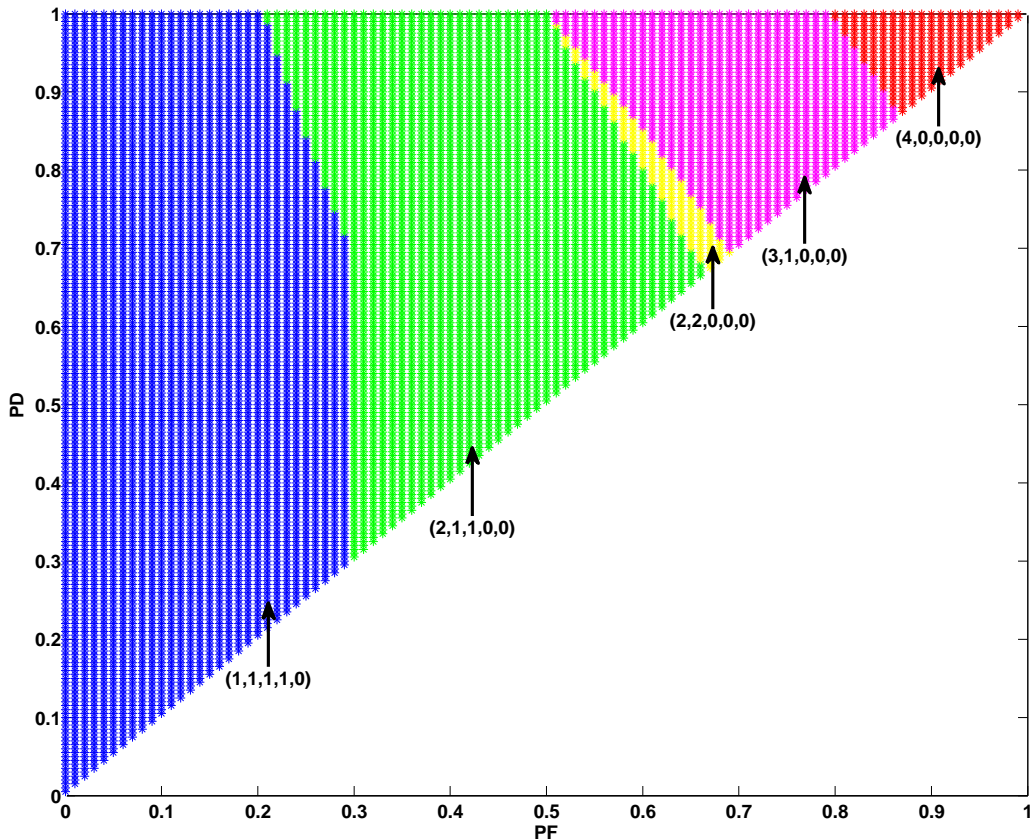}
\caption{Optimal placement structure over $(P_{_{F}},P_{_{D}})$ for $(M,N)=(4,5)$}
\end{figure}
In Fig. 4, we observe that uniform placement is strictly optimal along with the set of optimal placements for the case $(M,N)=(4,4)$. This was previously prescribed by Corollary 4.1 and these simulations provide a practical validation. As in the case $(M,N) = (4,4)$ the result of Proposition 5.1 holds for this case and can easily be verified using Fig. 4.    
\newpage
\par
\textbf{\textit{Example 6.3:}} $\displaystyle \bf M=5$
\par
Consider the case $(M,N)=(5,6)$, $P_{_{D}}>P_{_{F}}$. A $(P_{_{F}},P_{_{D}})$ plane using a set of optimal placements for this case is given in Fig. 5. 
\begin{figure}[h]
\centering
\includegraphics[trim=0cm 1.25cm 0cm 1.5cm, clip=true, width=1\textwidth]{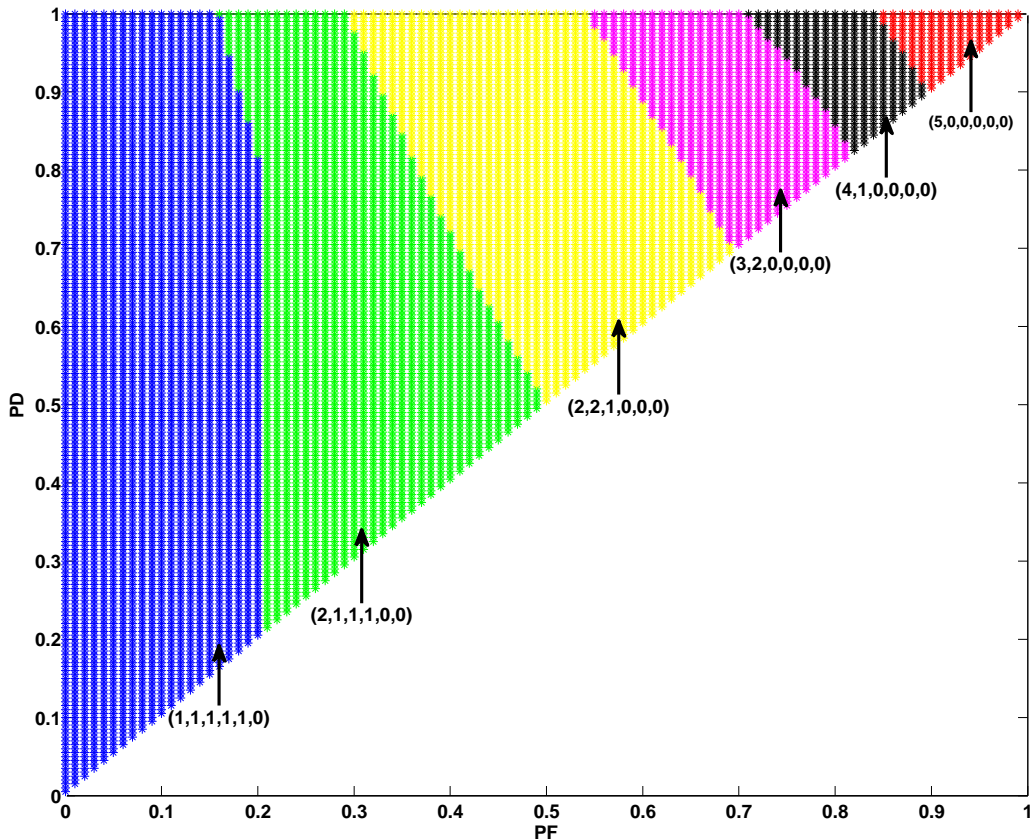}
\caption{Optimal placement structure over $(P_{_{F}},P_{_{D}})$ for $(M,N)=(5,6)$}
\end{figure}
In Fig. 5, by fixing either $P_{_{D}}$ or $P_{_{F}}$ and increasing the other parameter, we observe that the optimal placement occurs nondecreasingly on the majorization defined placement scale $(5,0)\succ (4,1) \succ (3,2) \succ (2,2,1) \succ (2,1,1,1) \succ (1,1,1,1,1)$. These simulations verify the statements and signifiy the practical importance of the Theorems as outlined in Section IV and Section V. 

\newpage
\section{Conclusions}
We have characterized several placement principles for sensors tasked with detecting the location of a randomly placed intruder. In particular, we have used notions from Majorization Theory, such as majorization-based placement scales, to formalize important sensor placement properties. The uniform placement of sensors has been thoroughly analyzed, culminating in an important result on the strict optimality of a uniform placement. In addition, changes in the optimal placement structure due to a variation in the number of points have been analyzed and related design principles have been presented. 


\end{document}